\RequirePackage{amsthm} 

\documentclass[sn-basic,iicol,pdflatex]{sn-jnl}

\usepackage[utf8]{inputenc}
\usepackage[T1]{fontenc}
\usepackage{float}
\usepackage{amsmath,amssymb,amsfonts}
\usepackage{mathrsfs}
\usepackage{theorem}
\usepackage{graphicx}
\usepackage{color}
\usepackage{wasysym}
\usepackage{hyperref} 
\usepackage{natbib}

\usepackage{mathtools}
\usepackage{dcolumn}
\usepackage{multirow}

\usepackage[title]{appendix}%
\usepackage{xcolor}%
\usepackage{textcomp}%
\usepackage{manyfoot}%
\usepackage{booktabs}%
\usepackage[switch]{lineno} 
\usepackage{datetime2}

\usepackage[finalnew]{trackchanges} 

\begin{document}

\newcommand{\ama}[1]{{\color{ama} #1}}
\newcommand{\vio}[1]{{\color{vio} #1}}
\newcommand{\blue}[1]{{\color{blue} #1}}
\newcommand{\green}[1]{{\color{green} #1}}
\newcommand{\red}[1]{{\color{red} #1}}
\newcommand{\marron}[1]{{\color{marron} #1}}
\newcommand{\verde}[1]{{\color{verde} #1}}
\newcommand{\celeston}[1]{{\color{celeston} #1}}
\definecolor{verde}{rgb}{0.,.5,0.4}
\definecolor{blue}{rgb}{0,0,1}
\definecolor{green}{rgb}{0.,0.65,0.25}
\definecolor{celeston}{rgb}{0.,0.65,0.75}
\definecolor{marron}{rgb}{0.7,0.2,0.1}
\definecolor{red}{rgb}{1,0,0}
\definecolor{vio}{rgb}{0.66,0,1}
\definecolor{ama}{rgb}{1,1,0}


\title[Sky localization and ...]{\sf Sky localization and polarization mode reconstruction of
	
	gravitational waves from GW170104 and GW150914
}

\author[1,2]{\fnm{Osvaldo M.} \sur{Moreschi}
	\href{https://orcid.org/0000-0001-9753-3820}{\includegraphics[scale=0.4]{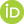}} }\email{o.moreschi@unc.edu.ar}

%

\affil[1]{\orgdiv{Facultad de Matemática, Astronomía, Física y Computación(FaMAF)}, 
	\orgname{Universidad Nacional de Córdoba}, 
	\orgaddress{\street{Ciudad Universitaria}, \city{Córdoba}, \postcode{(5000)}, 
		\country{Argentina}}}

\affil[2]{\orgdiv{Instituto de Física Enrique Gaviola(IFEG)}, 
	\orgname{CONICET}, 
	\orgaddress{\street{Ciudad Universitaria}, \city{Córdoba}, \postcode{(5000)}, 
		\country{Argentina}}}

\abstract{

The detections\citep{KAGRA:2023pio,LIGOScientific:2019lzm} and analysis of gravitational waves(GWs) 
have introduced us in a new era of our
understanding of the cosmos, providing new insights 
into astrophysical systems involving massive objects
as black holes and neutron stars.
Normally the precise sky localization of a GW source needs 
data from three or more observatories\citep{LIGOScientific:2016lio,Abbott:2017oio}.
However, the results presented in this article demonstrate that it is in fact
possible to obtain the position of a GW source in a small region of the celestial sphere
using data from just two GW observatories,
in this case LIGO Hanford and LIGO Livingston.
Furthermore, we are also able to reconstruct the gravitational-wave
polarization\citep{Poisson2014} modes(PMs) for the GW170104\citep{Abbott:2017vtc}
and GW150914\citep{TheLIGOScientific:2016qqj} events,
with data 
from only these two detectors. 
The procedure only uses the spin 2 properties of the GW,
so that it does not rely on specific assumptions on the nature of the source.
Our findings are possible through 
careful data filtering methods\citep{Moreschi:2019vxw},
the use of refined signal processing algorithms\citep{Moreschi:2024njx}, 
and the application of dedicated denoising\citep{Mallat2009} techniques.
This progress in the GW studies represents the first 
instance of a direct measurement of PMs using such a limited 
observational data.
We provide detailed validation through 
the reconstruction of PMs for different polarization angles,
and calculations of residuals for the GW170104 event.
We also test the procedure with synthetic data with ten different source locations
and polarization angles.

	
}


\maketitle

\date{\DTMnow} 

\newcommand{\linea}{\noindent\rule{\linewidth}{0.5mm} \,}


\tableofcontents


\section{Introduction}\label{sec:Introduction}

The localization of transient GW sources on the celestial sphere 
involves several significant challenges, coming from both the nature of the 
signals and the characteristics of current observational 
infrastructure\citep{KAGRA:2013rdx,Singer:2015ema,Veitch:2014wba}.
The sensitivity of GW detectors is affected by various sources 
of noise, including seismic activity, thermal fluctuations, and instrumental 
noise. These noises can hide the signals and complicate the process of 
determining the source location. For this reason, in our method, it is essential to handle
the strains with appropriate pre-processing filtering techniques\citep{Moreschi:2019vxw}.
Instead, for the detection problem, the Ligo-Virgo-Kagra(LVK) teams have elaborated a
series of methods including matched filtering techniques\citep{Allen:2005fk} 
and unmodeled methods\citep{Klimenko:2011hz,Klimenko:2015ypf}
that are used for LVK detection and localization
without employing the just mentioned pre-processing filtering techniques.
The accuracy of 
localizing a GW source improves with the number of detectors and their 
geographical spread. Currently, the primary detectors are the two LIGO, Hanford and Livingston,
the Virgo and the KAGRA observatories.
Recording signals on three or more observatories helps considerably in
determining the source location, since one can recur to 
the intersection of allowed sky regions,
by using the time delay of arrival among the observatories;
diminishing in this way the probable localization area on the celestial sphere.
Other issues that are involved in the precision to locate the sources
are the relative signal amplitude calibration and the phase consistency
between detectors\citep{KAGRA:2013rdx}. 
The amount of information that one can obtain from a GW observation
increases considerably when it can be related to multimessenger observations.
For this reason some efforts\citep{Chatterjee:2022ggk,Kolmus:2021buf,You:2021eeq,Hu:2021nvy,Tsutsui:2020sml,Singer:2015ema} 
are directed to optimize the
time required to indicate the localization of GW sources.
In general, all these techniques involve the Bayesian statistical framework
that require the choice of priors for the nature of the source;
nevertheless, the cWB pipeline\citep{Klimenko:2011hz,Klimenko:2015ypf}
uses a model-independent approach which is able to provide sky
localization based on coherent signal power.
The approach described below, does not use specific templates
neither we use information on individual masses,
but rather we use a universal chirp form for the inspiral phase, 
and so in this sense it is model independent.

The expression of a GW in terms of its 
polarization modes\citep{Poisson2014}
is one of the most
fundamental aspects involved in the detection process and its analysis.
Consequently it is the subject of constant discussion in the literature.
In particular in reference \cite{Eardley:1973br} and \cite{Eardley:1974nw} they treated 
PMs for the specific case of plane waves.
Some of the efforts in the literature have concentrated on the separability of the polarization modes 
with a limited number of gravitational-wave detectors\citep{Takeda:2018uai}.
In \cite{Abbott:2018utx} they have carried out a
search for a stochastic background of generically polarized
gravitational waves; where they found no evidence for a background of any polarization, placing
bounds on the contributions of vector and scalar polarizations to the stochastic background.
In \cite{Gursel:1989} the authors presented
a theoretical method for determining the source direction in the sky and the two polarization modes
of a GW that require to use the data from three detectors.
This approach has been extended in several directions\citep{Chatterji:2006nh}.
We improve on these types of approaches, since we here introduce a 
procedure for the sky localization
with two detectors and the gravitational-wave polarization modes reconstruction(L2D+PMR)
of the signal.
We thus provide first explicit measurements of the spin-2 polarization modes of a
gravitational wave.

This article is dedicated to the presentation of the procedure L2D+PMR,
and it is applied to the GW170104 and GW150914 events as a case studies.
We will devote future work to extract physical information from these events.

The GW170104 event has strong enough signals and low enough noise
which we can use to develop and apply
our procedure, and 
allows us to extract the spin-2 polarization modes of the detected GW.
Actually the reconstruction of the PMs can only be achieved if one can also localize
the direction in the sky of the source of the GW; it is for this reason
that both problems must be solved synchronously.
We do not mean here that one should carryout the localization and 
the reconstruction of PMs as a joint probability
distribution; instead we do the localization first by using partial properties
of the polarization modes and the delay time information.
In addition, in order to show how robust and effective is our method,
we also apply it to reconstruct the signal, the polarization modes
and to find the localization of the source for a simple injected signal,
starting from the corresponding synthetic spin-2 polarization modes
with 10 chosen positions for the hypothetic source and with different polarization 
frames for each of them.

We also apply the L2D+PMR procedure to the first GW detection, GW150914. 
This event was initially studied during 
the early stages of our work using a different measure, 
which presented several difficulties. 
To overcome these, we extended the analysis to later events, such as GW170104 
and others involving three detectors\citep{Moreschi:2025zde}. 
This expansion helped us to evolve and validate the method. 
The developed version now produces significant results for GW150914, 
which are presented below. 
Due to space constraints, we do not include a comprehensive analysis of GW150914 in this work; 
a detailed examination of this event's physical characteristics will be addressed in future research.

We here advance a sketch of the L2D+PMR procedure that is going to be presented
and applied along the article.

\vspace{2mm}
\noindent
{\sf The L2D+PMR procedure:}
\begin{itemize}
\item {\sf Selection of Strains:}
Choose the strains with the least amount of prior filtering for the event under study.

\item {\sf Pre-Processing Filtering:}
Apply pre-processing filtering techniques\citep{Moreschi:2019vxw}.

\item {\sf Window Selection:}
Select a window length $wl$ for the detailed study of the signals.

\item {\sf Time delay:}
Determine the appropriate working time delay between the observatories from a careful
study of the signal.(See section \ref{sec:timedelay}.)

\item {\sf Chirp times:}
Found the characteristic chirp times from a study of the signals
in the time-frequency domain.

\item {\sf Bandpass filter:} 
\add{Apply an appropriate bandpass filter to the strains data 
following time-frequency domain analysis.}

\item {\sf Denoising:}
Denoise each strain using wavelets techniques (Section \ref{sec:denoising}).

\item {\sf Source Localization:}
Determine the localization of the source using data from two observatories, 
as detailed in this article. 
If more than two observatories provide data for the event, 
iterate this procedure and calculate the final localization.

\item {\sf Polarization Mode Reconstruction:}
Reconstruct the polarization modes of the gravitational wave
using the algebra we present.

\end{itemize}

For the organization of this article, we present our results
and essential structure in the main text,
and relegate to the appendices a series of detailed explanations of some points
mentioned in the text.
In section \ref{sec:dec-pol-modes} we present the basic algebra used in the decomposition
of the signals in terms of polarization modes.
Section \ref{sec:denoising} is devoted to the presentation of the denoising of the signals.
In section \ref{sec:time-freq} we show the study of the GW in the time-frequency
domain.
We describe how to fit a universal chirp form to a signal in section \ref{sec:univ_fitt_chirp}.
To validate our methods, we study in section \ref{sec:synth} the localization
of ten sources and the corresponding reconstruction of the spin-2 polarization modes.
The result of the sky localization of the source of the GW for event GW170104
is presented in \ref{sec:localiz}.
In section \ref{sec:PM} we present explicitly, as time series, the polarization modes
of the GW of the GW170104 event.
We show that these spin-2 polarization modes account for the complete reconstruction of
the gravitational wave in section \ref{sec:signal-in-terms}; that is we measure no contribution
coming from polarization modes with spin 1 o 0 to GW170104.
The study of the nominal time shift of strain H with respect to L
for GW150914 is presented in 
section \ref{sec:nomianl-time-shift-GW150914}.
Section \ref{sec:time-freq_gw150914} is devoted to the analysis of GW150914 in the time-frequency domain.
In section \ref{sec:localizGW150914} we present the result of the localization of GW150914.
The reconstruction of the + and $\times$ polarization modes of GW150914 is presented in 
section \ref{sec:PM-GW150914}.
In section \ref{sec:final} we present some final comments.
And, as said above, in several appendices we present some details of topics
discussed in the main part. 

\section{Decomposition of a gravitational wave in terms of polarization modes}\label{sec:dec-pol-modes}

Having a detection of a GW in two observatories, in our case $H$(Hanford) and $L$(Livingston),
one has in each of them the recorded strain
\begin{equation}\label{eq:vX}
\begin{split}
v_X(t + \tau_X) &= n_X(t + \tau_X) + s_X(t + \tau_X) \\
&= n_X(t + \tau_X) + 
F_{+X}(\theta_X,\phi_X,\psi_X,t) s_+(t) \\
&\;\;\; +
F_{\times X}(\theta_X,\phi_X,\psi_X,t) s_\times(t)
,
\end{split}
\end{equation}
where $X$ stands for $H$ or $L$, $\tau_X$ is the time delay of detector $X$ with respect to
the chosen reference observatory, 
$(\theta_X,\phi_X)$ are the angular coordinates with respect
to detector $X$ of the direction of the source, $\psi_X$ is the angle of the
GW frame and $t$ is the time.
The strain is denoted by $v$, we use $n$ to refer to the noise, $s$ for the signal,
which is decomposed in the PMs $s_+$ and $s_\times$;
while $F_+$ and $F_\times$ are the detector pattern functions discussed in appendix \ref{sec:patt-func}.

Being a little bit more generic, let us denote with $D$ and $E$ two detectors
where one has recorded the signals of a GW.
Let $(\delta,\alpha)$ be an arbitrary direction in the celestial sphere,
and $(\delta_0,\alpha_0)$ the location of the source. 
Then, we define
{\small 
\begin{equation}\label{eq:v+DE}
\begin{split}
v_{+DE}(\delta,\alpha,t) &=
v_D(t + \tau_D) F_{\times E}(\delta,\alpha,t) \\
&\;\;\;-
v_E(t + \tau_E) F_{\times D}(\delta,\alpha,t) \\
&=
n_{\times DE}
+
( F_{\times E}F_{+D0} - F_{\times D}  F_{+E0} ) s_+(t) \\
&\;\;\;+
( F_{\times E}F_{\times D0} - F_{\times D}  F_{\times E0} ) s_\times(t)
,
\end{split}
\end{equation}
}where we are using the notation
$F_{\times D}\equiv F_{\times D}(\delta,\alpha,t)=
F_{\times D}( \theta_D(\delta,\alpha),\phi_D(\delta,\alpha),\psi_D(\delta,\alpha),t)$
and
$F_{\times D0}\equiv F_{\times D}(\delta_0,\alpha_0,t)$;
where we are considering $(\delta,\alpha)\neq (\delta_0,\alpha_0)$.
Then, 
when evaluated at the source position $(\delta_0,\alpha_0)$ one would obtain
a strain of the form:
{\small 
\begin{equation}\label{eq:v+DE_2}
\begin{split}
v_{+DE}(\delta_0,\alpha_0,t) &=
F_{\times E}(\delta_0,\alpha_0 ,t)
\\
&\quad
( n_D(t + \tau_D) + F_{+D}(\delta_0,\alpha_0,t) s_+(t) ) \\
& \quad -
F_{\times D}(\delta_0,\alpha_0,t)
\\
&\quad
( n_E(t + \tau_E) +
F_{+E}(\delta_0,\alpha_0,t) s_+(t) ) \\
&=
n_{\times DE}(t) + 
( F_{\times E0} F_{+D0} - F_{\times D0}  F_{+E0} ) s_+(t)\\
&=
n_{\times DE}(t) + f_{+0} s_+(t)
,
\end{split}
\end{equation}
}without the cross contribution,
and where 
$f_{+0}=F_{\times E0} F_{+D0} - F_{\times D0}  F_{+E0}$ and
{\small
\begin{equation}\label{eq:nxDE}
\begin{split}
n_{\times DE}(\delta,\alpha,t) =&
F_{\times E}(\delta,\alpha ,t)
\times  n_D(t + \tau_D) \\
&-
F_{\times D}(\delta,\alpha,t)
\times  n_E(t + \tau_E)
.
\end{split}
\end{equation}
}
In an analogous way we can select the $\times$ mode.
We define
{\small
\begin{equation}\label{eq:vxDE}
\begin{split}
v_{\times DE}(\delta,\alpha,t) &=
v_D(t + \tau_D)  F_{+ E}(\delta,\alpha,t) \\
&\;\;\;-
v_E(t + \tau_E)  F_{+ D}(\delta,\alpha,t) \\
&=
n_{+ DE}
+
( F_{+ E}F_{+D0} - F_{+ D}  F_{+E0} ) s_+(t) \\
&\;\;\;+
( F_{+ E}F_{\times D0} - F_{+ D}  F_{\times E0} ) s_\times(t)
,
\end{split}
\end{equation}
}where we have used the notation
$F_{+ D}(\delta,\alpha,t)=
F_{+ D}( \theta_D(\delta,\alpha),\phi_D(\delta,\alpha),\psi_D(\delta,\alpha),t)$.
Then, when evaluated at the source position $(\delta_0,\alpha_0)$ one would obtain
an strain of the form:
\vspace{1mm}
{\small
	\begin{equation}\label{eq:vxDE_2}
	\begin{split}
	v_{\times DE}(\delta_0,\alpha_0,t) &=
	n_{+DE}(t) + ( F_{+ E0}F_{\times D0} - F_{+ D0}  F_{\times E0} ) s_\times (t)\\
	&= n_{+DE}(t) + f_{\times0} s_\times(t) = n_{DE}(t) - f_{+0}  s_\times (t)
	,
	\end{split}
	\end{equation}
}without the plus contribution,
and where
{\small
\begin{equation}\label{eq:n+DE}
\begin{split}
n_{+ DE}(\delta,\alpha,t) =&
F_{+ E}(\delta,\alpha ,t)
\times  n_D(t + \tau_D) \\
&-
F_{+ D}(\delta,\alpha,t)
\times  n_E(t + \tau_E)
,
\end{split}
\end{equation}
}and where we have noted that
$f_{\times0}=-f_{+0}$.
It should be noted that all manipulations to obtain the polarization modes
come from the relations expressed in \eqref{eq:vX};
however, the approach we use through our definitions \eqref{eq:v+DE} and \eqref{eq:vxDE}
is very different from the inverse problem approach 
described in \cite{Gursel:1989} where they assume data
is recorded in three observatories.

\begin{figure}[H]
\centering
\includegraphics[clip,width=0.47\textwidth]{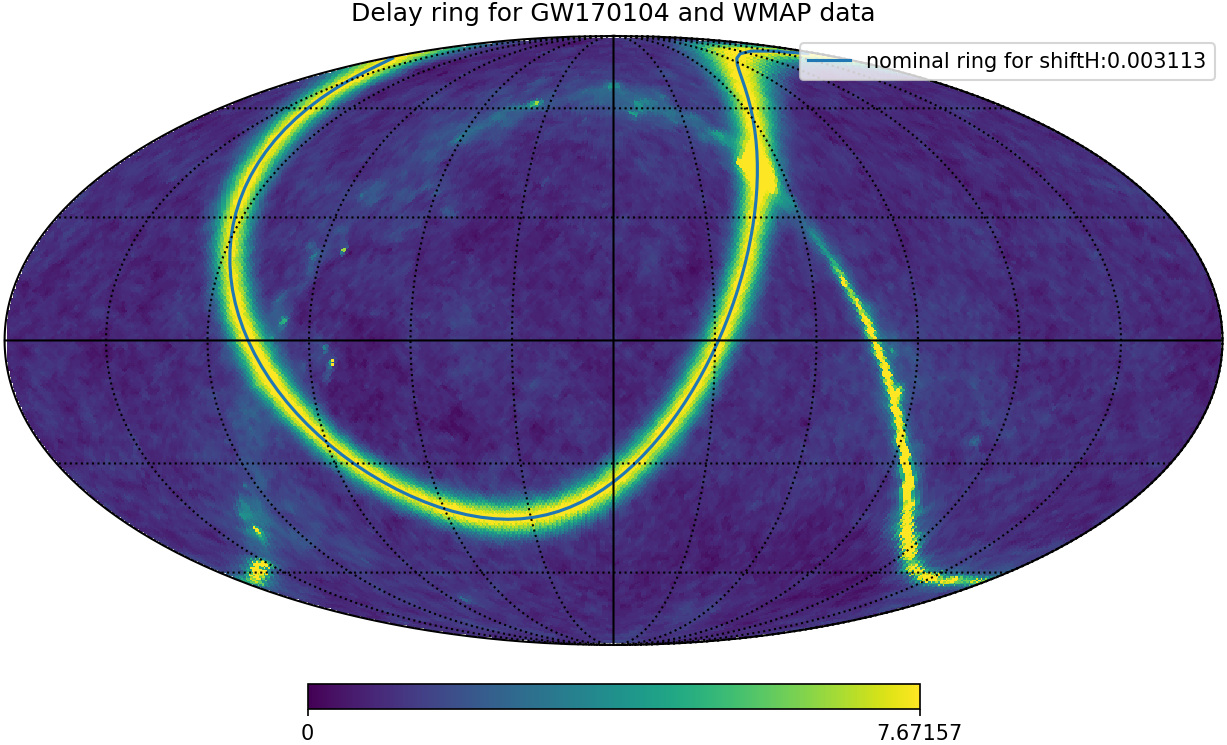}
\caption{Estimate of the probability distribution for the localization of the
	gravitational-wave source, considering the information encoded in 
	the measurement of the delay ring.
	The background with the WMAP data is included to shown the nature
	of the Mollweide projection being used; in equatorial coordinates
	with origin at center and East to the left.
}
\label{fig:delay-ring+gaussian+Wmap}
\end{figure}

Associated to a time delay between the observatories,
which is described in appendix \ref{sec:timedelay};
there is a `delay ring' in the sky showing the theoretical
possible locations of the source.
The determination of the delay ring involves an intrinsic error
which comes from the error in the time delay value.
In appendix \ref{sec:timedelay}
we discuss to possible origins for errors.
This in turn generates an uncertainty on the determination of the angle for the ring.
With this we can construct a Gaussian distribution around the delay ring
with the intention of giving a representation of the error
in the ring determination, as shown in Fig.
\ref{fig:delay-ring+gaussian+Wmap}.
\noindent
The Gaussian region of the delay ring represents an estimate of the probability distribution
for the sky localization of the source of the GW
based on time delay considerations.
	When an independent localization technique, such as the L2D+PMR procedure, 
is applied, one would expect to find the estimated source position near or 
within this Gaussian region. 
As we will demonstrate in sections \ref{sec:localiz} and \ref{sec:localizGW150914}, 
this is precisely what occurs 
when we apply the L2D+PMR method to two real events.
In Fig. \ref{fig:delay-ring+gaussian+Wmap} we also include the image of our Galaxy, through
a WMAP graph, in order to clarify the nature of the celestial coordinates and the location
of its origin.
We discuss further the time delay ring in appendix \ref{sec:timedelay}.

\section{Estimates of the signal by wavelet denoising of the strain}\label{sec:denoising}

Previously we have been working with the observed strains directly.
This has the advantage of dealing with filtered observed data,
but it has de disadvantage that it is difficult to estimate the
strength of the signals due to the ambient noise.
For this reason we also consider the estimate of the signal
at each detector, by applying wavelet denoising techniques.
In this way we arrive now at the estimates:
{\small
\begin{equation}\label{eq:wX}
\begin{split}
w_X(t + \tau_X) &= e_X(t + \tau_X) + s_X(t + \tau_X) \\
&= e_X(t + \tau_X) + 
F_{+X0}(\theta_X,\phi_X,\psi_X,t) s_+(t)\\
&\;\;\;+
F_{\times X0}(\theta_X,\phi_X,\psi_X,t) s_\times(t)
,
\end{split}
\end{equation}
}where $w_X$ are the estimates, and now $e_X$ stands 
for the error intrinsic to the estimates.
Contrary to the previous situation, discussed in the previous section, now we assume that the magnitude of the
errors are much smaller than the magnitude of the signals.
That is, we can neglect them in the localization process;
although we estimate them from the calculation of the sample standard deviation
with respect to $v_X$ on a chosen window. We chose this approach because
it provides time dependent values, which adapt to the characteristics
of the signal dynamically.
We will also assume that the scalar product of the error with the signals
are negligible.
These assumptions are supported from the results of 
Donoho and Johnstone\citep{Donoho:1994,Mallat2009} on thresholding estimators;
and also from the numerical estimation of the errors.

The technique for inferring the signals by
applying wavelet denoising methods is usual\citep{Mallat2009} in signal processing
and the use of wavelets for the representation of GW signals
has been mentioned in several LIGO/Virgo Collaboration articles\citep{Abbott:2017vtc,Abbott:2016blz,Cornish:2014kda}.
In our work we have used the discrete FIR approximation of Meyer wavelet;
which has the properties of being symmetric, orthogonal and biorthogonal.
More concretely, we have used the wavelet `dmey' included
in the package PyWavelets({\sf pywt}) in the {\sc python} language\citep{Lee:2019}.
Having chosen a convenient lapse of time for the study, we perform the decomposition of
the signals in terms of this wavelet basis and apply denoising methods\citep{Mallat2009}
to those coefficients; and then reconstruct the signal.

\begin{figure}[H]
\centering
\includegraphics[clip,width=0.48\textwidth]{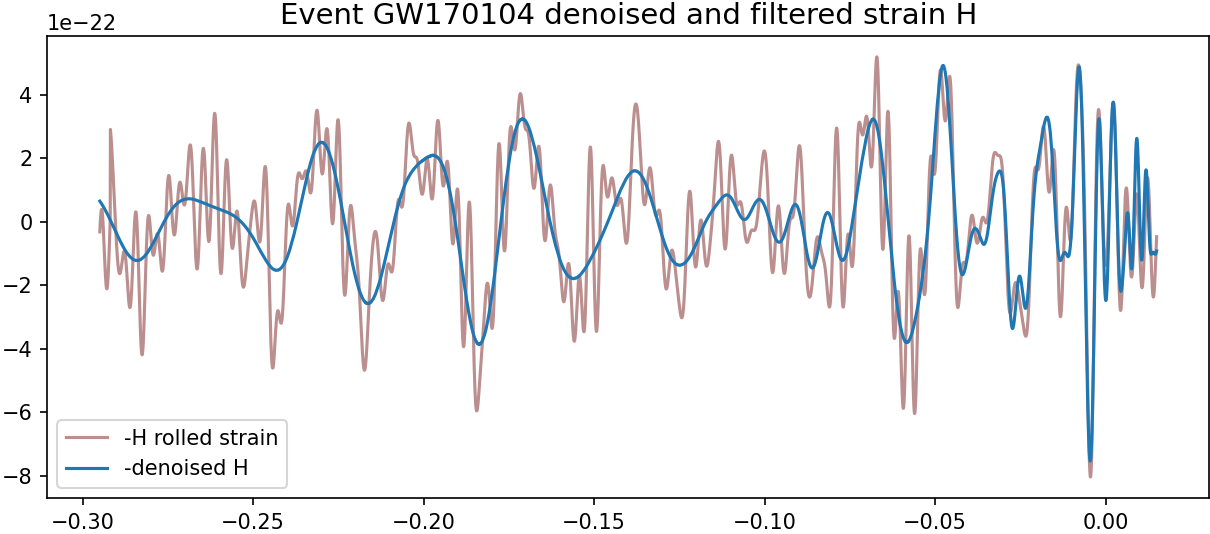}
\caption{Strain and denoised data for Hanford LIGO detector for GW170104 near the event time,
with bandpass filter of [30,350]Hz.
}
\label{fig:denoised-Hstrain}
\end{figure}

\begin{figure}[H]
\centering
\includegraphics[clip,width=0.48\textwidth]{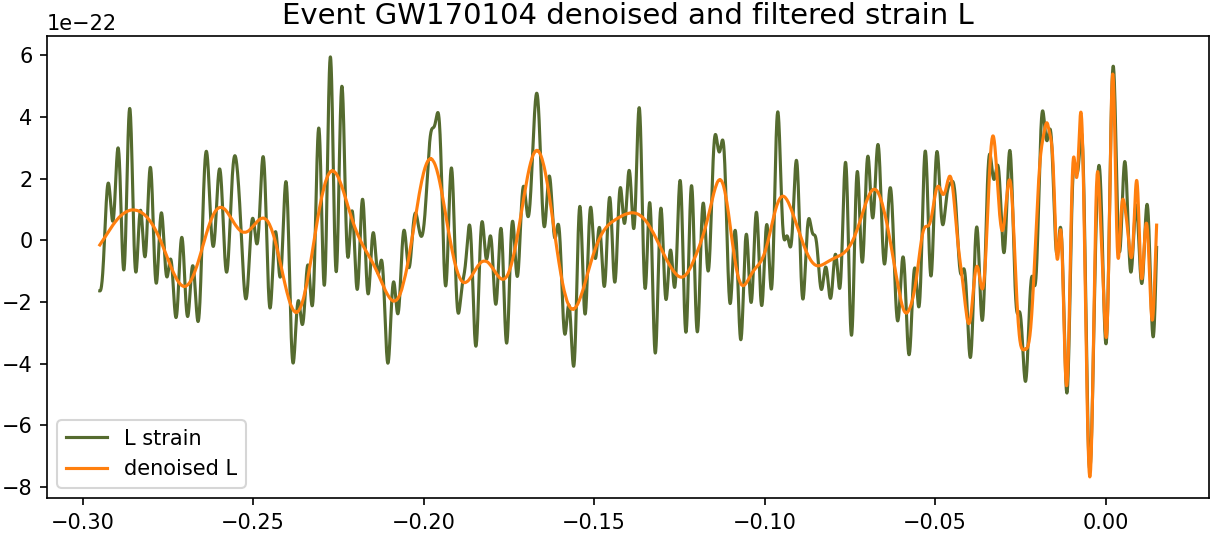}
\caption{Strain and denoised data for Livingston LIGO detector for GW170104 near the event time,
with bandpass filter of [30,350]Hz.
}
\label{fig:denoised-Lstrain}
\end{figure}

The denoised strains in the lapse of time of interest are shown in 
Figs. \ref{fig:denoised-Hstrain} and \ref{fig:denoised-Lstrain}.

The error in the denoised data depends on several factors and can not be perfectly calculated without
knowing the precise value of the signal, as it is our case. We can at most give estimates of them
based on the characteristics of the ambient present noise.
We have limited some of the factors inherent in the calculation by an appropriate 
choice of the wavelet basis.
Let us expand on this.
The wavelet transform comes in a variety of versions and the analyst has to
choose the framework that best suits the needs of the type of data under study.
To begin with, wavelet transforms comes in continuous and discrete versions.
We use continuous wavelet transforms to study the global behavior of the signal
in the scalograms we present in the section \ref{sec:time-freq}.
But for the detailed local analysis of the signal we use a discrete basis.
There is a great variety of them; for our purposes we require the basis to 
be orthogonal, since it is important for us to preserve the energy of the system.
Since ultimately we intend to make a decomposition of the signal in terms of its
polarization modes, their phase behavior is also important for us, which is
associated with the symmetry and biorthogonal\citep{Mallat2009,cohen1992biorthogonal}
 properties of the basis.
By choosing the Discrete Meyer (FIR Approximation), mentioned above,
we obtain all the desired properties for a wavelet basis;
thus mitigating factors that could contribute to errors in the analysis.
The crude estimate of the error is calculated from the local standard deviation 
of the difference between the strain and the 
denoised datum for each case; which we use.
These are understood as upper bounds on the possible errors of the denoised data.

It should probably be emphasized that our techniques for denoising the strains
do not assume the existence of the polarization modes;
which is a completely different approach from the one mentioned in
LIGO Scientific and Virgo Collaborations articles\citep{Abbott:2017vtc}
on the use of wavelet analysis.

\section{Study of the nature of the signals in the time-frequency domain}\label{sec:time-freq}

\begin{figure}[H]
\centering
\includegraphics[clip,width=0.48\textwidth]{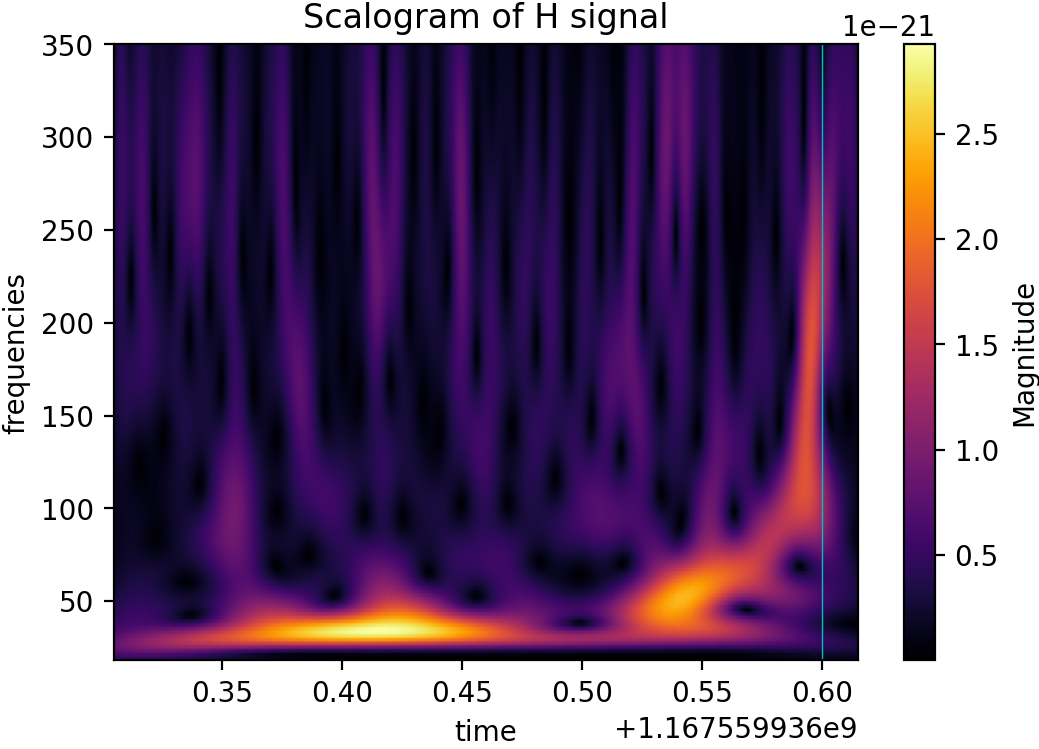}
\caption{Detail of the strain of LIGO detector H for GW170104 
	with a pass band of [30,350]Hz.
}
\label{fig:scalogramH}
\end{figure}

\begin{figure}[H]
\centering
\includegraphics[clip,width=0.48\textwidth]{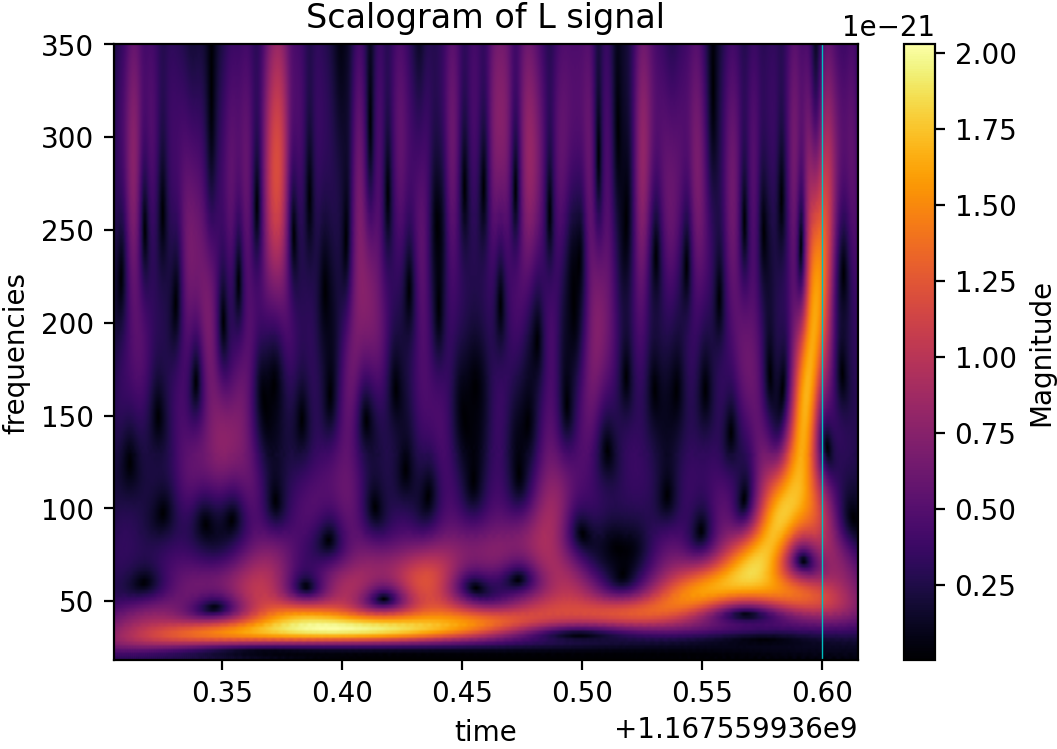}
\caption{Detail of the strain of LIGO detector L for GW170104
	with a pass band of [30,350]Hz.
}
\label{fig:scalogramL}
\end{figure}

As part of the systematic analysis of the strains of an event is the study
of the characteristics of the signal in the time-frequency domain.
It is then usual to study the spectrograms and scalograms of the strains.
We have found that it is better to resort to scalograms
for our purposes; which we show in Figs. \ref{fig:scalogramH} and \ref{fig:scalogramL}.
It is easy to observe a chirp like signal in both detectors.
We have indicated with a vertical line an approximate chirp time.

\section{Using a universal fitting chirp form for gravitational-wave polarization modes}
\label{sec:univ_fitt_chirp}

Due to the chirp like nature of the signal, we study here the possibility of
fitting the  gravitational-wave polarization modes with a couple of universal chirp shape
functions, that could handle a gross representation of the modes
during the inspiral phase.
In order to fit the observed data it is better to use the widest time range.
For early times it is expected that a universal fitting form could be
good enough to perform the adjustment of parameters.
In this work we choose the function $g(t) = 1/ \big( (t_f - t)^{p_a /4} + \epsilon_t^{p_a /4} \big)$
for fitting the amplitude time dependence of the modes,
and the function
$\Phi(t)=-2 \big(\frac{ t_f - t}{5 t_{ch}}\big)^{p_c 5/8}  + \phi_f$
for fitting the phase time dependence of the modes.
Then, we define mono-components polarization modes as:
\begin{equation}\label{eq:sf+}
\mathsf{P}_+(t) = A_+ g(t) 
\cos( \Phi(t)  )
,
\end{equation}
and
\begin{equation}\label{eq:sfx}
\mathsf{P}_\times(t) = A_\times g(t) \sin( \Phi(t) )
,
\end{equation}
with adjustable parameters $[A_+, A_\times, \phi_f  ]$;
while the other parameters $[t_f,p_a, \epsilon_t,t_{ch},p_c]$ are fixed from the time frequency studies. 
Note that one can use eq. \eqref{eq:X} for a fitting approximation
of the signal $w'_X$ in detector $X$.
More concretely, we define the corresponding fitting signals
\begin{equation}\label{eq:wpX}
w'_X = B_{+X} g(t) \cos( \Phi(t)) + B_{\times X} g(t) \sin( \Phi(t))
.
\end{equation}
This expression however has a degeneracy among the parameters $[B_+, B_\times, \phi_f  ]$ that is treated below.

In order to determine the original  $[A_+, A_\times]$ amplitudes 
we express $w'$ in terms of the orientation angles for each detector;
namely:
\begin{equation}\label{eq:wpH}
\begin{split}
w'_H =& F_{+H0} \mathsf{P}_+ + F_{\times H0} \mathsf{P}_\times \\
=& B_{+H} g(t) \cos( \Phi(t) ) + B_{\times H} g(t) \sin( \Phi(t) )
,
\end{split}
\end{equation}
and similarly
\begin{equation}\label{eq:wpL}
\begin{split}
w'_L =& F_{+L0} \mathsf{P}_+ + F_{\times L0} \mathsf{P}_\times \\
=& B_{+L} g(t) \cos( \Phi(t) ) + B_{\times L} g(t) \sin( \Phi(t) )
;
\end{split}
\end{equation}
from which one finds
\begin{equation}\label{eq:B+H}
B_{+H} =  F_{+H0} A_+
,
\end{equation}
\begin{equation}\label{eq:BxH}
B_{\times H} =  F_{\times H0} A_\times
,
\end{equation}
\begin{equation}\label{eq:B+L}
B_{+L} =  F_{+L0} A_+
,
\end{equation}
\begin{equation}\label{eq:BxL}
B_{\times L} =  F_{\times L0} A_\times
.
\end{equation}

Then, 
we have two equations for the single value of $A_+$
and another couple of equations for $A_\times$ which
should help us in determining the three angles  $(\delta,\alpha,\psi)$.
But we should also use the combination $A^2 \equiv A_+^2 + A_\times^2$,
due to its transformation properties.
More concretely, we can look for the zeros of
\begin{equation}\label{eq:AH-AL}
J_{HL}(\delta,\alpha,\psi) =  A_H^2 - A_L^2  ,
\end{equation}
which is expected to have multiple minima on the celestial sphere.

Note also that one has the relations
\begin{equation}\label{eq:B+}
B_{+H}  F_{+L0} = B_{+L} F_{+H0}
,
\end{equation}
and
\begin{equation}\label{eq:Bx}
B_{\times H} F_{\times L0} =  B_{\times L} F_{\times H0} 
.
\end{equation}
So that we can study the zeros of
\begin{equation}\label{keq:C+}
C_+(\delta,\alpha,\psi) = B_{+H}  F_{+L} - B_{+L} F_{+H}
,
\end{equation}
and
\begin{equation}\label{keq:Cx}
C_\times(\delta,\alpha,\psi) = B_{\times H}  F_{\times L} - B_{\times L} F_{\times H}
,
\end{equation}
or better the minima of their squares.
Then, for each choice of $\psi$ we also study the minima of
\begin{equation}\label{eq:C+2_Cx2}
N(\delta,\alpha,\psi) =  C_+^2 + C_\times^2
,
\end{equation}
in terms of the location angles; with the difficulty that the minimum
of one is washed by the other.
Alternatively, we can study the maxima of
\begin{equation}\label{eq:invC+2_invCx2}
\mathsf{N}(\delta,\alpha,\psi) = \frac{1}{C_+^2}  + \frac{1}{C_\times^2}
;
\end{equation}
where each minimum of $C_{+,\times}$ contributes independently.

We will use as initial measure the function
\begin{equation}\label{eq:M}
M_i= \frac{1}{\sqrt{N}} = \frac{1}{\sqrt{ C_+^2 + C_\times^2}}
;
\end{equation}
where the location would be indicated by the maximum values.

\begin{figure}[H]
\centering
\includegraphics[clip,width=0.48\textwidth]{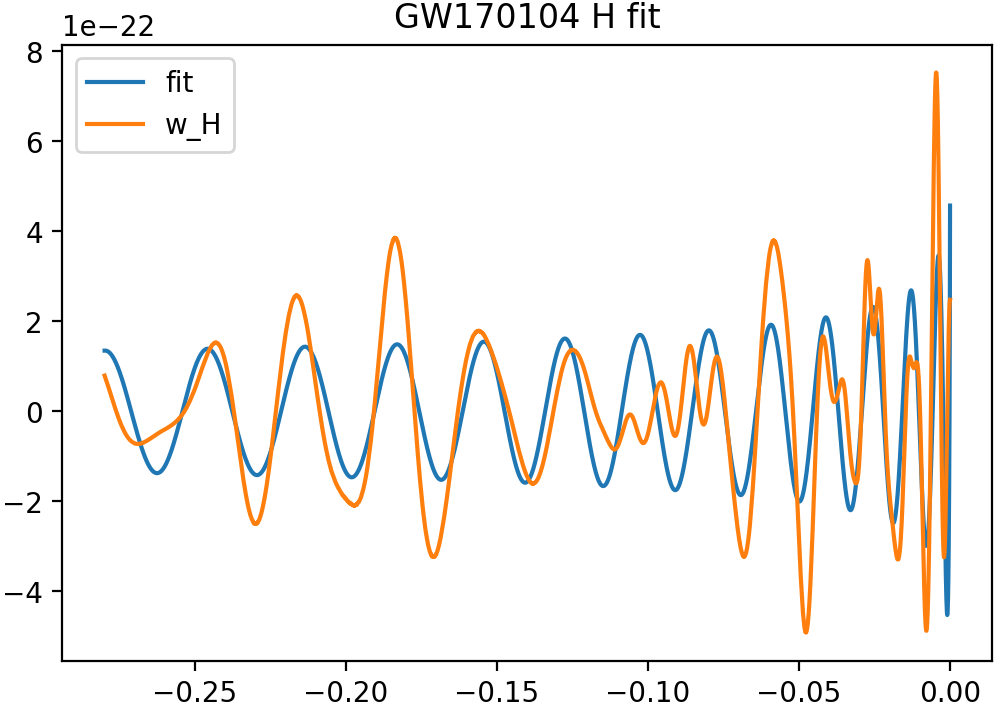}
\caption{Fitting result to the denoised signal H.
}
\label{fig:fitH}
\end{figure}

\begin{figure}[H]
\centering
\includegraphics[clip,width=0.48\textwidth]{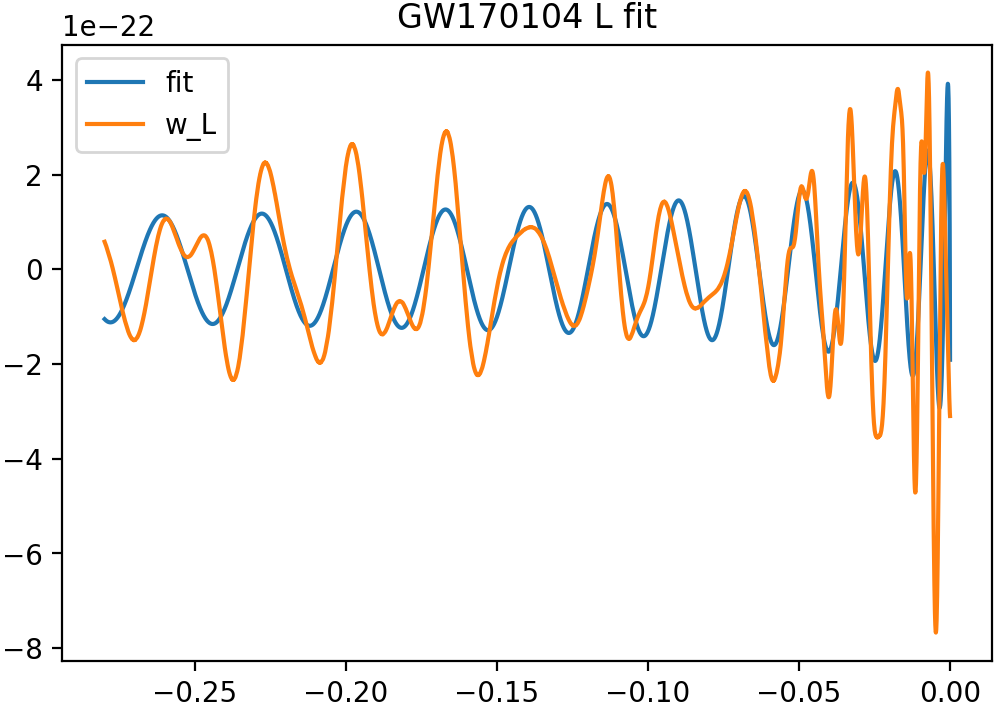}
\caption{Fitting result to the denoised signal L.
}
\label{fig:fitL}
\end{figure}

The results of fitting a universal chirp form for the polarization
of the GW to the denoised signals are shown in 
Figs. \ref{fig:fitH} and \ref{fig:fitL}.
\noindent
It is obvious that this is a gross approximation to the signals;
but the remarkable result is that even with this universal fitting chirp forms
we can obtain excellent results; as shown below.

It is probably worthwhile to comment on the role of the angle $\psi$.
At this stage $\psi$ is a hyperparameter which is adjusted by
using the measure
$\mathsf{N}$
so that the crossing appears on the delay ring.	
Since our intention is to present a procedure to localize the source, and to reconstruct the spin-2 polarization modes,
we are trying to obtain these goals as far as we can without entering into the detail astrophysical
description of the source system; that is, in terms of inclination of the orbital plane
and other parameters. For this reason we do not intend in this article to relate our $\psi$
to the detailed description of the source.
But with a more accurate prescription of the proposed polarization modes,
this procedure would employ a $\psi$ that would probably give a good first
estimate to relate to the natural GW base as described in terms
of the source orbit parameters\citep{Poisson2014}.

\section{Injected signals from synthetic spin-2 polarization modes}\label{sec:synth}

\subsection{Injected signals and denoised reconstruction}

\begin{figure}[H]
\centering
\includegraphics[clip,width=0.48\textwidth]{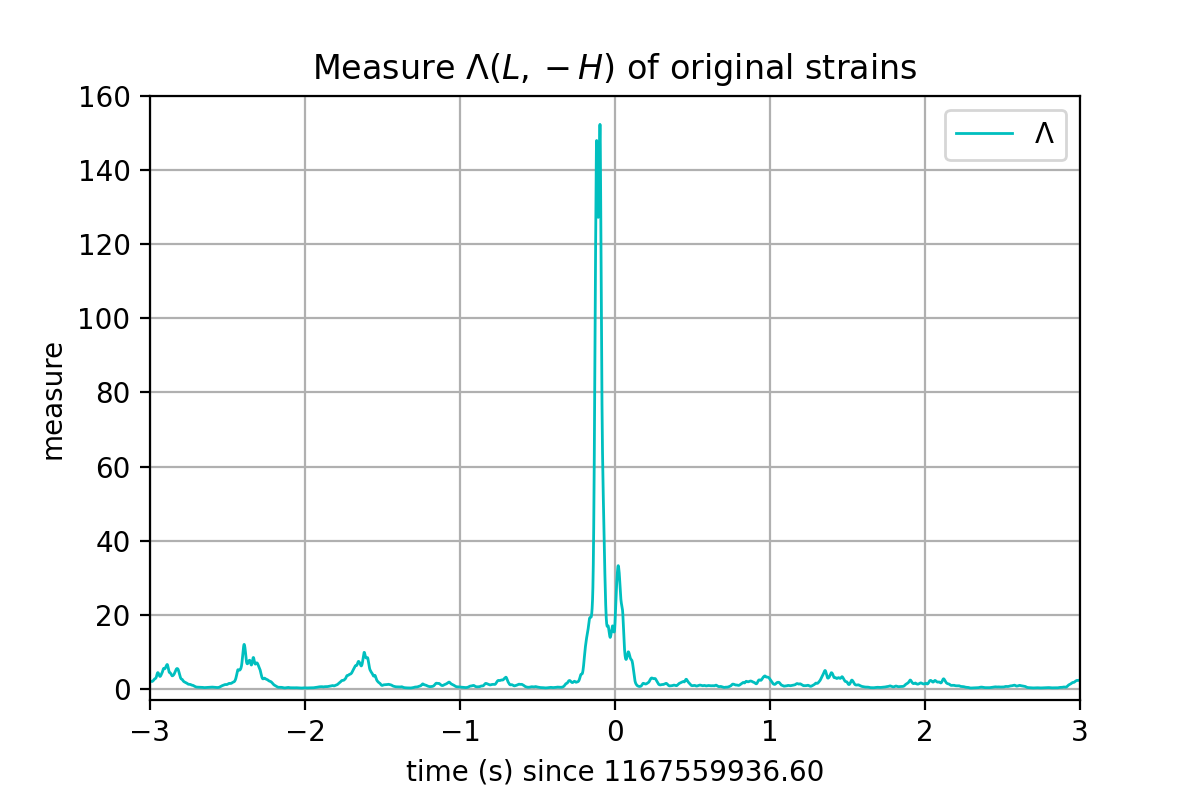}
\includegraphics[clip,width=0.48\textwidth]{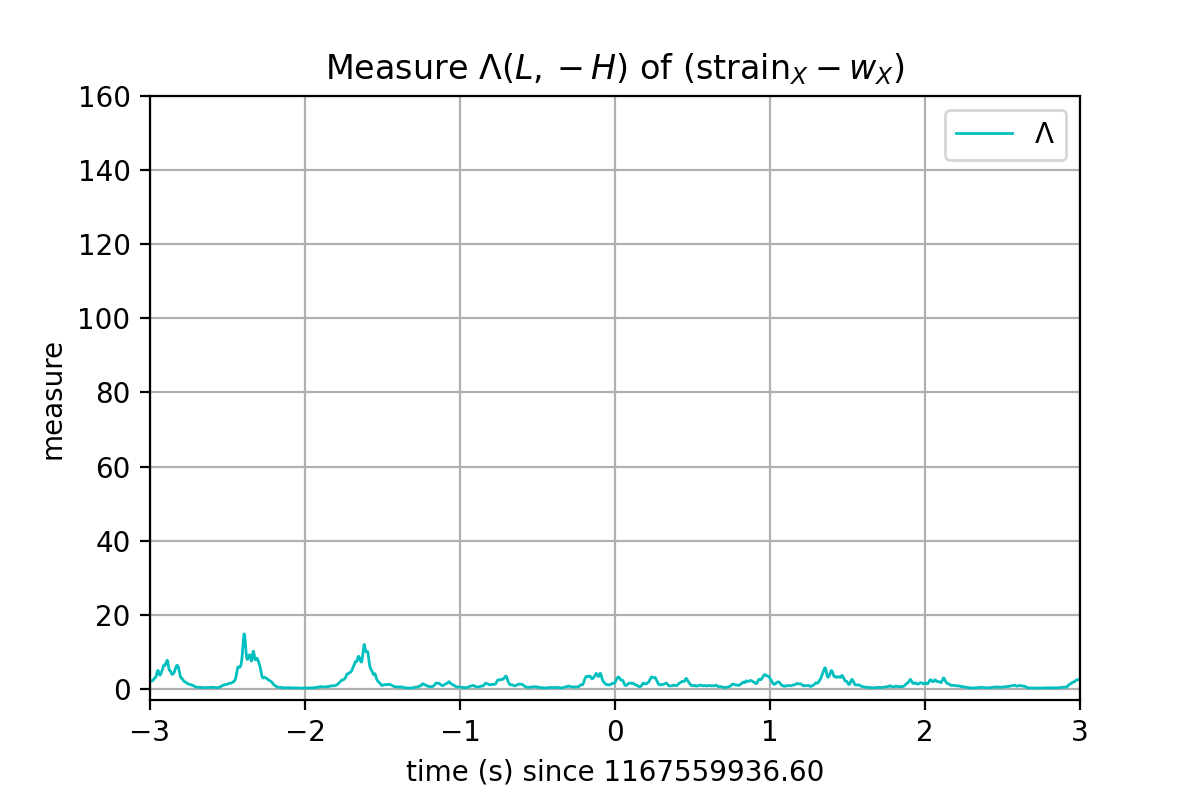}
\caption{On the top graph, the values of the measure $\Lambda$ close to the reference event time
	for the original filtered strains of GW170104,
	and on the bottom for the strains after the subtraction of the denoised 
	signals $w$ from the spin-2 polarization modes. The residual is ambient noise.
}
\label{fig:strain-w}
\end{figure}

In this section we apply the L2D+PMR procedure to the case of injected signals that
have been generated from spin-2 polarization modes at different locations in the celestial sphere
and with distinct polarization frames.
To incorporate realistic noise conditions, we perform signal injections 
on the strains from the GW170104 event after subtracting the corresponding denoised signals. 
This process allows us to construct ten synthetic variations of the event, which we refer to as GW170104synth.
To evaluate the impact of this subtraction, we compare the H and L strains before and after removing 
the denoised signals $w$. 
The difference is illustrated through the OM measure  $\Lambda$ introduced in  \cite{Moreschi:2024njx},
as shown in Fig. \ref{fig:strain-w}.

As shown in Fig.  \ref{fig:strain-w},
the comparison between the H and L strains of GW170104 after 
	subtracting the denoised signals reveals no significant coincidence within the studied time window.
It must be stressed that these graphs are generated with the initial bandpass filter 
in the range [27,1003]Hz; that is, the injected signals have the intrinsic initial noise.
	
\begin{figure}[H]
\centering
\includegraphics[clip,width=0.48\textwidth]{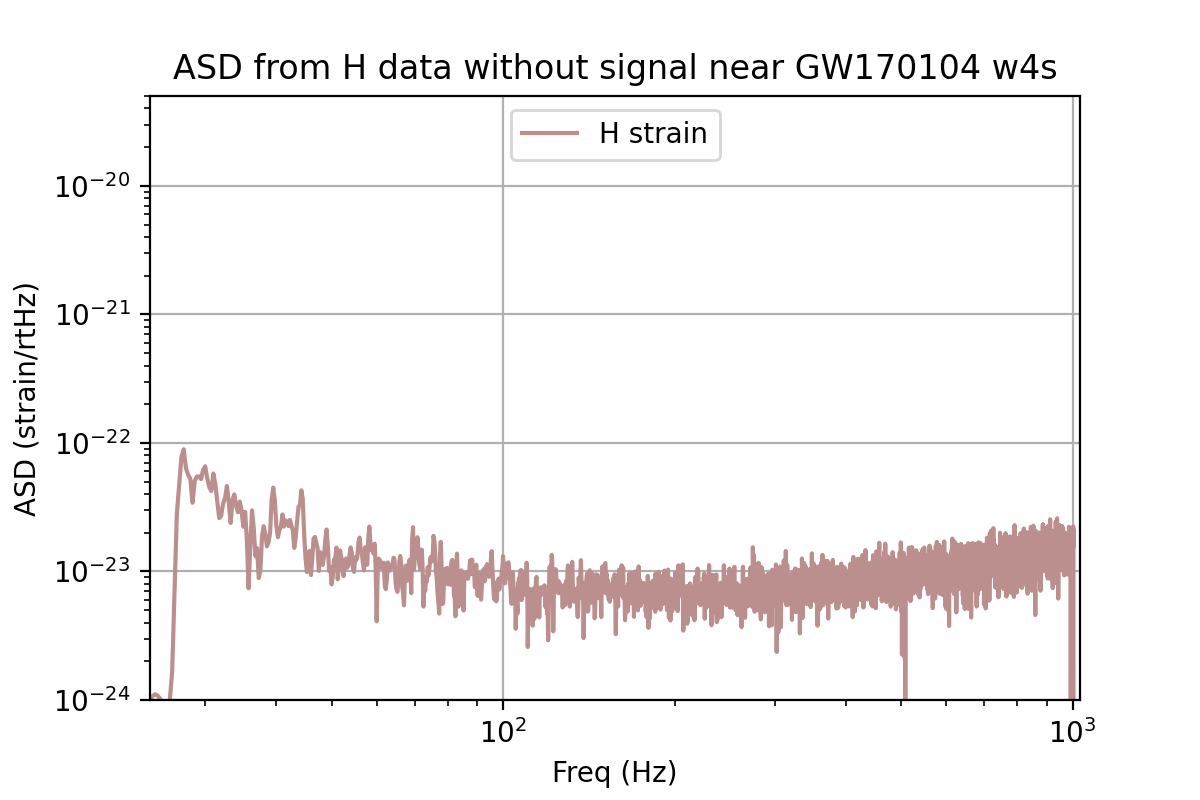}
\includegraphics[clip,width=0.48\textwidth]{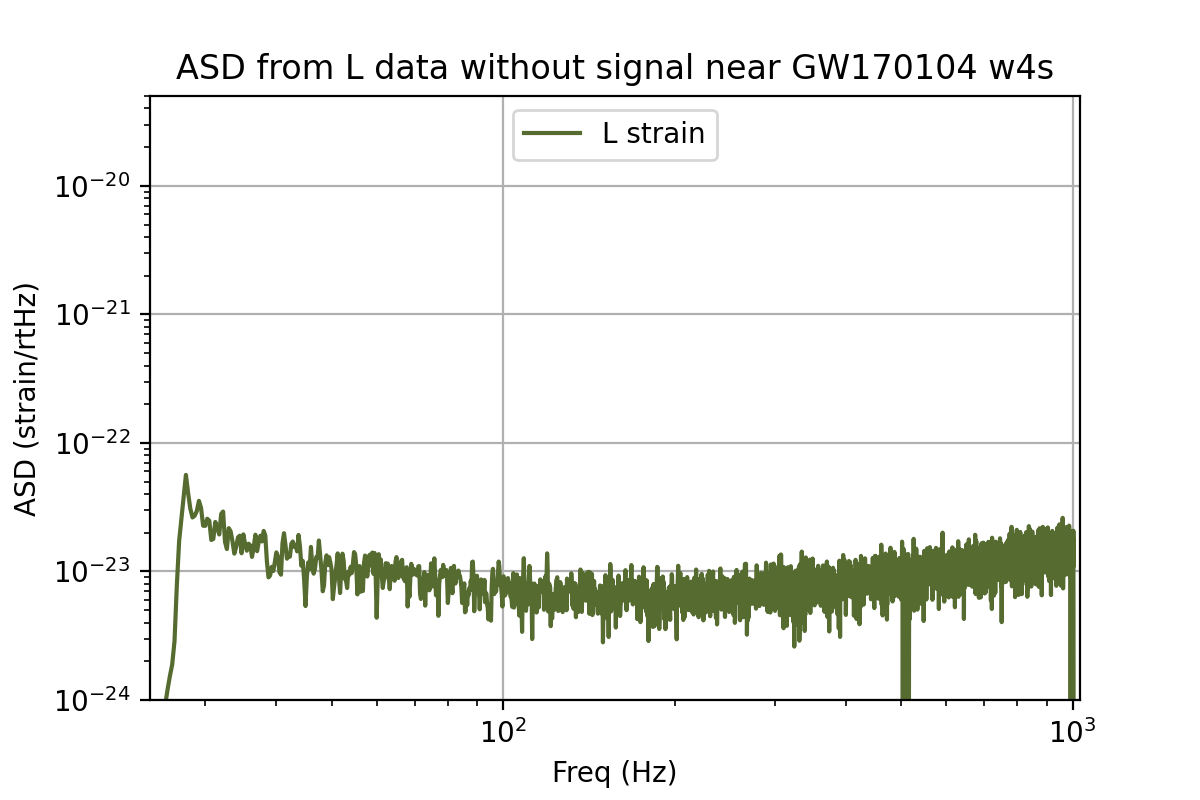}
\caption{On the top graph, the ASD of strain H
	of the 22s around the time of the event GW170104,
	and on the bottom graph the corresponding ASD for the L strain.
}
\label{fig:ASD_sinsenial}
\end{figure}

To  complement this observation, we also present the Amplitude Spectral Density (ASD) 
graphs of these strains after subtraction in Fig. \ref{fig:ASD_sinsenial}, 
which provide insight into the spectral characteristics of the remaining noise.
In Fig. \ref{fig:ASD_sinsenial} one can see the nature of the state of the noise
for the event GW170104, in a lapse of time of 22s centered at the event time.
This is of course after we have applied the preprocessing filtering techniques\citep{Moreschi:2019vxw}.

For reasons of time and space we present in this occasion ten simulations. 
Since our methods do not rely on the astrophysical information of the signal,
but rather on the spin-2 behavior, we consider for this purpose
a signal, generated by very simple polarization modes;
which details are described in appendix \ref{sec:simul-signal}.
They correspond to a post-Newtonian\citep{Jaranowski:2009zz,Poisson2014} system with radiation reaction as described by \cite{Peters:1964zz}
with initial $0.35$ eccentricity.
We study ten different locations of this signal on a different delay ring
corresponding to a time difference of $0.005737$s.
We denote the cases from 1 to 10.
The polarization frames were chosen at random from 10 different options, 
for each case by taking
$\Delta \psi = 0, 0.6891$, $0.7854$, $0.6108$, $0.1745$, 
$0.4363$, $0.3490$, $0.0873$, $0.5236$ and $0.2618$ respectively;
while the relative amplitudes were chosen as
$A_{ch}=7 \times 10^{-22}$.
With these choices we intend to cover a variety of situations one could encounter
in the location and reconstruction tasks.

In order to verify the denoising procedure quantitatively we here calculate the 
correlation coefficient\citep{Ferguson67,Helstrom75,McDonough1995} 
between the two signals 
in terms of the natural inner product
as described in \cite{Moreschi:2024njx}.

\begin{figure}[H]
\centering
\includegraphics[clip,width=0.48\textwidth]{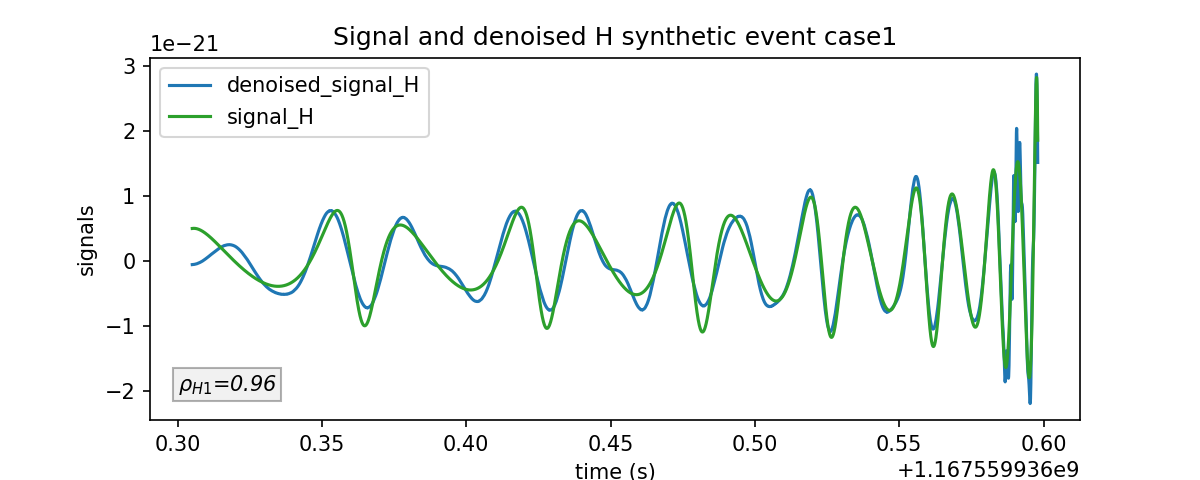}
\includegraphics[clip,width=0.48\textwidth]{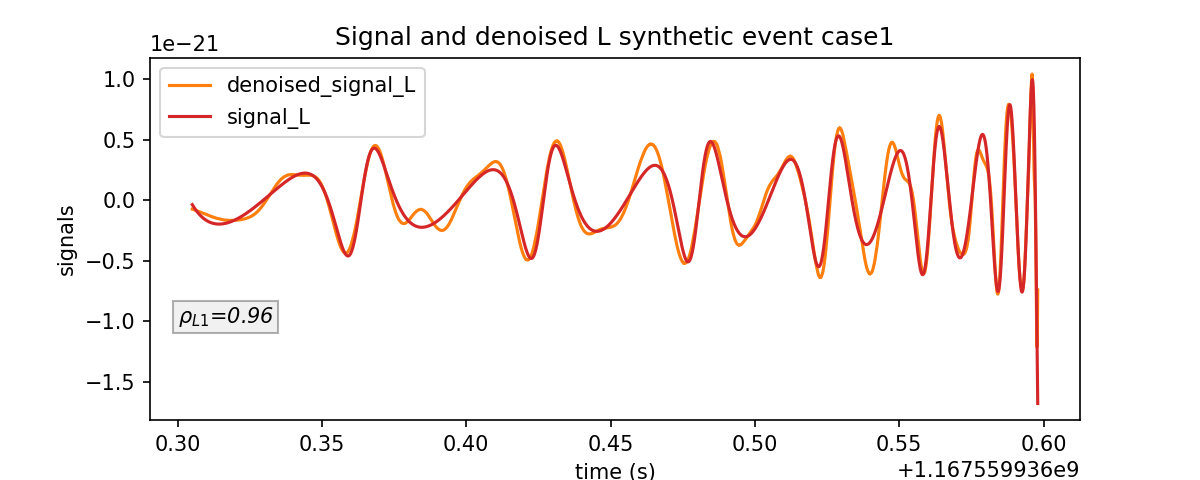}
\caption{Top panel, comparison of original synthetic signal\_H with the extracted denoised signal
	from H strain with the noise of the GW170104 event.
	In the bottom panel the corresponding graphs for L.
	The inset shows the value of the correlation coefficient for both curves.
	}
\label{fig:signal+denoised-1}
\end{figure}

\begin{figure}[H]
	\centering
	\includegraphics[clip,width=0.48\textwidth]{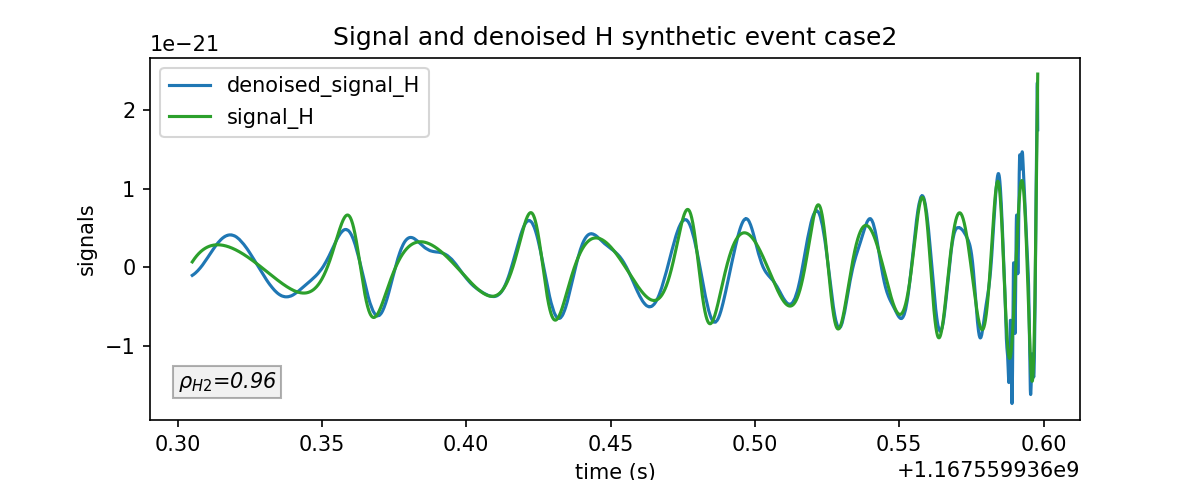}
	\includegraphics[clip,width=0.48\textwidth]{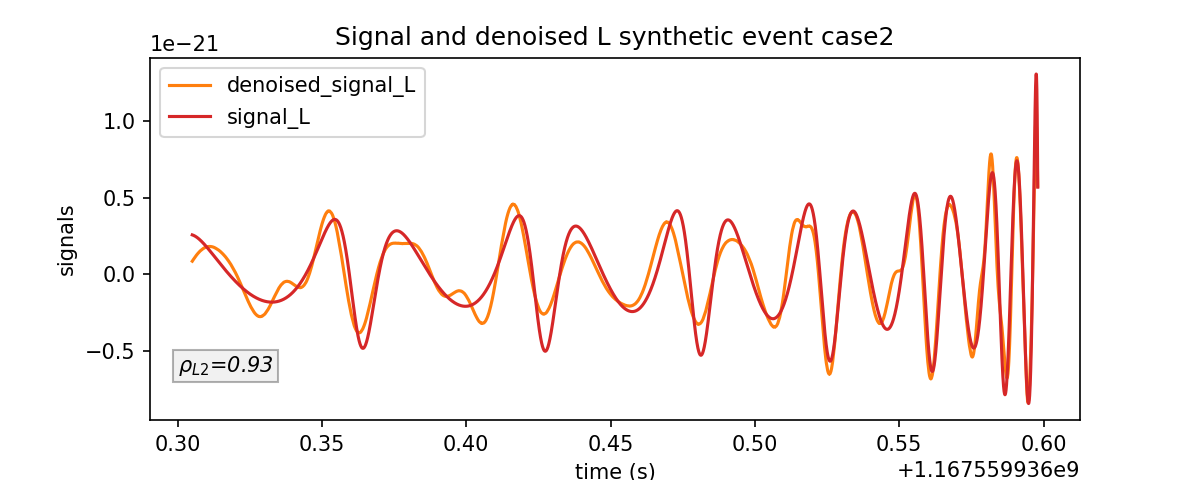}
	\caption{Top panel, comparison of original synthetic signal\_H with the extracted denoised signal
		from H strain with the noise of the GW170104 event.
		In the bottom panel the corresponding graphs for L.
		The inset shows the value of the correlation coefficient for both curves.
	}
	\label{fig:signal+denoised-2}
\end{figure}

\begin{figure}[H]
	\centering
	\includegraphics[clip,width=0.48\textwidth]{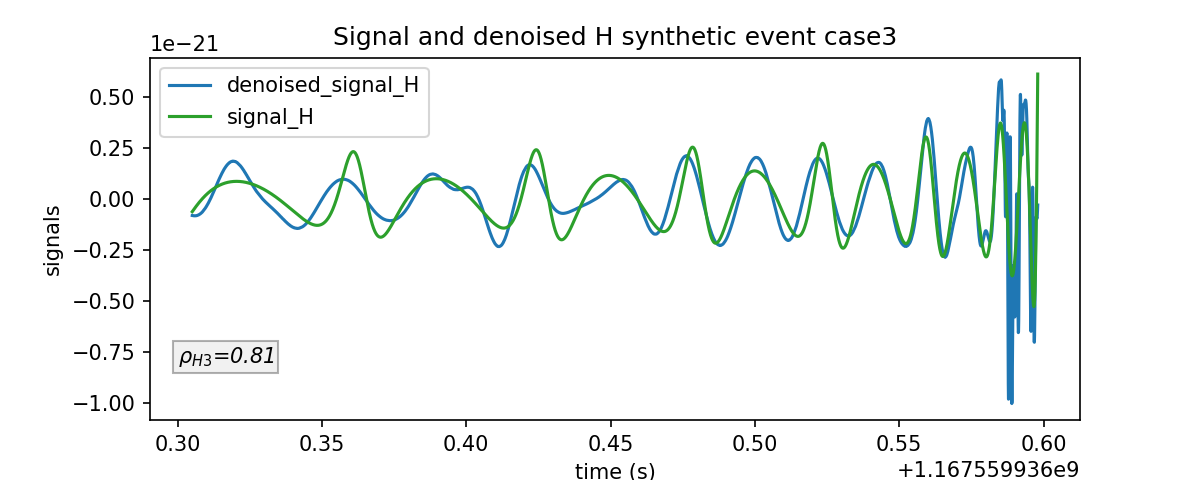}
	\includegraphics[clip,width=0.48\textwidth]{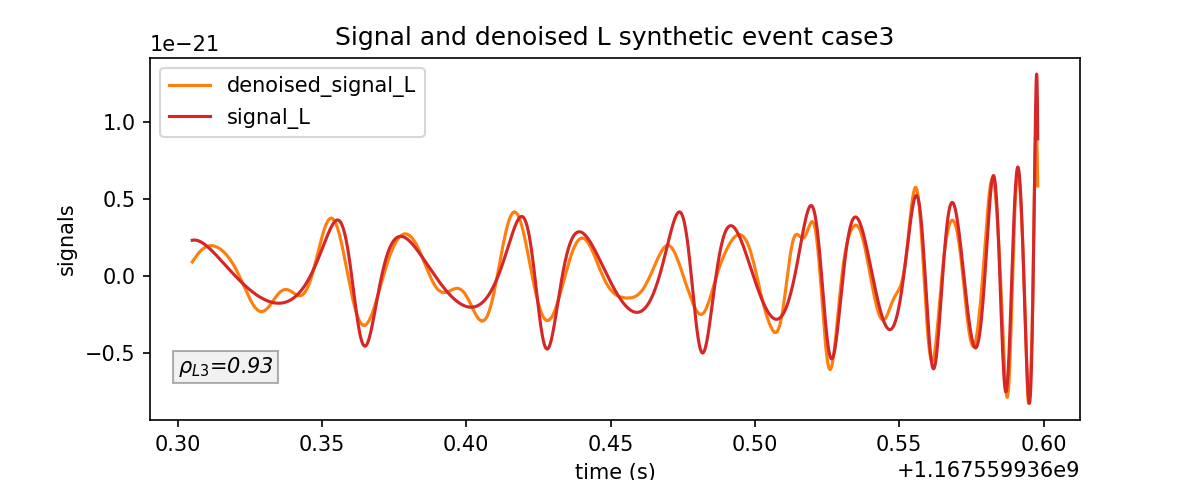}
	\caption{Top panel, comparison of original synthetic signal\_H with the extracted denoised signal
		from H strain with the noise of the GW170104 event.
		In the bottom panel the corresponding graphs for L.
		The inset shows the value of the correlation coefficient for both curves.
	}
	\label{fig:signal+denoised-3}
\end{figure}

\begin{figure}[H]
	\centering
	\includegraphics[clip,width=0.48\textwidth]{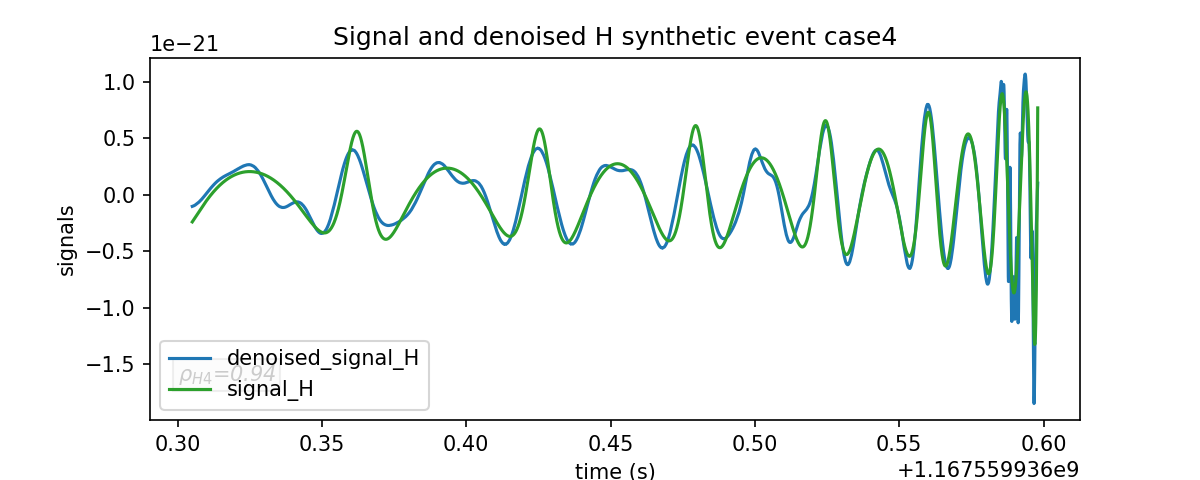}
	\includegraphics[clip,width=0.48\textwidth]{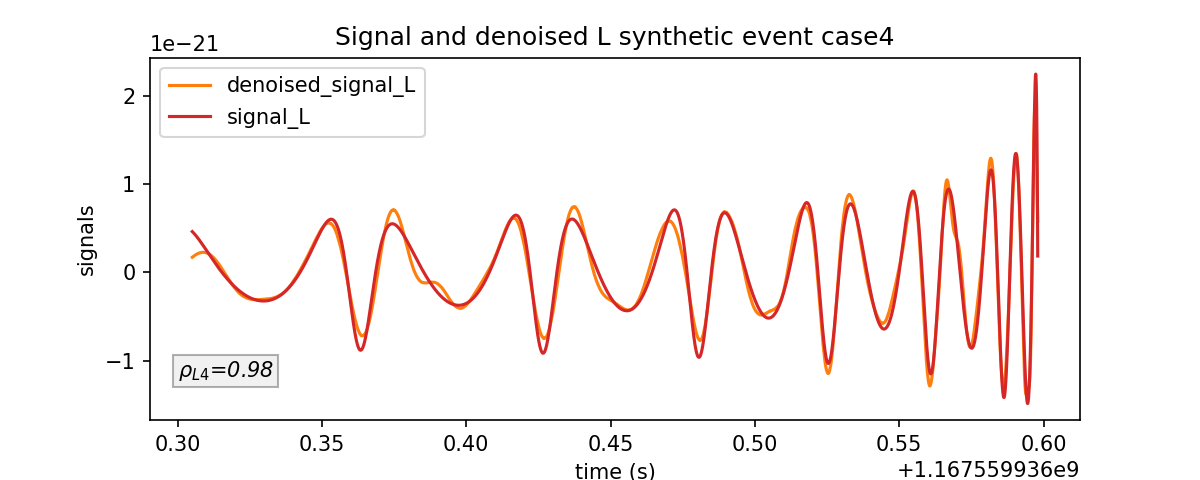}
	\caption{Top panel, comparison of original synthetic signal\_H with the extracted denoised signal
		from H strain with the noise of the GW170104 event.
		In the bottom panel the corresponding graphs for L.
		The inset shows the value of the correlation coefficient for both curves.
	}
	\label{fig:signal+denoised-4}
\end{figure}

\begin{figure}[H]
	\centering
	\includegraphics[clip,width=0.48\textwidth]{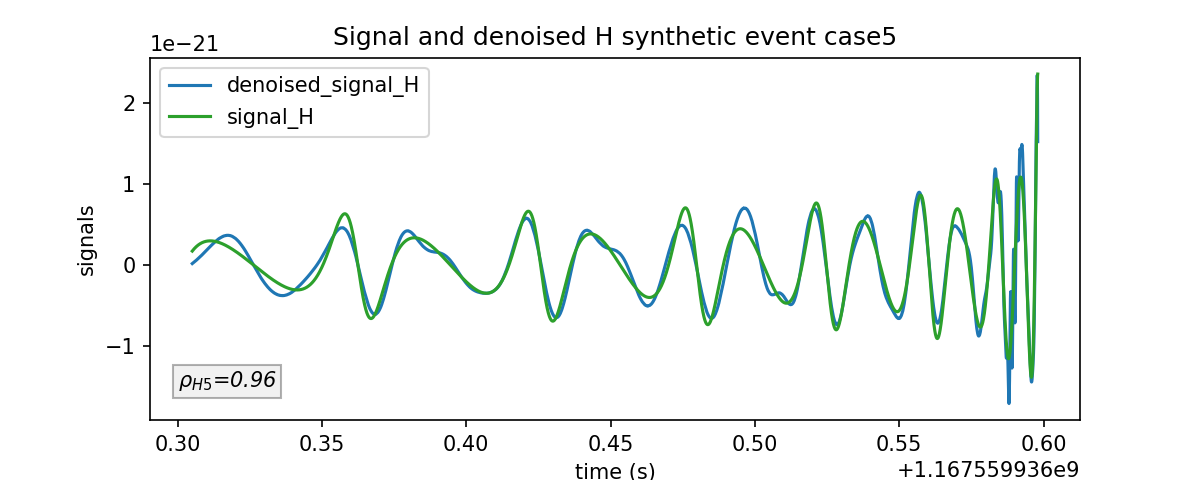}
	\includegraphics[clip,width=0.48\textwidth]{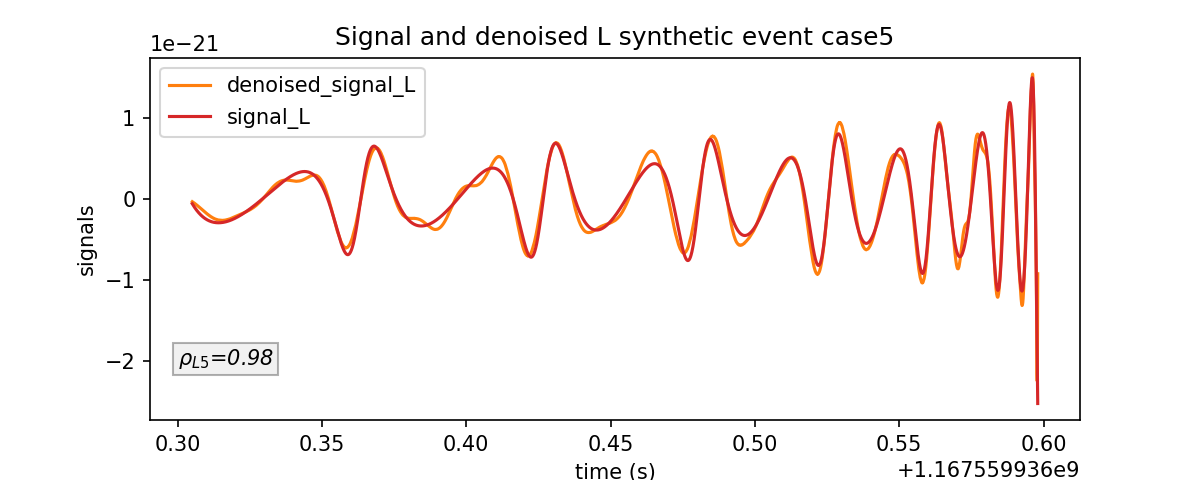}
	\caption{Top panel, comparison of original synthetic signal\_H with the extracted denoised signal
		from H strain with the noise of the GW170104 event.
		In the bottom panel the corresponding graphs for L.
		The inset shows the value of the correlation coefficient for both curves.
	}
	\label{fig:signal+denoised-5}
\end{figure}

\begin{figure}[H]
	\centering
	\includegraphics[clip,width=0.48\textwidth]{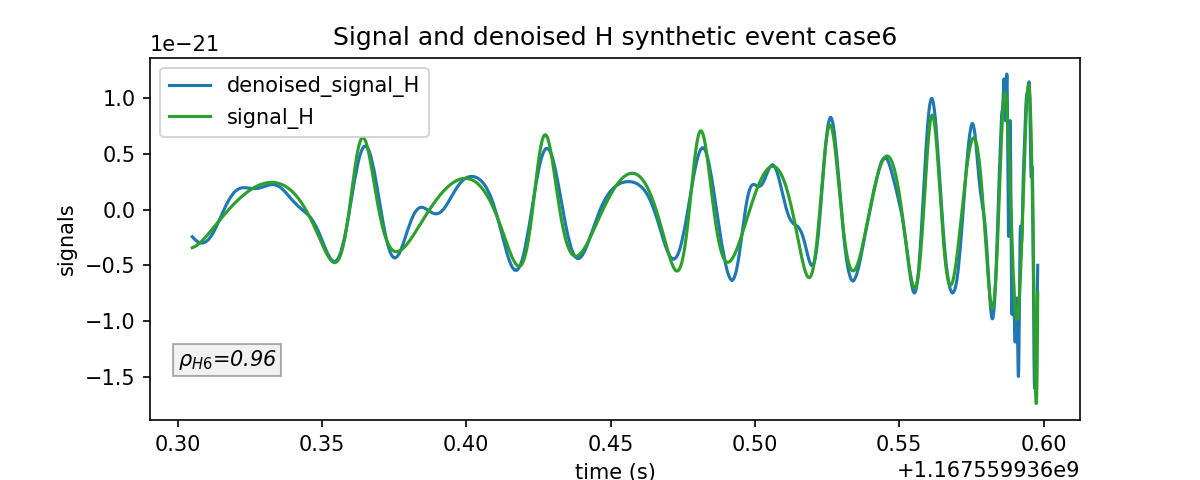}
	\includegraphics[clip,width=0.48\textwidth]{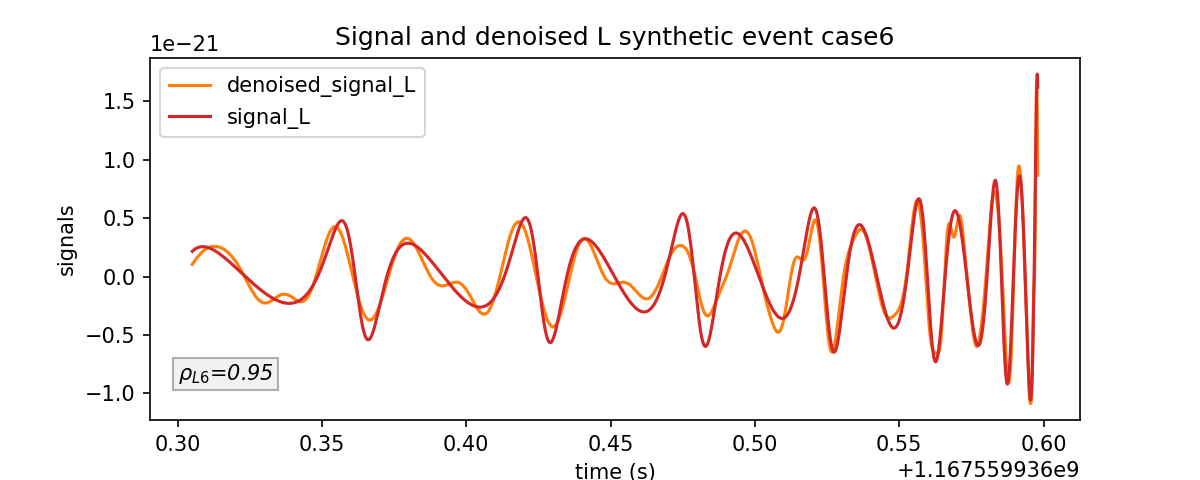}
	\caption{Top panel, comparison of original synthetic signal\_H with the extracted denoised signal
		from H strain with the noise of the GW170104 event.
		In the bottom panel the corresponding graphs for L.
		The inset shows the value of the correlation coefficient for both curves.
	}
	\label{fig:signal+denoised-6}
\end{figure}

\begin{figure}[H]
	\centering
	\includegraphics[clip,width=0.48\textwidth]{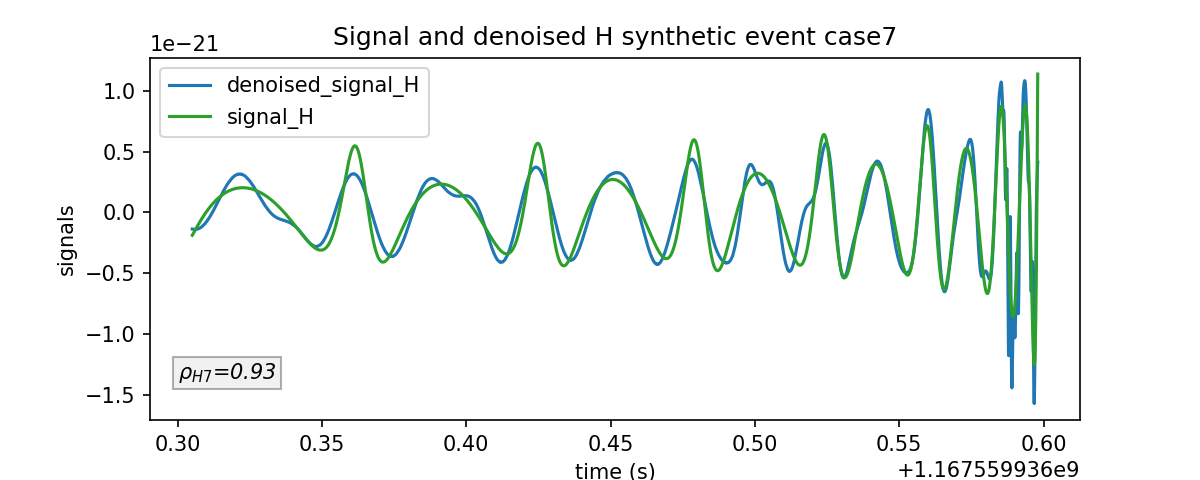}
	\includegraphics[clip,width=0.48\textwidth]{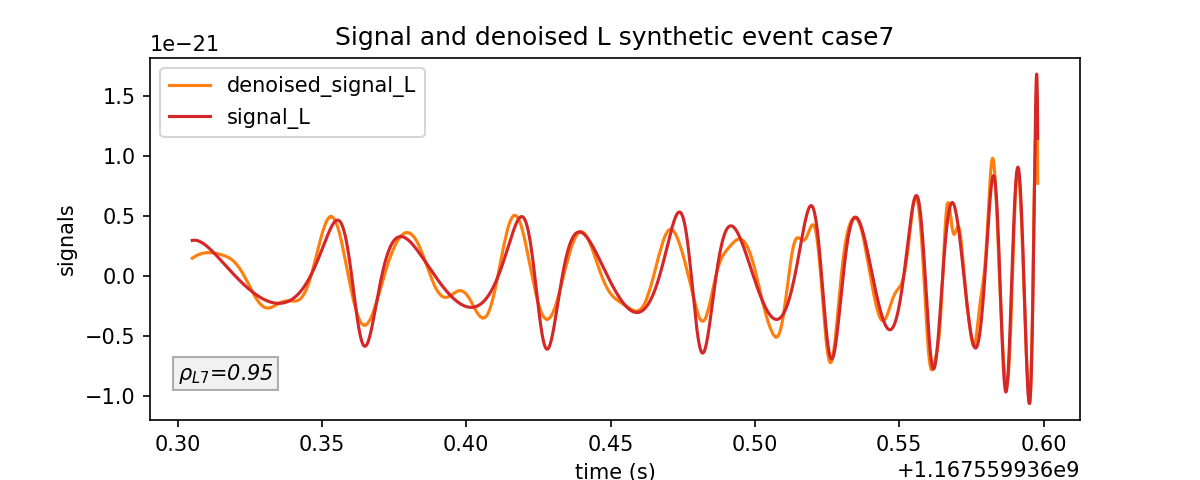}
	\caption{Top panel, comparison of original synthetic signal\_H with the extracted denoised signal
		from H strain with the noise of the GW170104 event.
		In the bottom panel the corresponding graphs for L.
		The inset shows the value of the correlation coefficient for both curves.
	}
	\label{fig:signal+denoised-7}
\end{figure}

\begin{figure}[H]
	\centering
	\includegraphics[clip,width=0.48\textwidth]{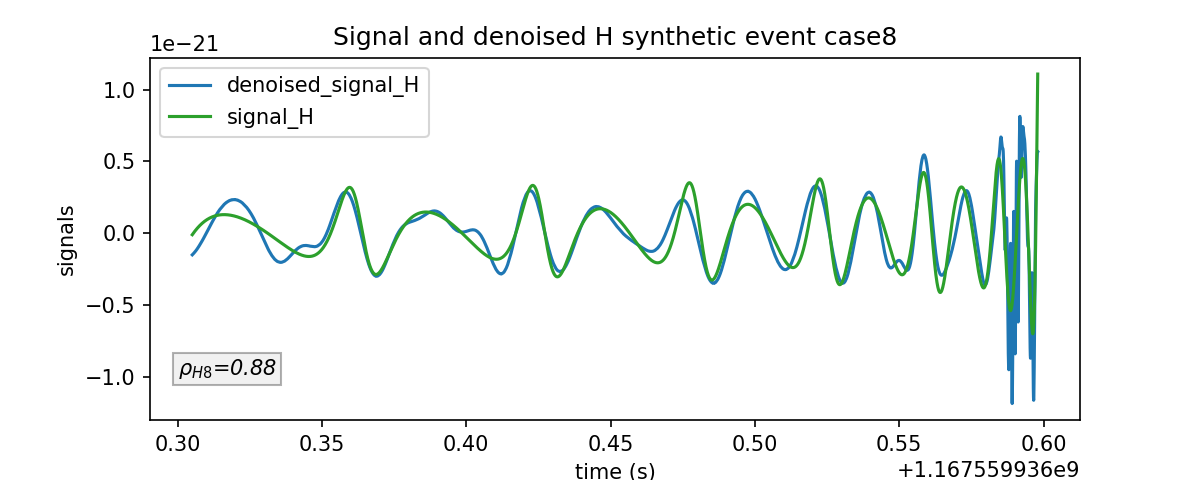}
	\includegraphics[clip,width=0.48\textwidth]{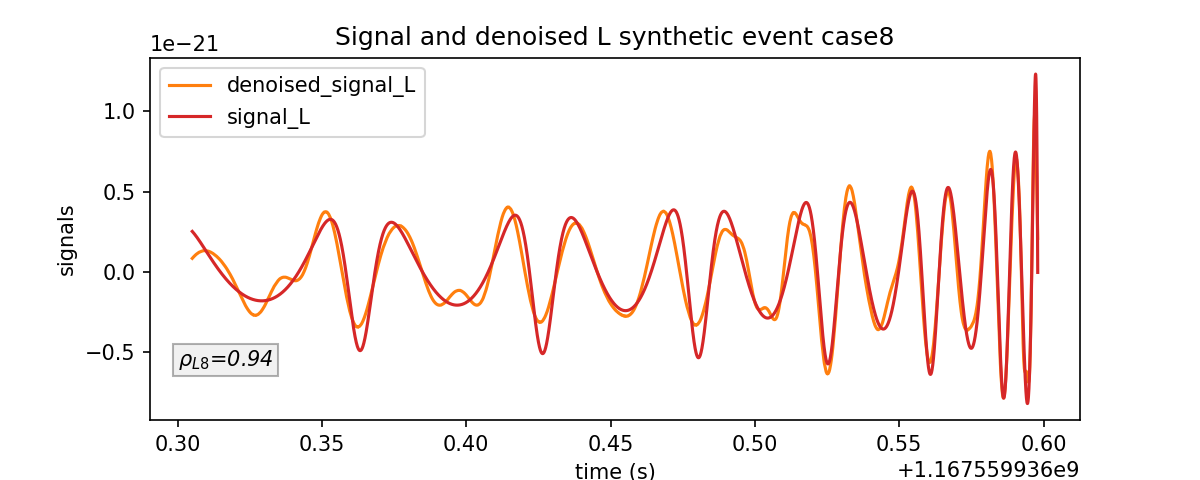}
	\caption{Top panel, comparison of original synthetic signal\_H with the extracted denoised signal
		from H strain with the noise of the GW170104 event.
		In the bottom panel the corresponding graphs for L.
		The inset shows the value of the correlation coefficient for both curves.
	}
	\label{fig:signal+denoised-8}
\end{figure}

\begin{figure}[H]
	\centering
	\includegraphics[clip,width=0.48\textwidth]{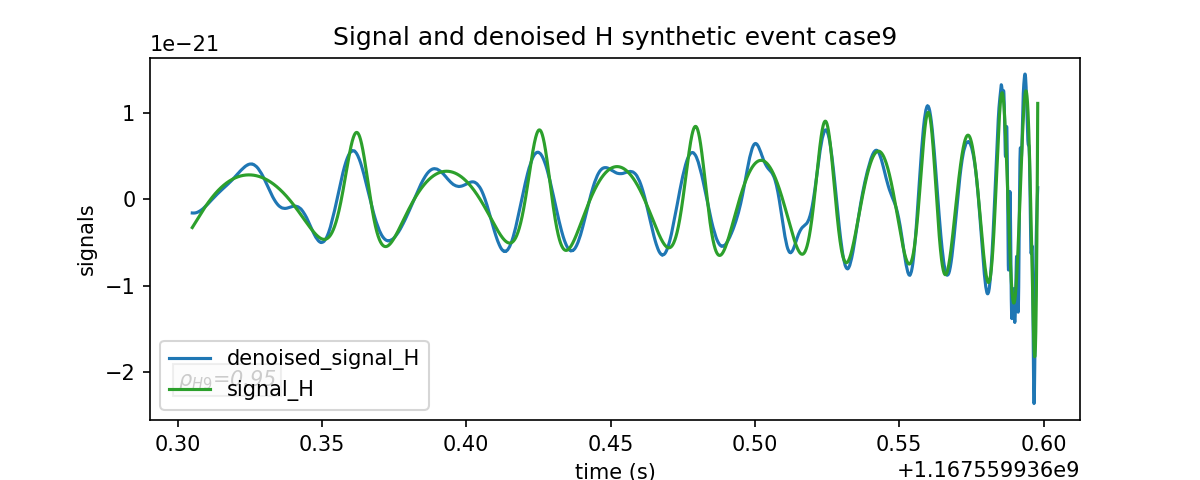}
	\includegraphics[clip,width=0.48\textwidth]{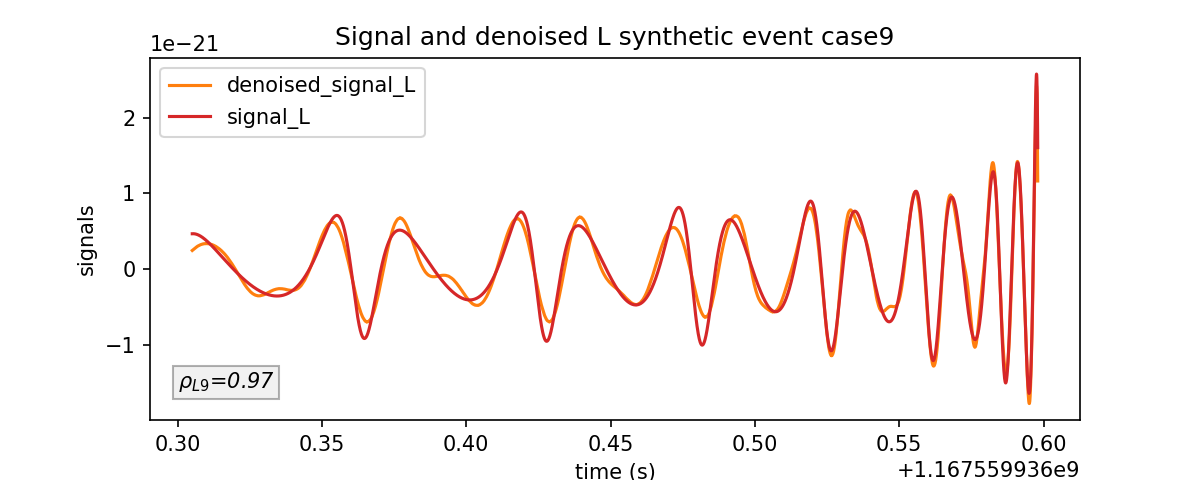}
	\caption{Top panel, comparison of original synthetic signal\_H with the extracted denoised signal
		from H strain with the noise of the GW170104 event.
		In the bottom panel the corresponding graphs for L.
		The inset shows the value of the correlation coefficient for both curves.
	}
	\label{fig:signal+denoised-9}
\end{figure}

\begin{figure}[H]
	\centering
	\includegraphics[clip,width=0.48\textwidth]{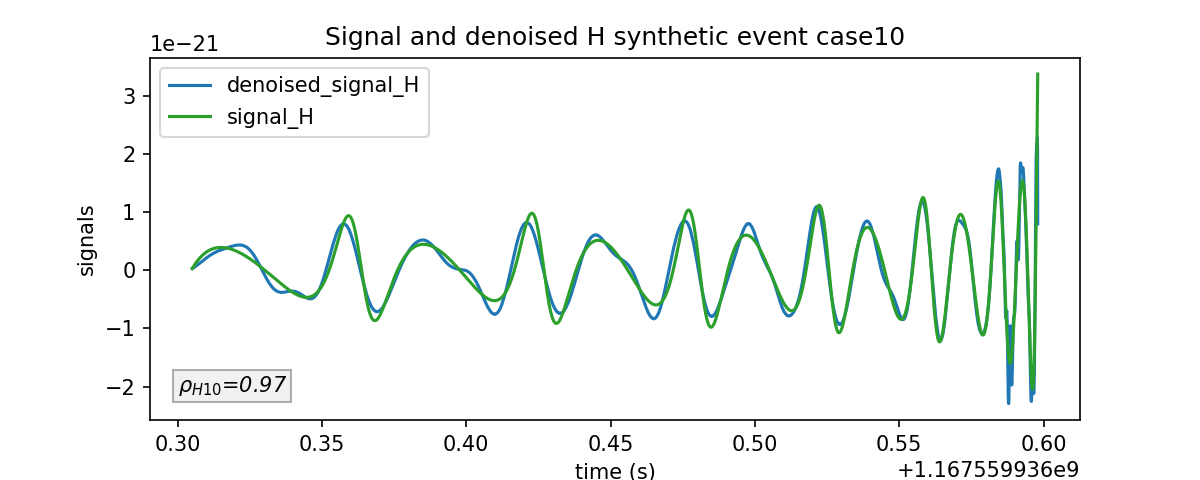}
	\includegraphics[clip,width=0.48\textwidth]{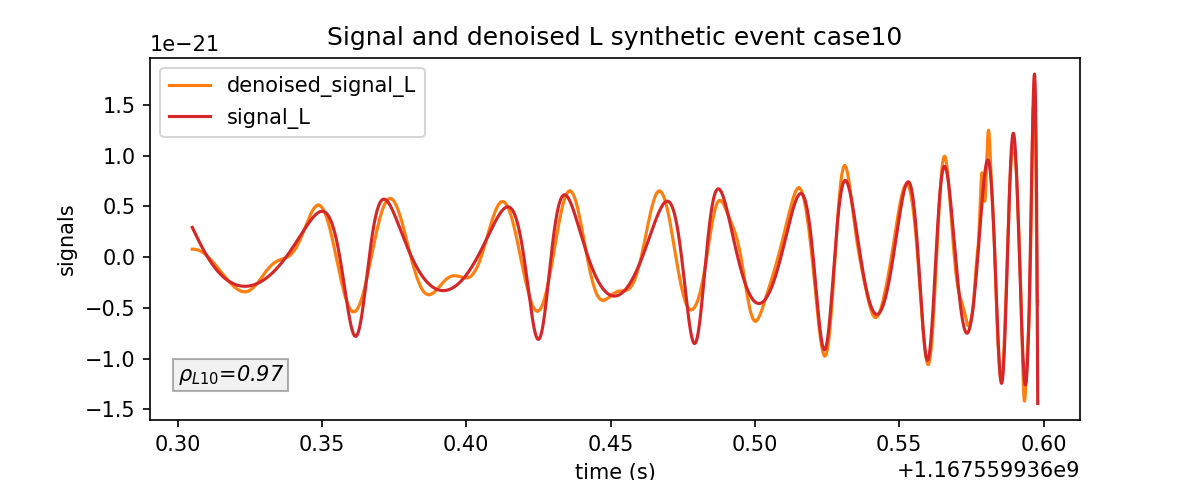}
	\caption{Top panel, comparison of original synthetic signal\_H with the extracted denoised signal
		from H strain with the noise of the GW170104 event.
		In the bottom panel the corresponding graphs for L.
		The inset shows the value of the correlation coefficient for both curves.
	}
	\label{fig:signal+denoised-10}
\end{figure}

The comparison of the original synthetic signals with the
corresponding denoised signals extracted from the strains with the noise of the GW170104 event
are shown in Figs. \ref{fig:signal+denoised-1}--\ref{fig:signal+denoised-10}.


\begin{figure}[H]
\centering
\includegraphics[clip,width=0.48\textwidth]{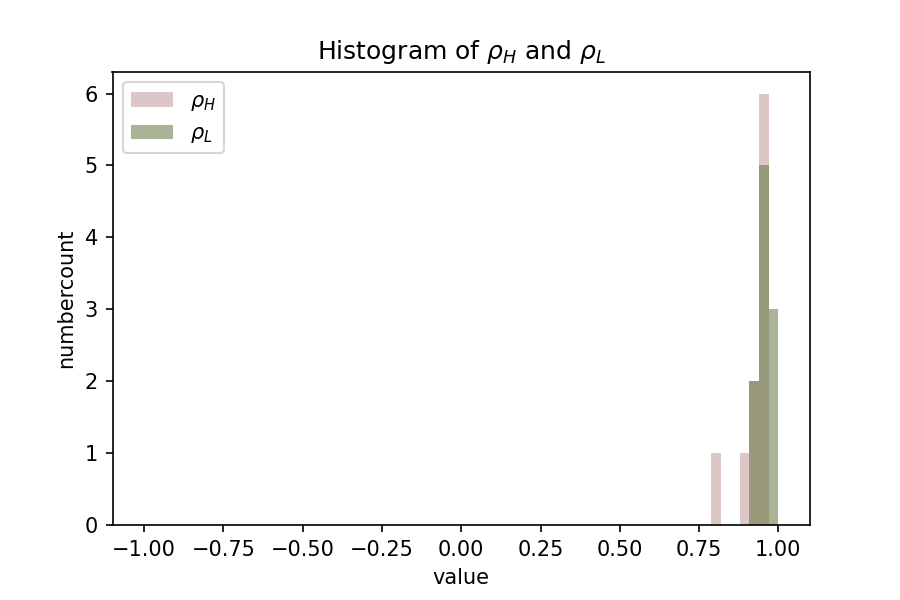}
\caption{Histograms of $\rho_H$ and $\rho_L$ superimposed.
The darker regions indicates same covering for both graphs.
}
\label{fig:rhoH_rhoL_histogram}
\end{figure}

It can be seen that the denoising techniques make an excellent job in
finding the signals in the strains with real noise, in both detectors.
This can be observed from the behavior of the curves in these graphs;
but it can also be reflected from the results in the calculation of
the correlation coefficients for the pair of curves whose histograms
are presented in Fig. \ref{fig:rhoH_rhoL_histogram}.
Although the sample size is relatively small, Fig. \ref{fig:rhoH_rhoL_histogram}, 
which presents histograms of $\rho_H$ and $\rho_L$,
shows that their values are predominantly concentrated near their respective maxima. 
This indicates that the denoising techniques effectively recover very well
the original signal across all tested scenarios.

To see other approaches for denoising and the reconstruction of gravitational-wave signals
and their result on injected simulated waveform please see \cite{Bacon:2022lsm,Chatterjee:2024alf}.

\subsection{Localization}

It is important to emphasize that localization based on time delay studies 
is the first and most fundamental method available in gravitational-wave observations. 
As evident from equation
\ref{eq:vX},
the presence of time delays between observatories 
is essential for the study
of the wave's content across multiple detectors.

However, the time delay between two observatories does not define a precise triangular 
localization but rather a delay ring in the sky. 
Therefore, the first estimate for the source's location is always a region surrounding this delay ring.

Given this, any additional localization method is expected to yield results that 
are consistent with and situated near the delay ring. 
In this study, we construct a Gaussian region around the nominal delay ring, 
providing an estimate of the probability distribution of the sky localization based solely 
on time delay considerations.

In methods where probability density maps are available, constructing confidence 
level regions is straightforward. 
However in our case, we do not have probability density maps, nor have we estimated any from our approach.
Instead, we construct sky maps based on measures derived from observational noisy data after appropriate processing.
To estimate confidence level regions, we apply Chebyshev’s inequality \citep{Casella2002}, treating our measure 
$M_i$ as an indicator of a signal within a random variable process. 
This approach allows us to define confidence regions in a rigorous manner, 
even in the absence of conventional probability distributions.
In this way, high values of $M_i$ give us indications of a signal, as compared with respect
to the mean value $\mu$ in units of the standard deviation $\sigma$.
So for example to estimate a 0.9 region, one looks for the boundary value
$v_0 = \mu + t \sigma$; with $\frac{1}{t^2}=0.1=1-0.9$. It should be noted that $M_i$ is positive definite,
and so $v_0$ is greater than $\mu$.
The difficulty with this estimate for the boundary value is that although $M_i$ is giving us a signal
over some noise; it does not provide a single signal, but also yields several phantom
and additional non physical maxima. This has as a consequence, that the sample mean value
overestimates the noise mean value, and the sample standard deviation overestimates
the noise standard deviation.
To cope with this situation, we could select a diminishing parameter $c_{90}$ that,
in order to deal with the phantom repetition, should be $\leqslant 0.5$.
In the previous discussion, we did not explicitly account for the fact that 
we already have an initial estimate of the probability distribution for 
the source's sky localization; this estimate is provided by the Gaussian regions. 
These regions serve as a crucial starting point, refining our understanding of 
the source's potential location. 
In the following analysis, we incorporate this information 
to further improve our localization accuracy.

The need for a diminishing factor is naturally resolved if we instead use 
the ultimate measure $M_u = \{| G_\text{ring} * M_i |\}$;
This measure represents the normalized product of the Gaussian distribution,
interpreted as a probability estimate, associated with the delay ring 
(as shown in Fig. \ref{fig:delay-ring+gaussian+Wmap}) and the initial measure $M_i$.
By incorporating this approach, we ensure a more refined and reliable localization estimate.
It is important to note that the more precise measure $M_u$ 
naturally emerges from the inferable knowledge about the source location 
derived from time delay considerations. In this formulation, $M_u$ 
effectively suppresses phantom signals when they are positioned away from the delay ring. 
As a result, we can apply Chebyshev's inequalities directly, 
without requiring any additional corrective factors.

\begin{figure}[H]
\centering
\includegraphics[clip,width=0.44\textwidth]
{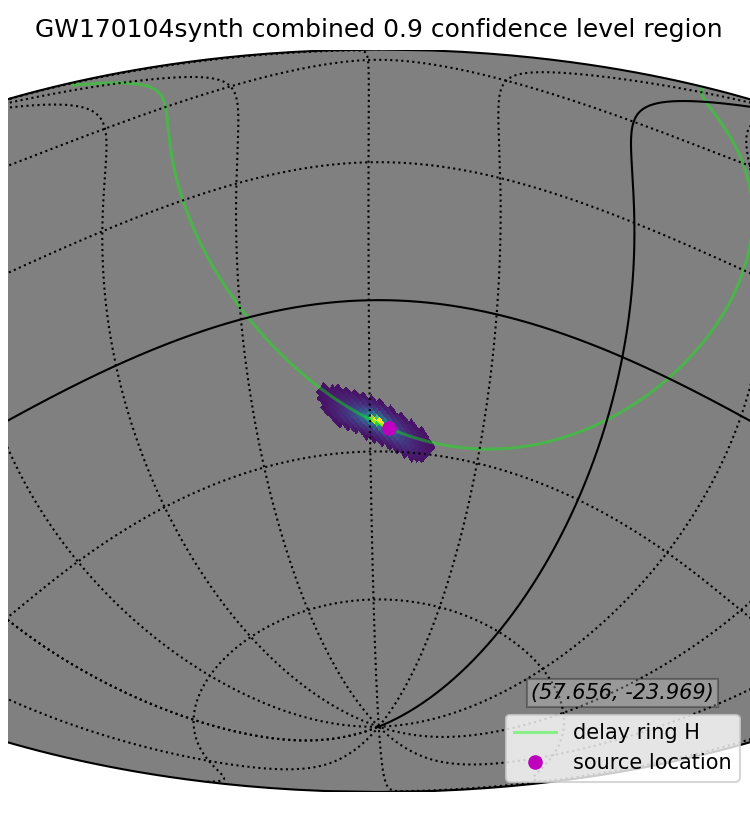}
\caption{This Mollweide view, shows the final 0.9 region for the GW170104synth synthetic signal case 1.
	The magenta circle indicates the position of the synthetic signal.
	The two numbers between parentheses denote the preferred central position as longitude and latitude.
}
\label{fig:synth_v3_c1}
\end{figure}

\begin{figure}[H]
\centering
\includegraphics[clip,width=0.44\textwidth]
{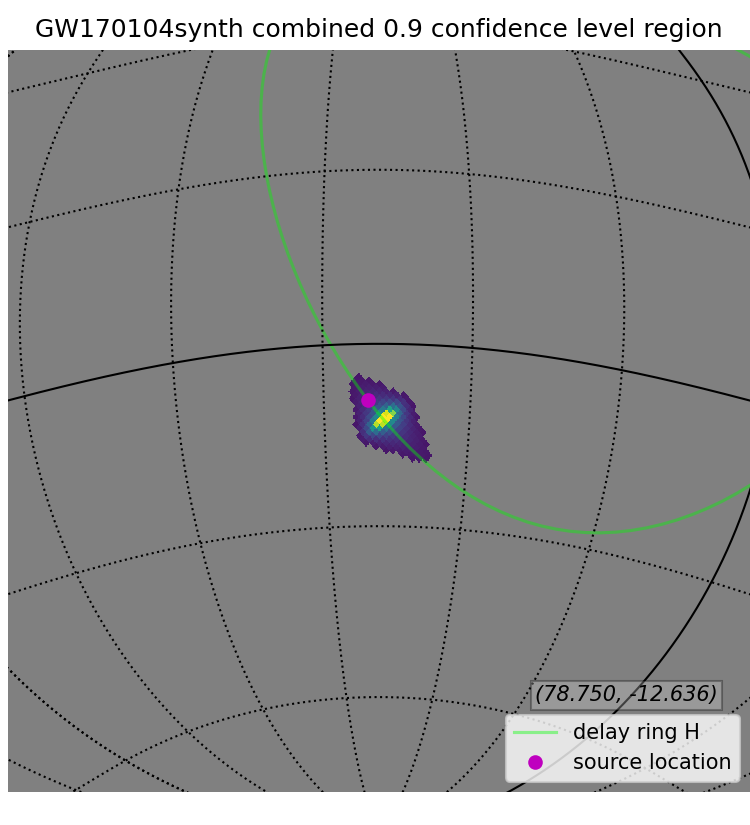}
\caption{This Mollweide view, shows the final 0.9 region for the GW170104synth synthetic signal case 2.
	The magenta circle indicates the position of the synthetic signal.
	The two numbers between parentheses denote the preferred central position as longitude and latitude.
}
\label{fig:synth_v3_c2}
\end{figure}

\begin{figure}[H]
\centering
\includegraphics[clip,width=0.44\textwidth]
{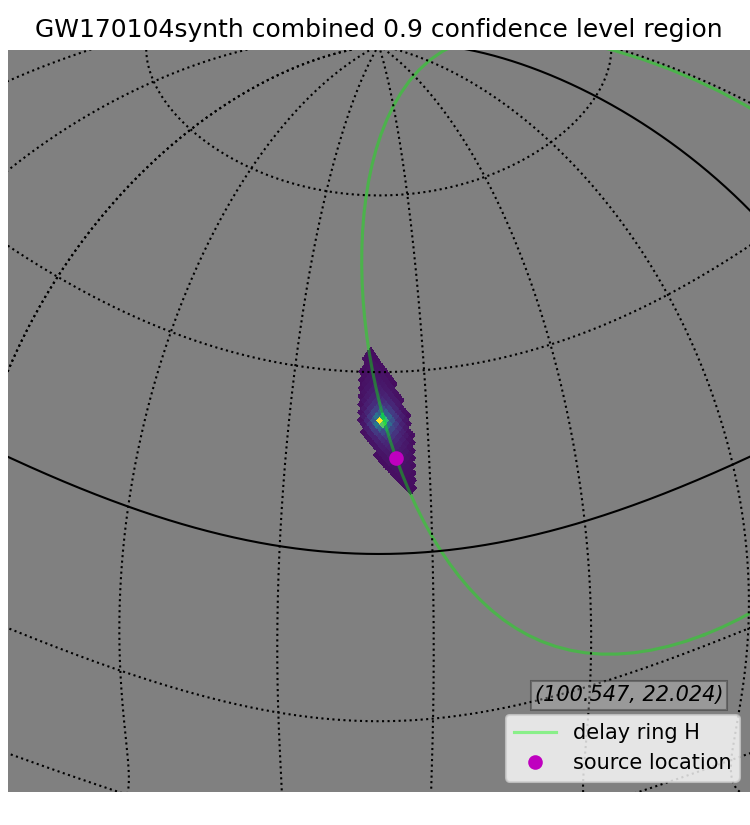}
\caption{This Mollweide view, shows the final 0.9 region for the GW170104synth synthetic signal case 3.
	The magenta circle indicates the position of the synthetic signal.
	The two numbers between parentheses denote the preferred central position as longitude and latitude.
}
\label{fig:synth_v3_c3}
\end{figure}

\begin{figure}[H]
\centering
\includegraphics[clip,width=0.44\textwidth]
{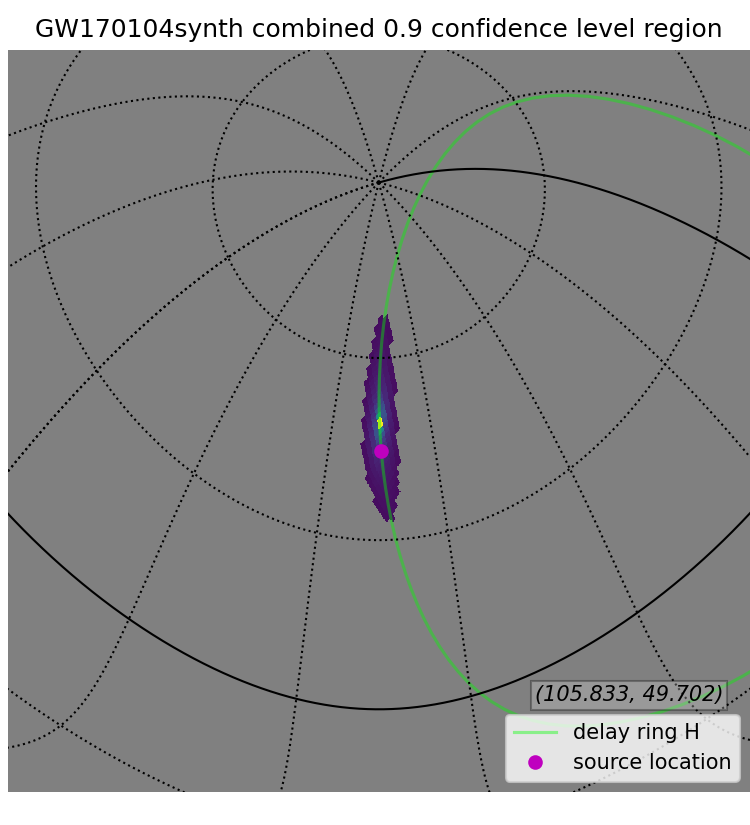}
\caption{This Mollweide view, shows the final 0.9 region for the GW170104synth synthetic signal case 4.
	The magenta circle indicates the position of the synthetic signal.
	The two numbers between parentheses denote the preferred central position as longitude and latitude.
}
\label{fig:synth_v3_c4}
\end{figure}

\begin{figure}[H]
\centering
\includegraphics[clip,width=0.44\textwidth]
{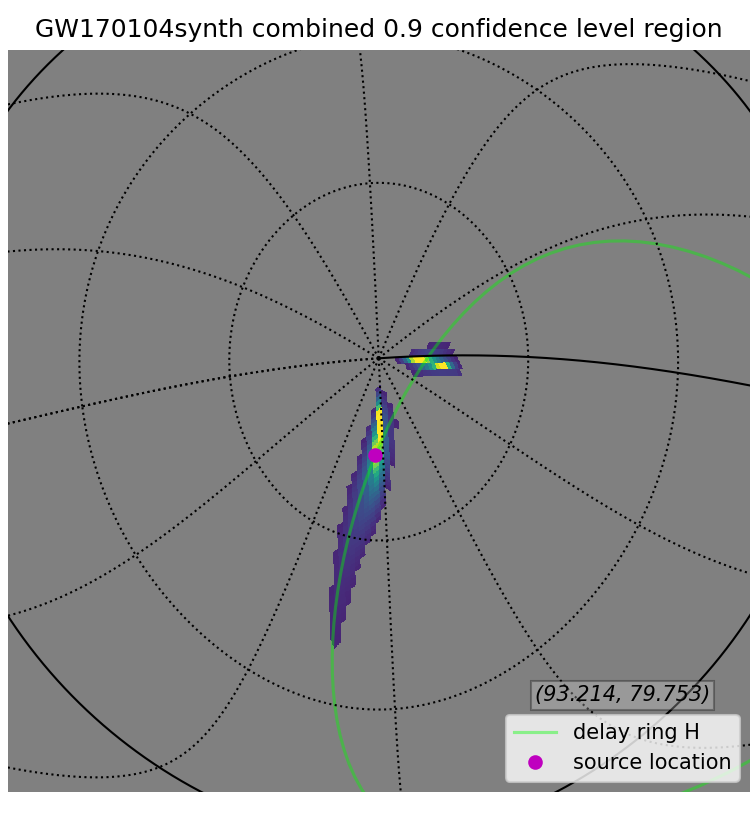}
\caption{This Mollweide view, shows the final 0.9 region for the GW170104synth synthetic signal case 5.
	The magenta circle indicates the position of the synthetic signal.
	The two numbers between parentheses denote the preferred central position as longitude and latitude.
}
\label{fig:synth_v3_c5}
\end{figure}

\begin{figure}[H]
\centering
\includegraphics[clip,width=0.44\textwidth]
{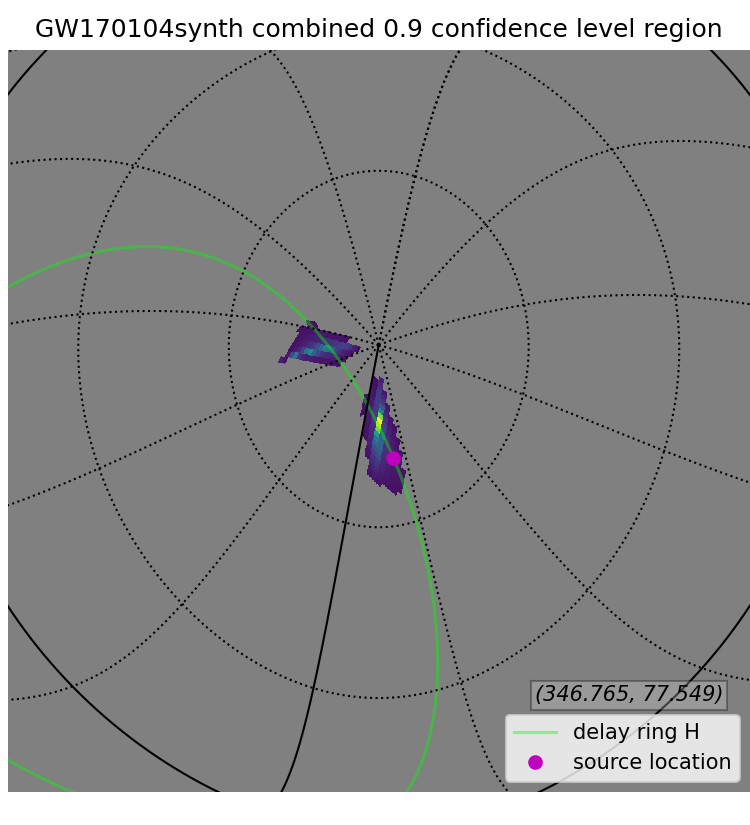}
\caption{This Mollweide view, shows the final 0.9 region for the GW170104synth synthetic signal case 6.
	The magenta circle indicates the position of the synthetic signal.
	The two numbers between parentheses denote the preferred central position as longitude and latitude.
}
\label{fig:synth_v3_c6}
\end{figure}

\begin{figure}[H]
\centering
\includegraphics[clip,width=0.44\textwidth]
{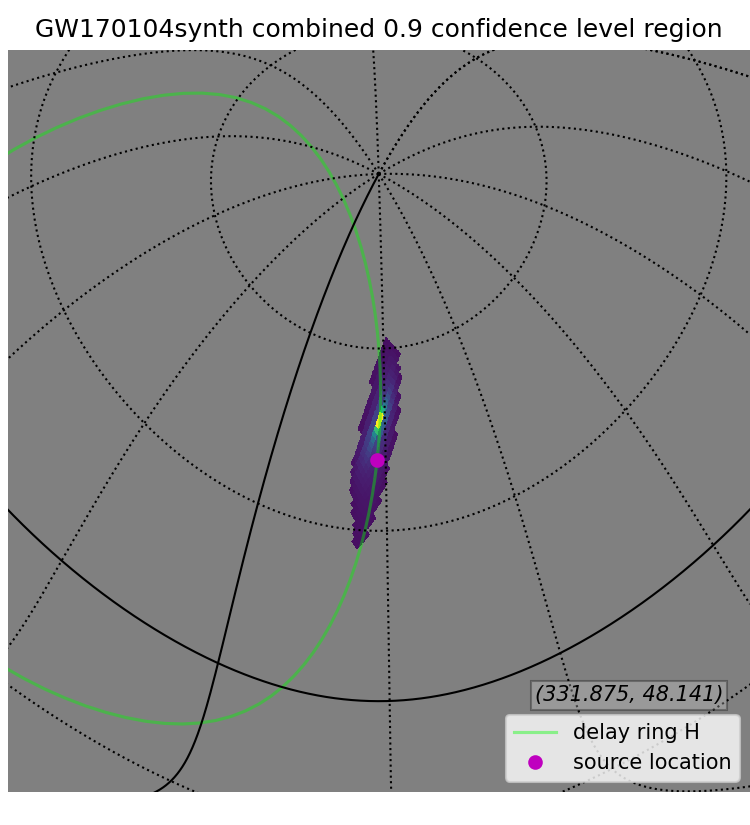}
\caption{This Mollweide view, shows the final 0.9 region for the GW170104synth synthetic signal case 7.
	The magenta circle indicates the position of the synthetic signal.
	The two numbers between parentheses denote the preferred central position as longitude and latitude.
}
\label{fig:synth_v3_c7}
\end{figure}

\begin{figure}[H]
\centering
\includegraphics[clip,width=0.44\textwidth]
{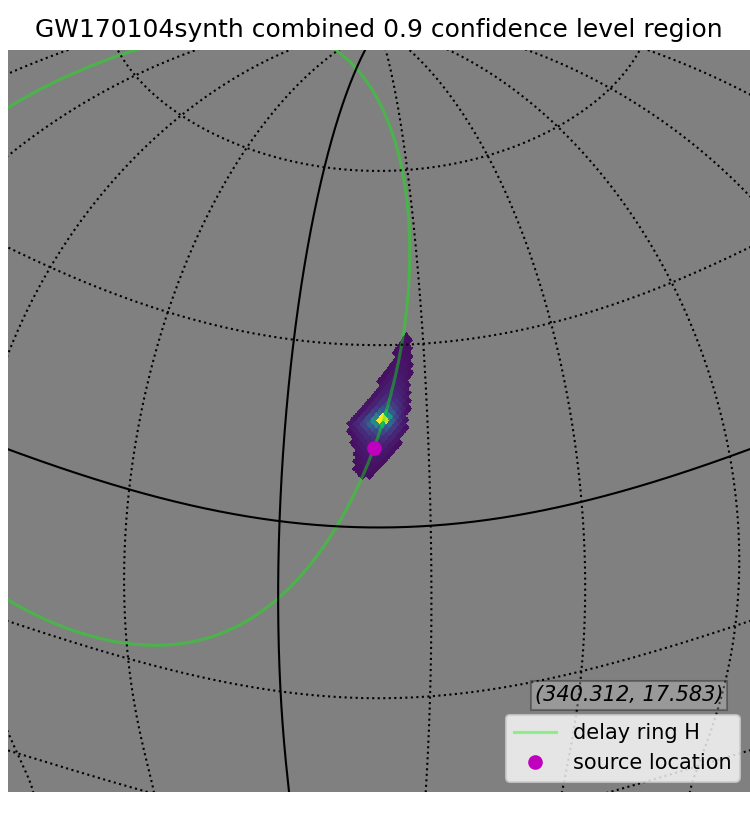}
\caption{This Mollweide view, shows the final 0.9 region for the GW170104synth synthetic signal case 8.
	The magenta circle indicates the position of the synthetic signal.
	The two numbers between parentheses denote the preferred central position as longitude and latitude.
}
\label{fig:synth_v3_c8}
\end{figure}

\begin{figure}[H]
\centering
\includegraphics[clip,width=0.44\textwidth]
{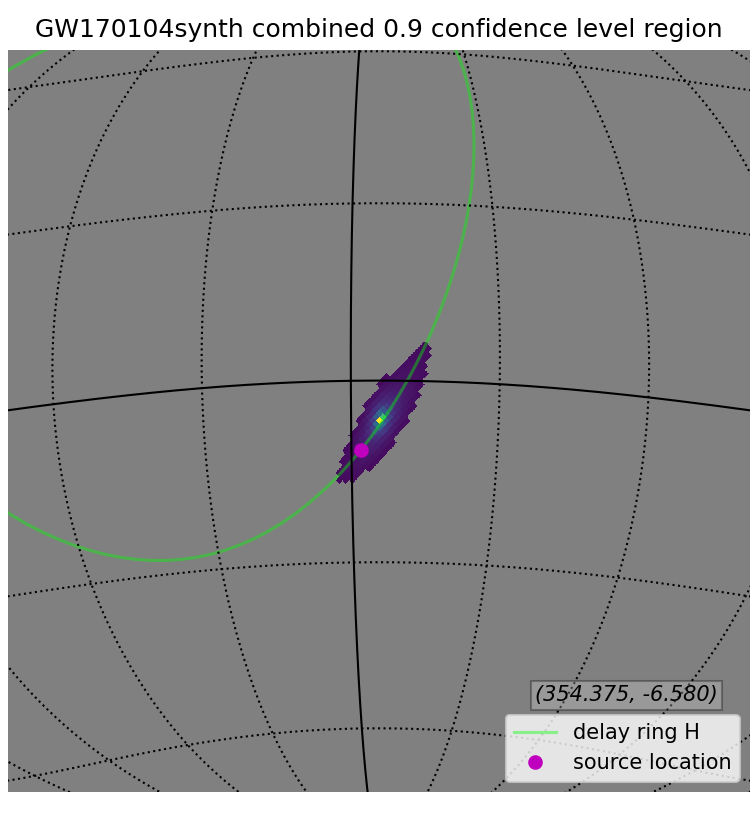}
\caption{This Mollweide view, shows the final 0.9 region for the GW170104synth synthetic signal case 9.
	The magenta circle indicates the position of the synthetic signal.
	The two numbers between parentheses denote the preferred central position as longitude and latitude.
}
\label{fig:synth_v3_c9}
\end{figure}

\begin{figure}[H]
\centering
\includegraphics[clip,width=0.44\textwidth]
{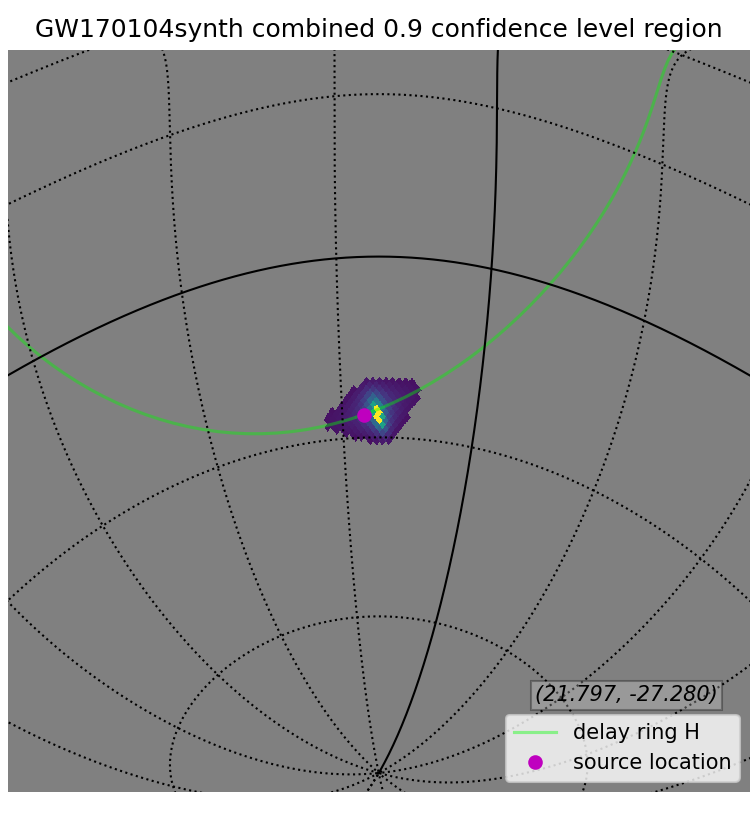}
\caption{This Mollweide view, shows the final 0.9 region for the GW170104synth synthetic signal case 10.
	The magenta circle indicates the position of the synthetic signal.
	The two numbers between parentheses denote the preferred central position as longitude and latitude.
}
\label{fig:synth_v3_c10}
\end{figure}

The final $M_u$ 0.9 region constructed from the direct application of the Chebyshev inequality to case 1
is shown in Fig.  \ref{fig:synth_v3_c1}.
The final 0.9 regions for the other cases 2--10 are shown in Figs. \ref{fig:synth_v3_c2}-\ref{fig:synth_v3_c10}.

The irregular shapes and saw type edges are due to the discretization effects on the sphere.

The area of the 0.9 regions in square degrees for each case are 
248, 193, 165, 183, 410, 174, 196, 169, 175 and 162
respectively.
The actual errors in the preferred pixel's position and the exact location of the synthetic sources 
are in all cases few degrees; 
which shows reasonable precision properties of this localization method.

There are some works studying theoretical predictions on the precision of localization
for gravitational wave detectors.
For example in \cite{Fairhurst:2009,Fairhurst:2010is},
the author examines 
the accuracy of source localization using a network of detectors 
by analyzing how gravitational-wave detectors measure transient signals 
based solely on timing information at each site. 
However, these studies do not take into account the polarization mode content 
of the GW. As a result, we anticipate that our method will 
yield different uncertainty estimates due to its incorporation of polarization mode information.
%
Instead in \cite{Wen:2010cr} 
the authors dealt with the problem of obtaining
general geometrical expressions for the angular resolution of an arbitrary
network of interferometric GW detectors when the arrival time of a GW is unknown;
which is different from our case because we use the delay time information
in our procedure. In any case, in this last reference they propose that for 
the two LIGO detector case one would expect an error of about $5^\circ$;
which is consistent with our results that we have just presented.

\subsection{Reconstruction of the spin-2 polarization modes}

\begin{figure}[H]
\centering
\includegraphics[clip,width=0.48\textwidth]{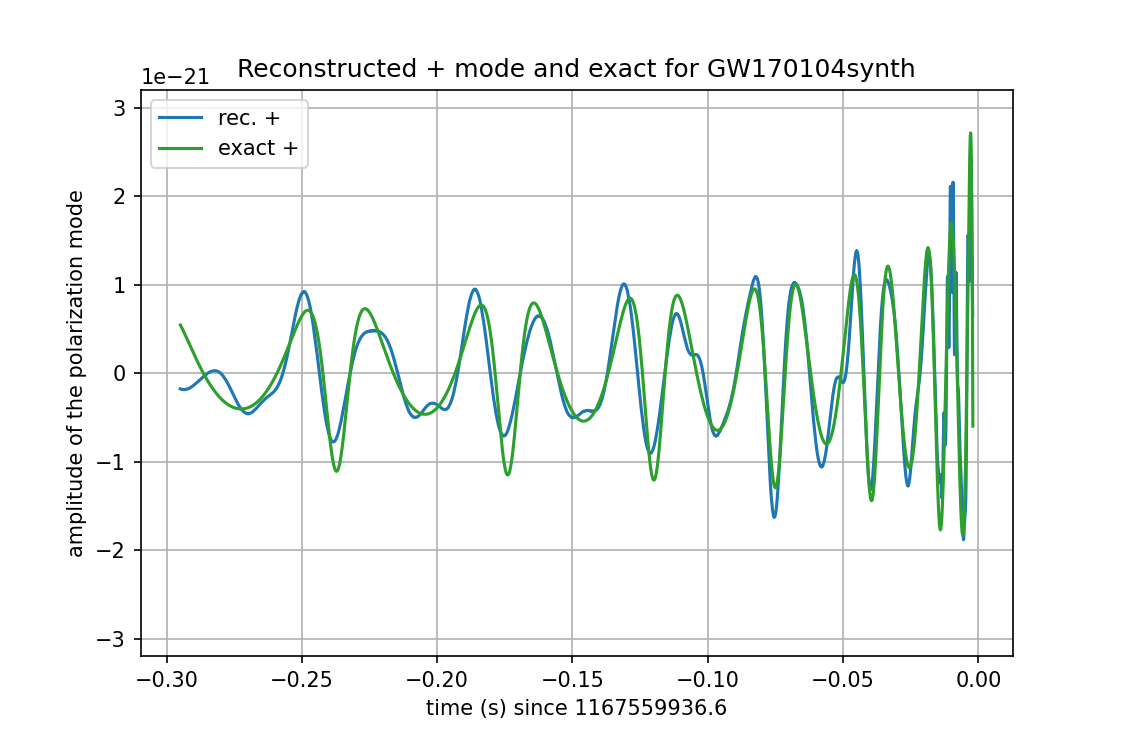}
\caption{Reconstructed + polarization mode and exact initial corresponding mode;
	used in the injected signal for first case. }
\label{fig:synth_+mode}
\end{figure}

\begin{figure}[H]
\centering
\includegraphics[clip,width=0.48\textwidth]{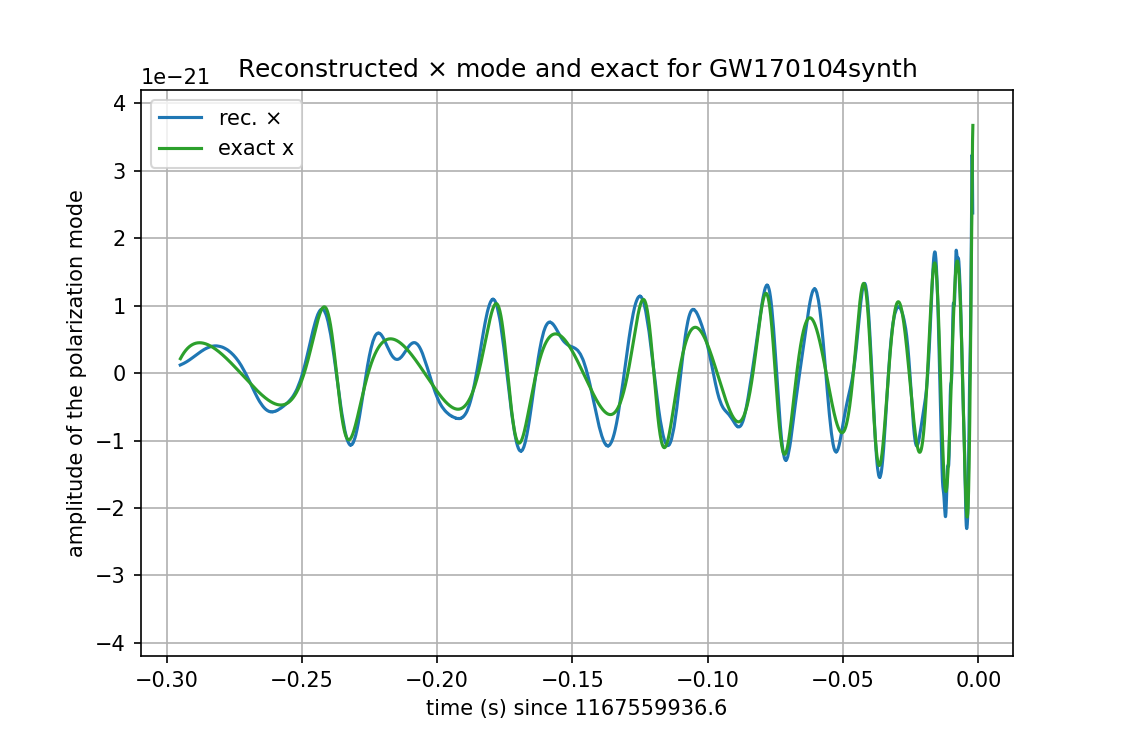}
\caption{Reconstructed $\times$ polarization mode and exact initial corresponding mode;
	used in the injected signal for first case. }
\label{fig:synth_xmode}
\end{figure}

Regarding the measurement of the polarization modes, we show
in Figs. \ref{fig:synth_+mode} and \ref{fig:synth_xmode}
the reconstructed polarization modes (in blue) along with the 
exact injected version (in green), that were used to build the synthetic
signal at the new source position for the case 1.
It can be noted that the slight error in the determination of the sky localization of the synthetic source
does not preclude an excellent reconstruction of the original polarization modes.

\section{Localization of GW170104}\label{sec:localiz}

By `localization' we mean the sky localization, that is,
the celestial position of the source.

\begin{figure}[H]
\centering
\includegraphics[clip,width=0.48\textwidth]{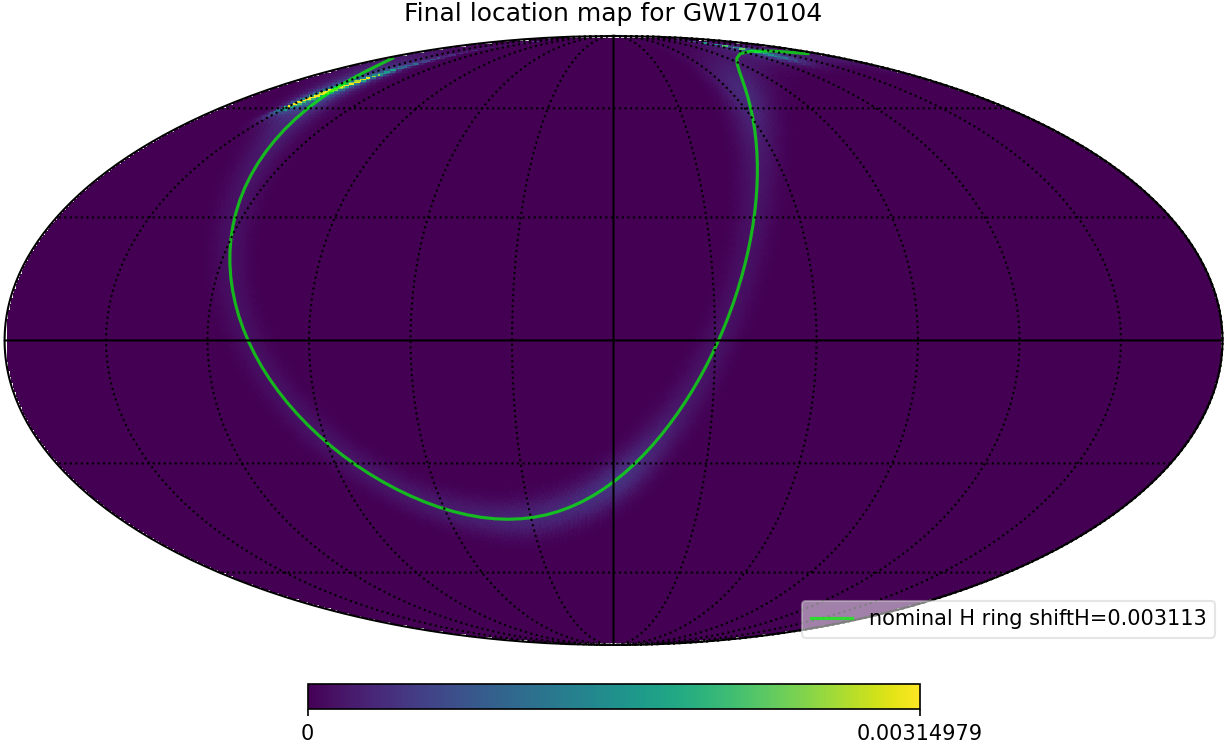}
\caption{Final sky localization for the source of the GW170104 event,
where one can see a signal in the northern east part of the nominal ring.
}
\label{fig:loc-gw170104}
\end{figure}

The result of our procedure  for the localization of event GW170104 is presented 
in Fig. \ref{fig:loc-gw170104}.
This final location is obtained 
from the measure $M_u$ described above.
The preliminary location, is shown in Fig. \ref{fig:loc_1-gw170104} in appendix \ref{sec:loc-chirp}.

\begin{figure}[H]
\centering
\includegraphics[clip,width=0.44\textwidth]{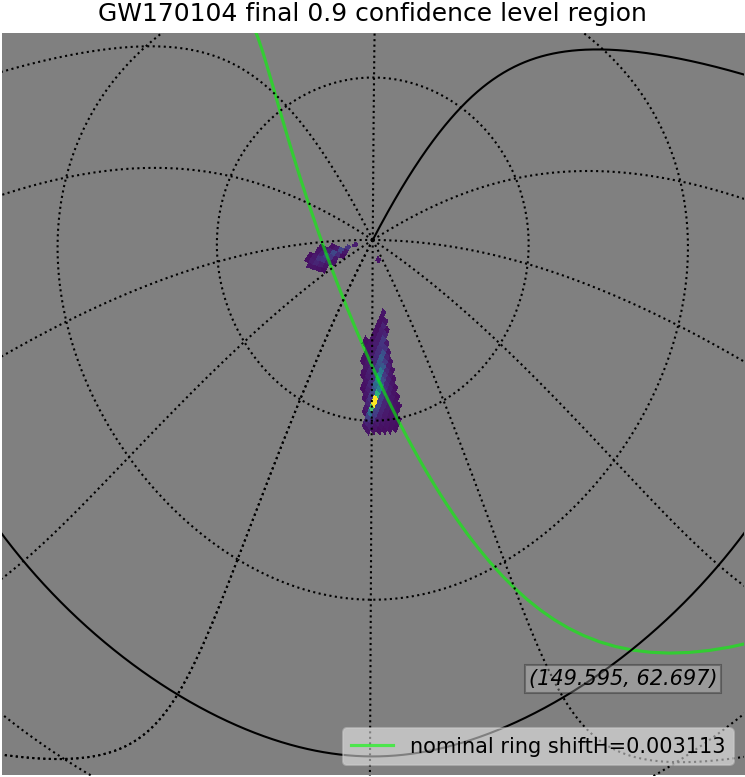}
\caption{Final sky localization for the source of the GW170104 event,
	with 0.9 confidence level region of measure $M_u$.
	The two numbers between parentheses denote the preferred central position as longitude and latitude.
}
\label{fig:loc-gw170104-v3}
\end{figure}

To manipulate and show data on the celestial sphere we use
the {\sc python} package {\sf healpy} employing 49152 pixels for the sphere.
In our approach we do not use probabilities, but instead we use measures
on the sphere $(M_u)$ 
whose maximum indicates the location of the source
in the sky. We have not related this function to any probability distribution.
For this reason, we estimate regions associated to level of significance
using the Chebyshev inequality
\citep{Casella2002} 
to our measure.
Choosing a level of significance
$\alpha = 0.1$%
, we can select
the region in the sky; where this criteria is satisfied, and so
we characterize the confidence level region of
$\gamma = (1-\alpha)=0.9$.
The final region of interest at the 0.1 level of significance,
is obtained from $M_u$; which is shown in Fig. \ref{fig:loc-gw170104-v3}.
Further discussion is presented in section \ref{sec:synth};
but we should probably note here that the meaning
or our confidence level region, based on the values of a measure on the sphere 
(where the location is indicated by its maximum values over noise)
is different from the notion of confidence area (or region) defined on the
values of a probability density function.

The measure shown in Fig. \ref{fig:loc-gw170104-v3}
has maximum at celestial coordinates, longitude and latitude, (150, 63) degrees;
and the area covered by pixels with a confidence level of 0.9 or higher is 142 square degrees.
The global maximum is found in a single distinct pixel.

\begin{figure}[H]
\centering
\includegraphics[clip,width=0.48\textwidth]{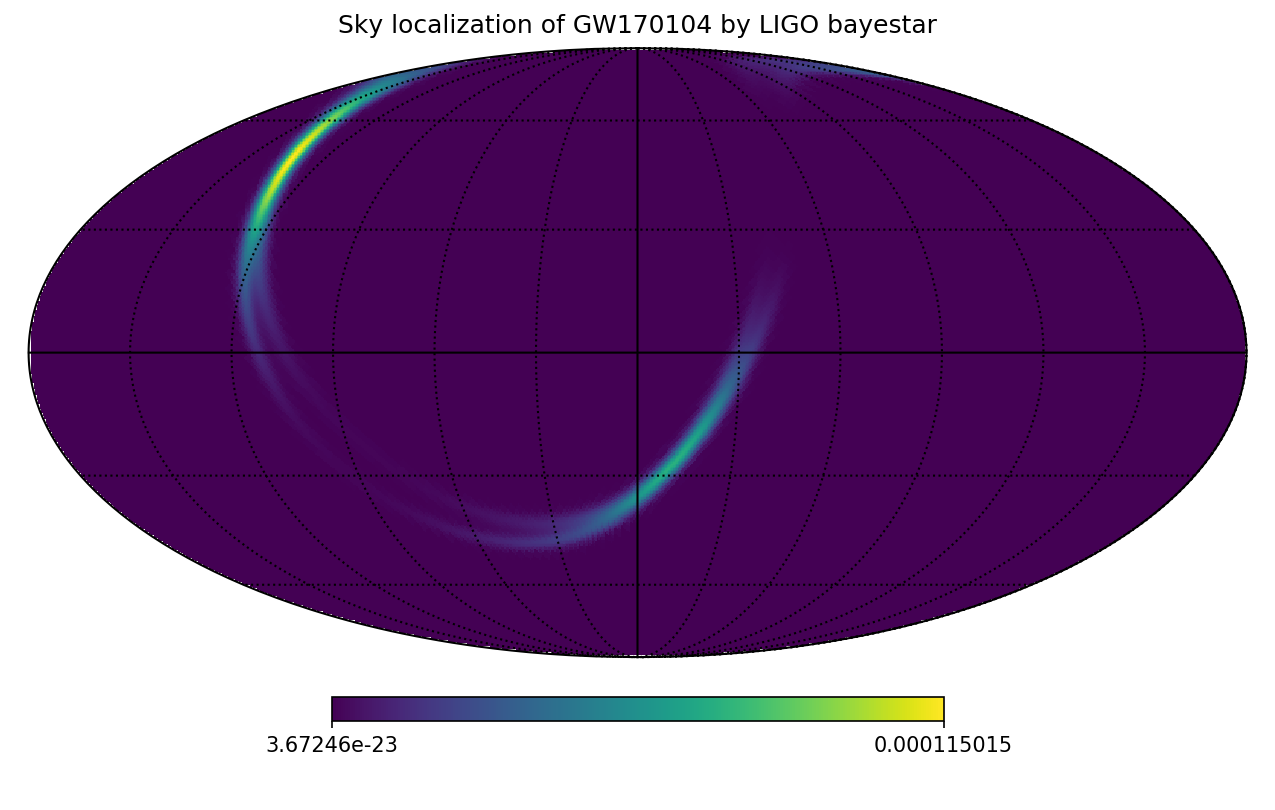}
\caption{Sky localization from LIGO team with  Bayestar method.
}
\label{fig:loc-ligo-bayestar}
\end{figure}

\begin{figure}[H]
\centering
\includegraphics[clip,width=0.48\textwidth]{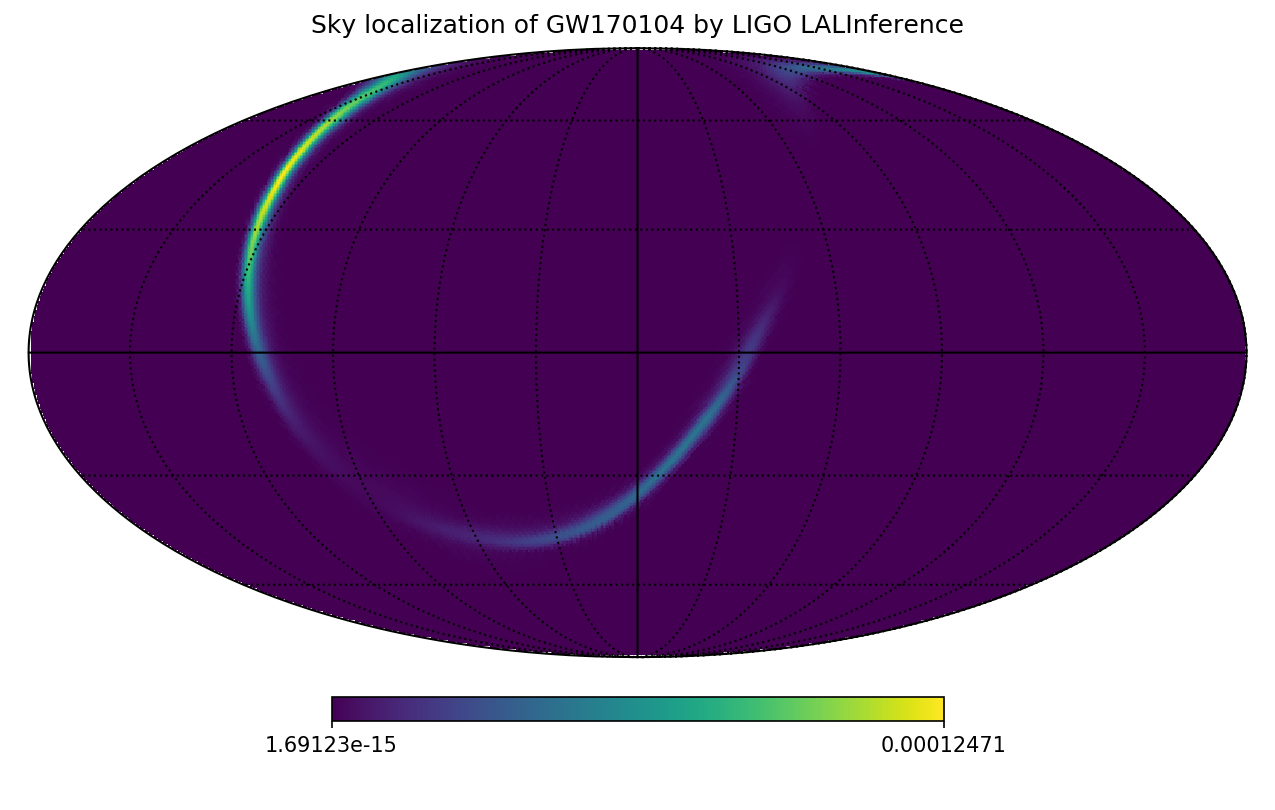}
\caption{Sky localization from LIGO team with LALInference method.
}
\label{fig:loc-ligo-LALI}
\end{figure}

Our final localization is close to one of the local maxima communicated by
the LIGO Scientific Collaboration\citep{LIGO_loc_GW170104} that we reproduce 
in Figs. \ref{fig:loc-ligo-bayestar} and \ref{fig:loc-ligo-LALI}.
Getting into more detail, it can be seen that both LIGO approaches
yield two local maxima on the delay ring; the absolute maximum in the northeastern sector
and the second local maximum in the southwest sector.
The procedure we are presenting here is very different from the LIGO approaches
and so probably this is the main reason that our result does not coincide
\add{exactly}
with their absolute maximum,
\add{although our maximum is within the region of their main maximum area}. 
At this time, studying a single event one can not state which of these approaches
is more accurate; mainly because for this event there are no electromagnetic counter parts.
We should not compare the surface area of our confidence level region
derived from a measure, with 
the confidence limit regions defined by probability density functions
as previously mentioned; since they have fundamentally different meaning.
It would be convenient to perform further studies to clarify these issues. 
To address this, we plan to apply our procedure to a diverse set of 
real gravitational-wave events in future analyses.

It is  worth highlighting that in order to discuss the polarization mode content
of a GW one needs to refer to a frame.
In the next section we present the reconstructed polarization modes
in terms of the basis obtained in the localization procedure;
but later in appendix
\ref{sec:spin2-polar}
we also present 
the PMs in other frames related by a 
differential polarization angle
$\Delta\psi$.

\section{Reconstruction of the spin-2 polarization modes of GW170104}\label{sec:PM}

\begin{figure}[H]
\centering
\includegraphics[clip,width=0.48\textwidth]{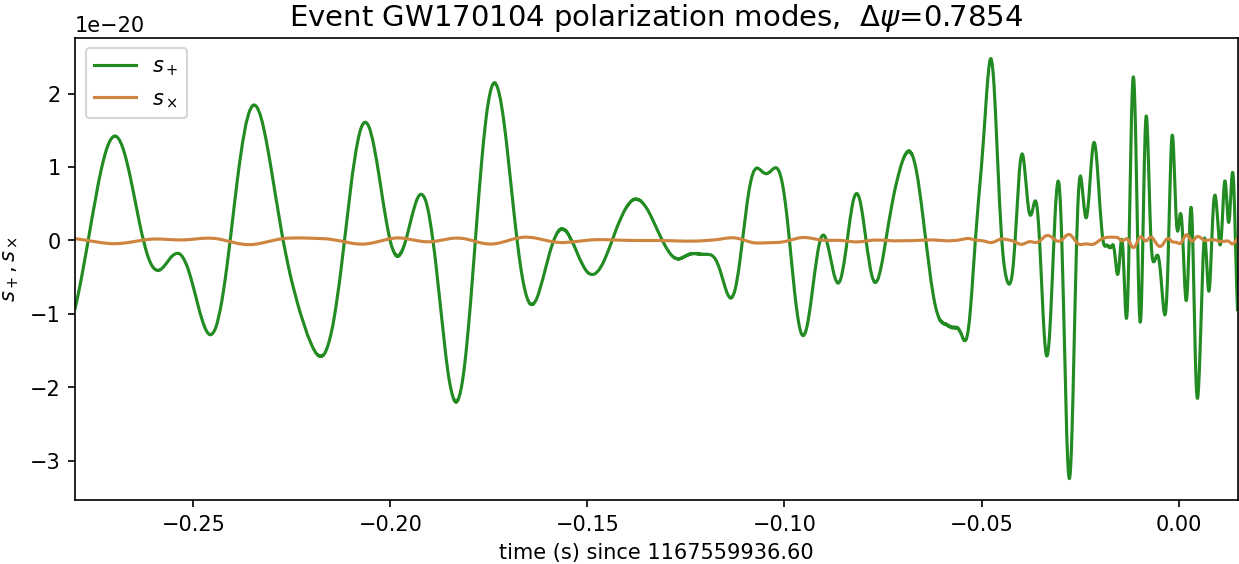}
\caption{Polarization modes + and $\times$ close to the reference time of the GW170104 event.}
\label{fig:s+_sx}
\end{figure}

In the previous section we have presented the results on the localization
of the source. Although the 0.9 region covers several square degrees,
our measure has maximum at a single pixel; which we take
as the position of the source for this part of our work.
Then, knowing the localization in the celestial sphere of the source 
and applying denoising techniques as explained above
we can perform the reconstruction of the PMs which we show
in Fig. \ref{fig:s+_sx}.
\add{The graph in Fig.} \ref{fig:s+_sx}
\add{shows that the $\times$ component is significantly
smaller than the + component; 
although small variations in the $\times$ component remain visible. 
This demonstrates that both observatories recorded nearly identical spin-2 
polarization mode components, and that the selected polarization frame 
happened to be one in which this effect is clearly observable.}

\add{It is noteworthy that the time series for $s_+$ exhibits an irregular 
shape that does not resemble the denoised signals shown in Figs.}
\ref{fig:denoised-Hstrain} and \ref{fig:denoised-Lstrain}
\add{which more closely match expected chirp-like signals. 
To understand this behavior, it is important to recognize that in cases 
where one detector has essentially recorded the negative of the signal 
recorded by the other detector, the algebraic reconstruction of polarization modes 
involves a subtraction between both signals, as demonstrated by the equations in Section}
 \ref{sec:dec-pol-modes}.
\add{Specifically, we have verified that the factors $F_{\times E}$ and $F_{\times D}$ 
have opposite signs at the source location, and similarly, the factors $F_{+E}$ 
and $F_{+D}$ also have opposite signs at this location. 
This is expected when the signals at the observatories are anticorrelated 
and consequently results in cancellation of both signals during 
polarization mode calculation. 
This property is not a limitation of our procedure but rather a general characteristic 
of the algebra involving data from two observatories, as derived from eq.}
 \ref{eq:vX}.
\add{We therefore interpret the preceding time series for $s_+$ as exhibiting 
a high noise contribution, where the signal-to-noise ratio is insufficient 
to allow physically meaningful reconstruction of the polarization modes (PMs). 
Consequently, one expects significantly better results for cases where the 
source localization produces signals at the two observatories 
that are less strongly (anti)correlated.}

In any case, there are no other
comparable explicit reconstruction
of the polarization modes + and $\times$ of a GW signal in the literature,
as those shown with the time series in Fig. \ref{fig:s+_sx}\,
close to the reference time of the GW170104 event.
\add{The subject of estimating errors for the reconstructed polarization modes
for these special cases of great correlation between the signals of the observatories
is complex and deserves an specific study,
at present we concentrate in the presentation of the results of applying this
procedure to this event and postpone for future works the study of additional contributions
to errors due to correlation of the signals.
}

\begin{figure}[H]
\centering
\includegraphics[clip,width=0.48\textwidth]{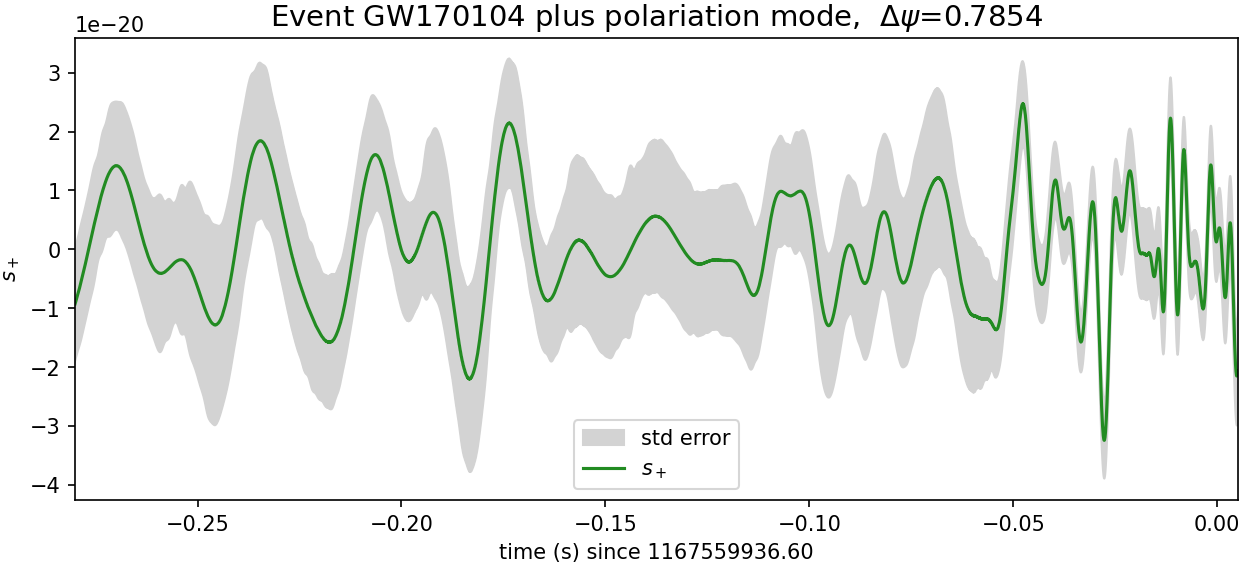}
\includegraphics[clip,width=0.48\textwidth]{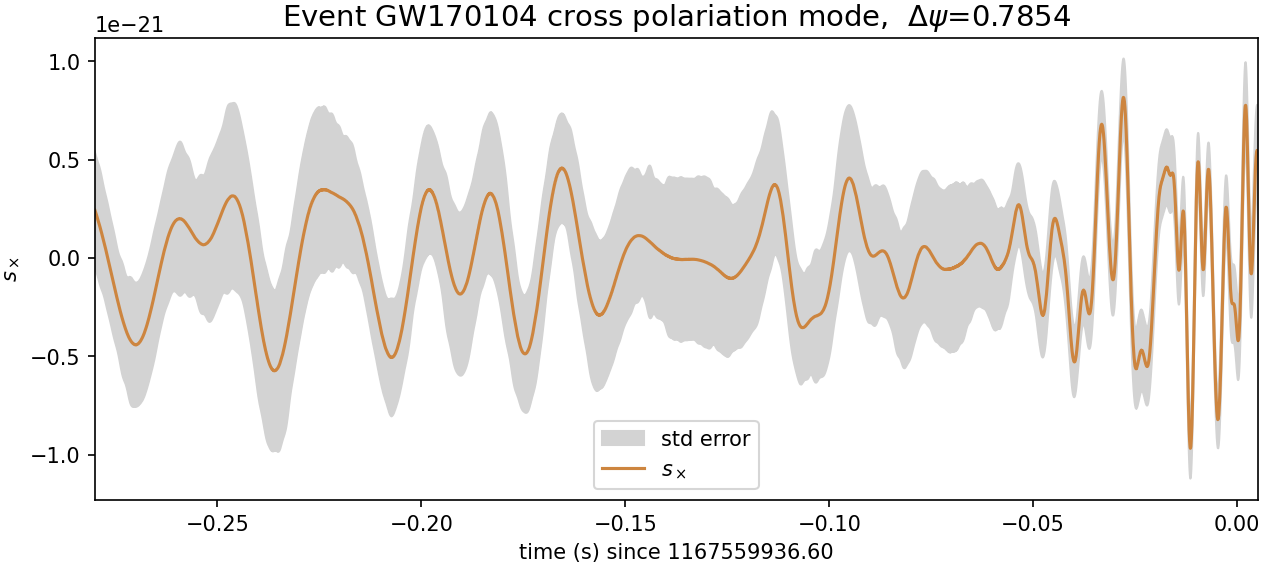}
\caption{Polarization modes of the event GW170104 for $\Delta\psi=0.7854$ with estimated error bands,
	in the region close to the nominal event time.
}
\label{fig:s+_sx_errores_aum}
\end{figure}

We however carryout first estimates of the error for the spin-2 polarization modes,  
using the information of the denoised signals, which are presented in Fig. \ref{fig:s+_sx_errores_aum}.
In order to check that the polarization modes behave as spin-2 quantities,
we introduce in appendix \ref{sec:spin2-polar} the calculation of the + and $\times$
PMs for the frames with $\Delta\psi=0, \frac{\pi}{16}, \frac{2\pi}{16}, \frac{3\pi}{16}$;
and we also calculate the consistency behavior of the PMs for $\Delta\psi=\frac{\pi}{4}$
with the $\Delta\psi=0$ frame.

An estimate of the possible errors
in the denoised data, can be calculated from the local standard deviation 
of the difference between the strain and the 
denoised datum for each case; carried out on an appropriate window length.
Let us call this upper bound estimate $\sigma_{X-w_X}$; where we use $X$ to denote
the strain of a detector and $w_X$ its denoised signal.

When performing the reconstruction of the polarization modes,
we use linear combinations of the denoised signals, of the form
$P = a X + b Y$.
So that to estimate the
possible error in their calculation we can apply the previous construction,
and use the relation
$\sigma_P^2 = a^2 \sigma_{X-w_X}^2 + b^2 \sigma_{Y-w_Y}^2 + 2 a b \, {\sf cov}(X-w_X,Y-w_Y)$
to estimate them; where ${\sf cov}$ means covariance.

In Fig. \ref{fig:s+_sx_errores_aum}
we show the graphs of the polarization modes with their respective
estimated error bands.
The calculation for the time series polarization errors presented above
provides estimates based on the statistics of the denoised signals.
The differences between both graph is due to the way the detector pattern functions
contribute to the final expressions.
\add{We expect that these error estimates will increase when a detailed treatment 
of the high correlation between signals from both observatories 
is taken into account; a study that we defer to future work}

\section{Complete reconstruction of the signal in terms of the spin-2 
	polarization modes for GW170104}\label{sec:signal-in-terms}

\begin{figure}[H]
\centering
\includegraphics[clip,width=0.48\textwidth]{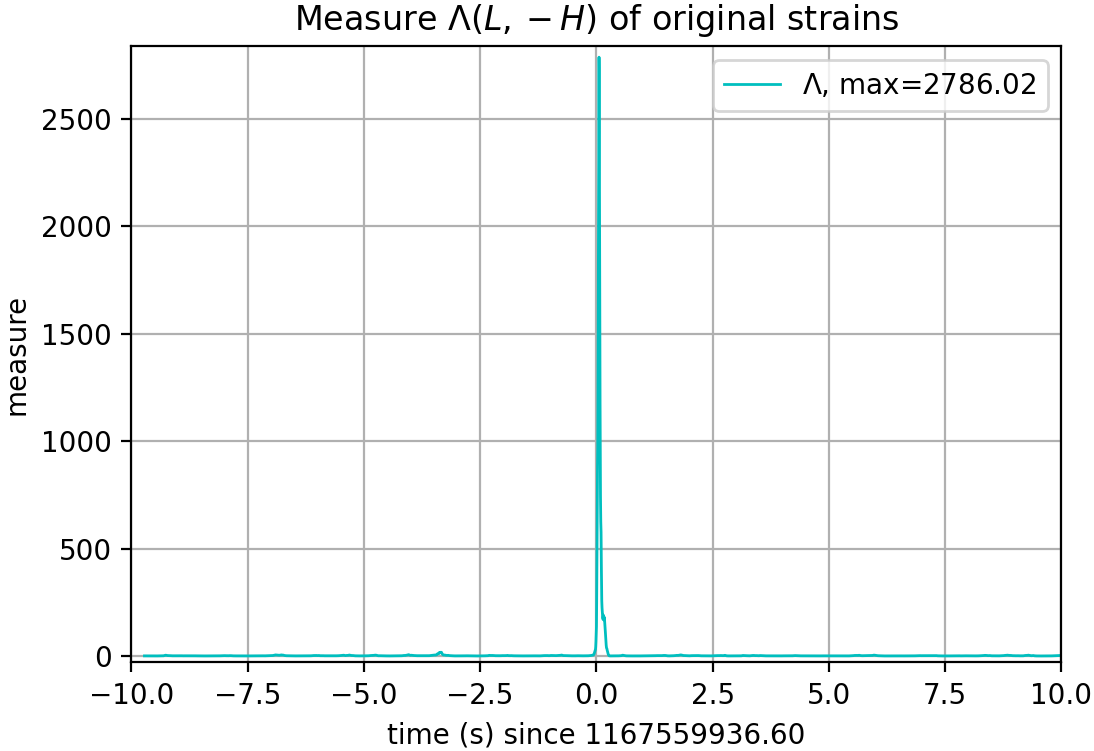}
\includegraphics[clip,width=0.48\textwidth]{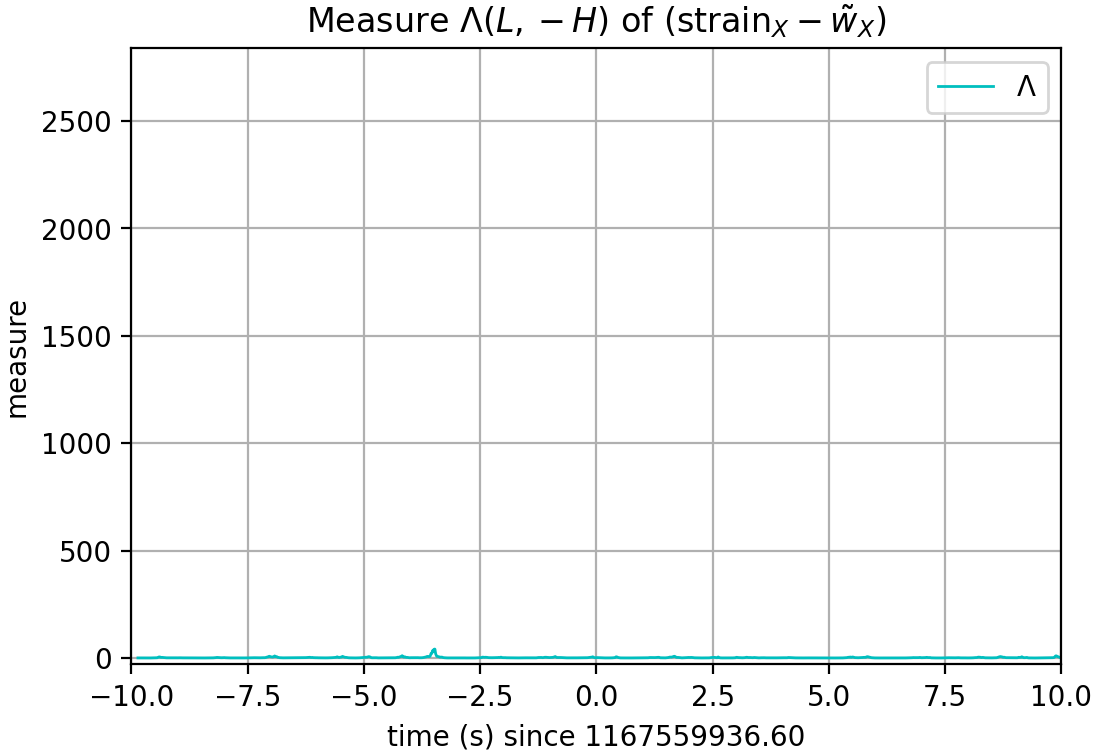}
\caption{On the top graph, the values of the measure $\Lambda$ close to the reference event time
	for the original filtered strains of GW170104,
	and on the bottom for the strains after the subtraction of the reconstructed
	signals from the spin-2 polarization modes. The residual is undetected.
}
\label{fig:strain-wtilde}
\end{figure}

We study here the situation in which the GW signal is decomposed only in terms
of the spin-2 polarization modes.
From the decomposition of the signal in terms of these PMs and
having obtained previously the sky localization of the source, 
we can use a similar equation to \eqref{eq:vX}
to reconstruct the signal of the GW at each detector.
Explicitly, we define the tilde signals in terms of the measured PMs
from
\begin{equation}\label{eq:wtildeX}
\begin{split}
\tilde{w}_X(t + \tau_X) &=  s_X(t + \tau_X) \\
&= F_{+X0}\, s_+(t) 
+
F_{\times X0}\, s_\times(t)
;
\end{split}
\end{equation}
where the notation for $F_{+X0}$ and $F_{\times X0}$ was presented in section
\ref{sec:dec-pol-modes}
and here $s_+$ and $s_\times$ denote the measured PMs.
Then, we can subtract this reconstructed signal from the original strains
to check whether the spin-2 polarization modes account for the observed GW.
An appropriate tool to study this situation is the optimized measure $\Lambda$
introduced in \cite{Moreschi:2024njx}%
; which allows to compare two strains.
In each of the following two graphs, we compare the strains $-H$ and $L$ 
on the two observatories, Hanford and Livingston
\add{after applying a bandpass filter of [30,350]Hz}.
In Fig. \ref{fig:strain-wtilde} we present the graph of the measure $\Lambda$ 
for the filtered strains and for the strain after the subtraction of the reconstructed
signals from the spin-2 polarization modes.
If the original GW had a significant contribution
from spin-1 and/or spin-0 polarization modes then there should appear some
residual in the bottom graph of Fig. \ref{fig:strain-wtilde}.
The fact that the measure $\Lambda$ behaves as ambient noise
close the the reference event time, after the subtraction 
of the reconstructed signals from the spin-2 polarization modes, indicates that
the GW signals for GW170104 can be completely understood in terms of the
spin-2 polarization modes.
That is, no other polarization modes of spin-1 or spin-0 
contribute to this detected GW signal.

For this reason we claim that there is no need for more observatories to
make a measurement of the polarization modes of this gravitational wave.

\section{Nominal time shift for GW150914}\label{sec:nomianl-time-shift-GW150914}

\begin{figure}[H]
\centering
\includegraphics[clip,width=0.48\textwidth]{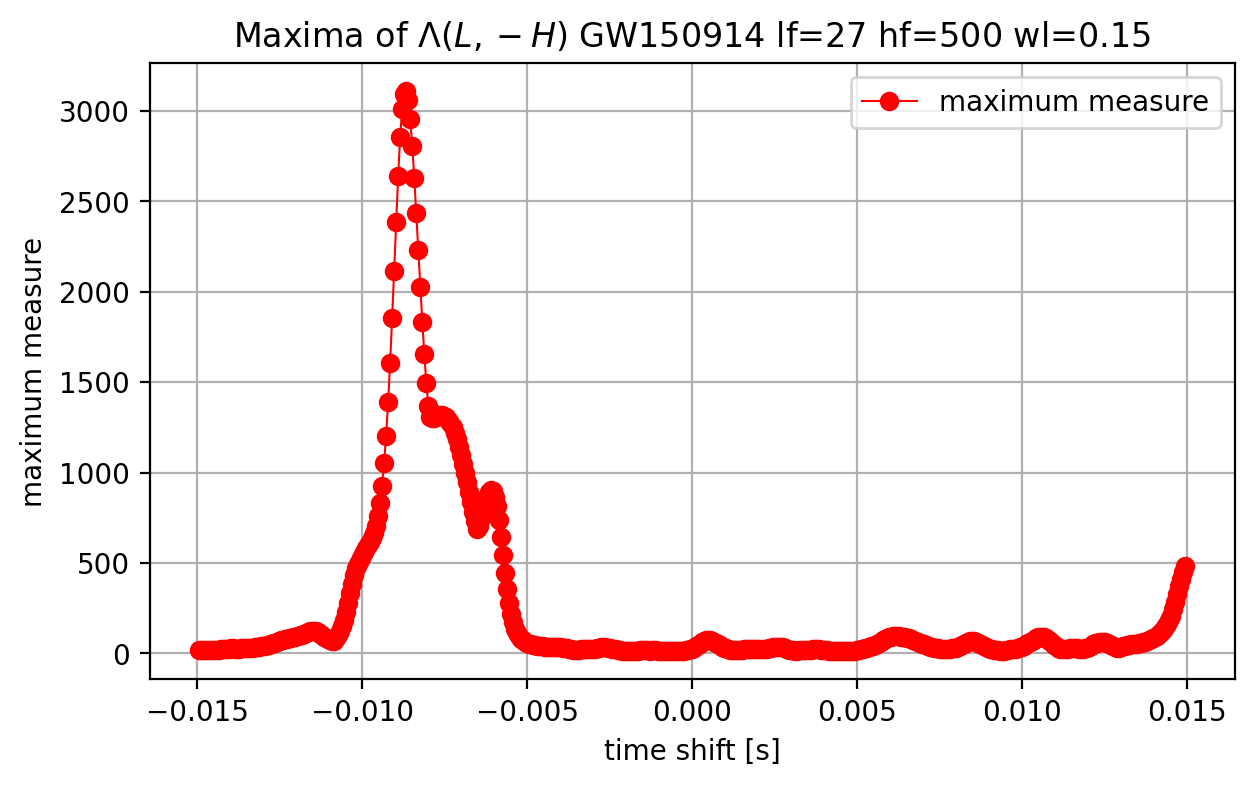}
\caption{This graph shows the behavior of mathematical evaluation $\Lambda$ of
	the OM measure as a function of shift
	of the Hanford data with respect to the Livingston data for a window of 0.15s and
	a band limiting frequency filter of [27,500]Hz.
}
\label{fig:likelihoodH-gw150914}
\end{figure}

\begin{figure}[H]
\centering
\includegraphics[clip,width=0.48\textwidth]{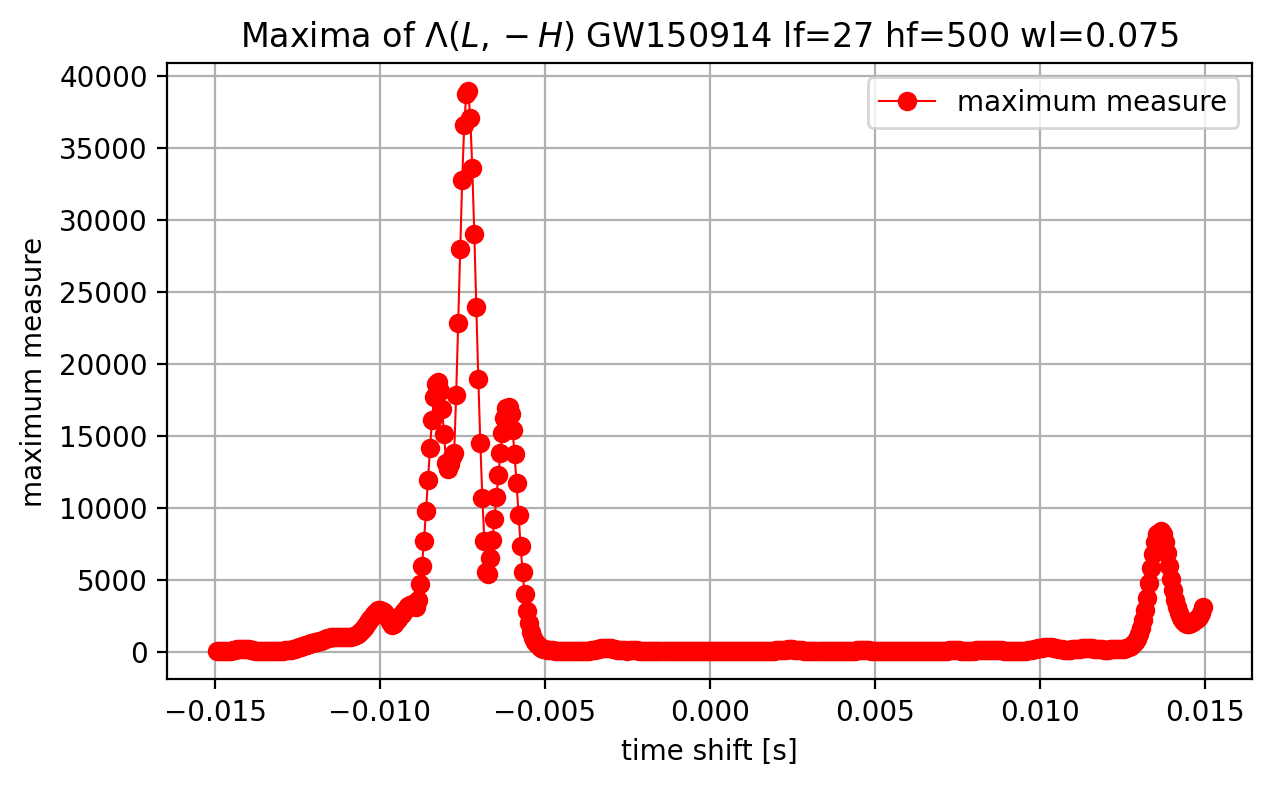}
\caption{This graph shows the behavior of mathematical evaluation $\Lambda$ of
	the OM measure as a function of shift
	of the Hanford data with respect to the Livingston data for a window of 0.075s and
	a band limiting frequency filter of [27,500]Hz.
}
\label{fig:likelihoodH-gw150914_2}
\end{figure}

To determine a nominal time shift for strain H with respect to strain L for the GW150914 event, 
we used the original filtered strains and applied additional bandpass filtering, 
testing several window widths for each analysis. 
We initially studied the strain in the frequency band [22,1024]Hz, 
where a natural signal duration of 0.5s appears due to low-frequency contributions at early times. 
However, the study of the $\Lambda$ measure\citep{Moreschi:2024njx} proved to be excessively noisy. 
Therefore, we tested several frequency ranges and ultimately chose to work in the [27,500]Hz band. 
This choice suggests an optimal window length of 0.15s for the relevant signal. 
Applying the window width $wl = 0.15$s yields the behavior for the $\Lambda$ measure 
shown in Fig. \ref{fig:likelihoodH-gw150914}.
When we apply this analysis to a window width of $wl = 0.075$s, 
the $\Lambda$ measure exhibits the behavior shown in Fig. \ref{fig:likelihoodH-gw150914_2}.
The fact that these two window widths yield different results 
suggests that the two observatories may have recorded different components 
of the spin-2 polarization modes of the gravitational wave. 
Using this preliminary information, we estimate the time shift for H as the 
weighted average of these two values, based on the relationship between 
the highest frequency in the short window strain and the dominant frequency contribution 
in the wide window strain, obtaining $sh_H = -0.007568$s. 

\begin{figure}[H]
\centering
\includegraphics[clip,width=0.5\textwidth]{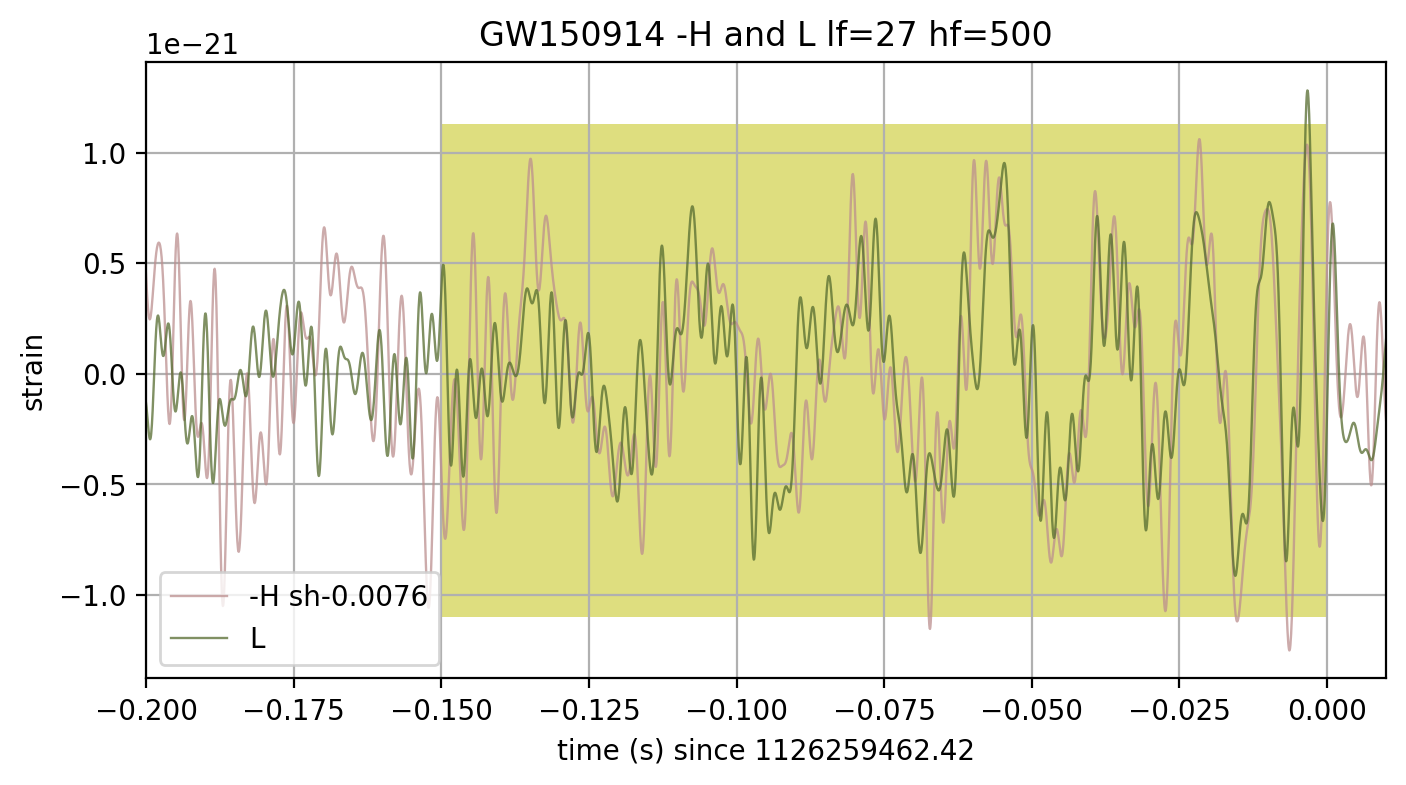}
\caption{Strains of -H and L with a	band limiting frequency filter  of [27,500]Hz and with $sh_{H}=-0.007568$s.
	The colored rectangular region indicates the initial width of the window used 
	in the measure. 
}
\label{fig:-HyL_gw150914}
\end{figure}

To examine the relationship between both strains during the relevant 
time interval near the event time, Fig. \ref{fig:-HyL_gw150914} shows 
both the -H and L strains when H is shifted by $sh_H$.
An impressive coincidence in both form and phase can be observed between the two signals 
in the colored region, which has the width of the initial window length $wl$. 

It should be noted that in the original presentation of GW150914\citep{TheLIGOScientific:2016qqj}, 
a nominal time shift of -0.0069s was reported for strain H with respect to L, 
which represents a difference of $6.7 \times 10^{-4}$s between the two estimates. 
While this may appear to be a small separation, and indeed is small with regard to source localization, 
it is significant for polarization mode decomposition. 
For this reason, we strive to be as precise as possible with this estimate.

\section{Study of GW150914 in the time-frequency domain}\label{sec:time-freq_gw150914}

\begin{figure}[H]
\centering
\includegraphics[clip,width=0.48\textwidth]{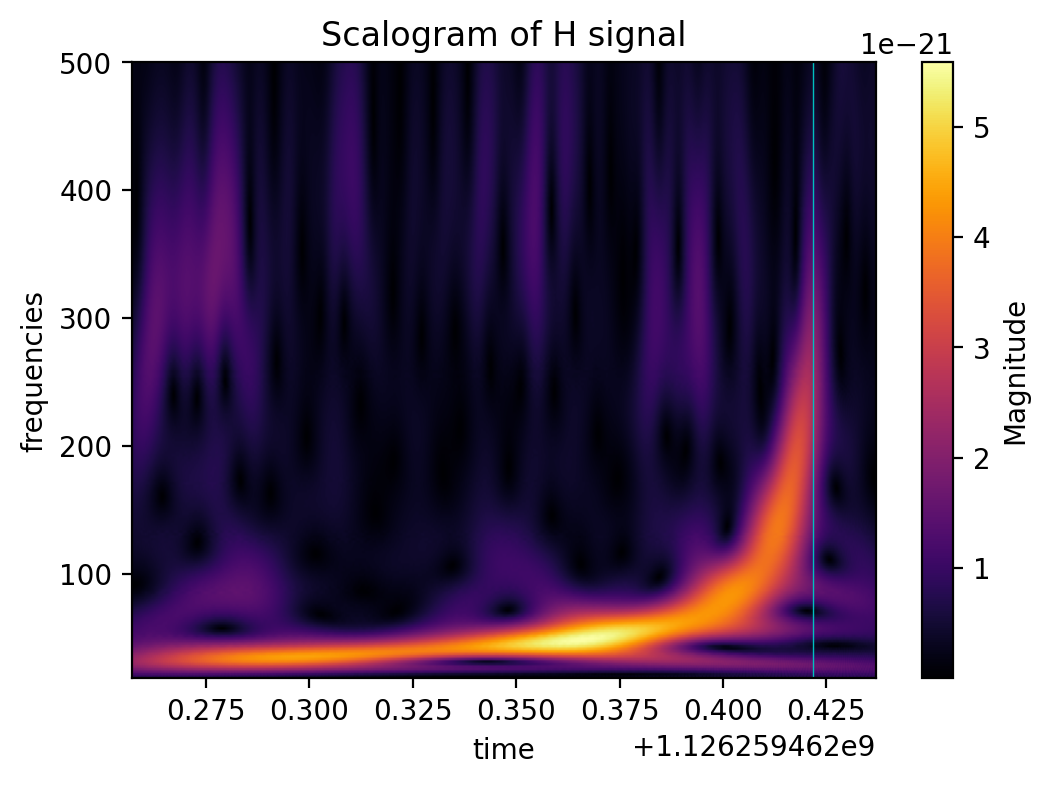}
\caption{Detail of the strain of LIGO detector H for GW150914 
	with a pass band of [27,500]Hz.
}
\label{fig:scalogramH-gw150914}
\end{figure}

\begin{figure}[H]
\centering
\includegraphics[clip,width=0.48\textwidth]{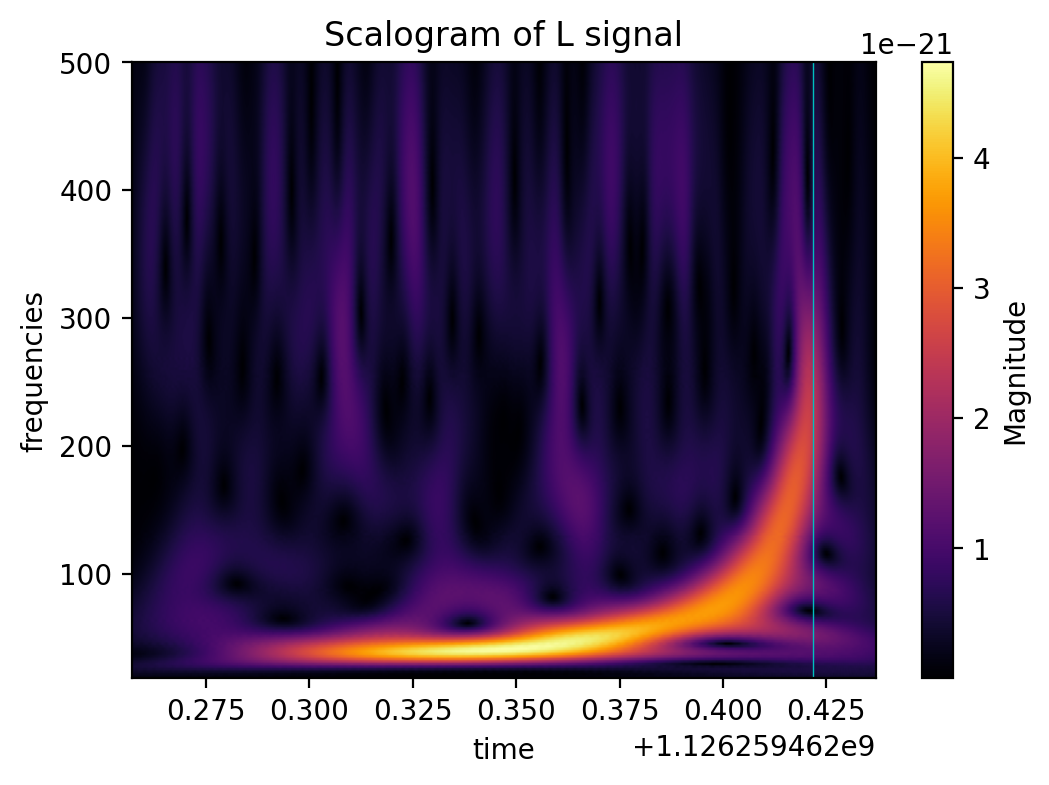}
\caption{Detail of the strain of LIGO detector L for GW150914
	with a pass band of [27,500]Hz.
}
\label{fig:scalogramL-gw150914}
\end{figure}

Figures \ref{fig:scalogramH-gw150914} and \ref{fig:scalogramL-gw150914} show 
the signals in the time-frequency domain through their corresponding scalograms.

These two graphs show clear chirp-like signals observed by both observatories.

\section{Localization of GW150914}\label{sec:localizGW150914}
\begin{figure}[H]
\centering
\includegraphics[clip,width=0.48\textwidth]{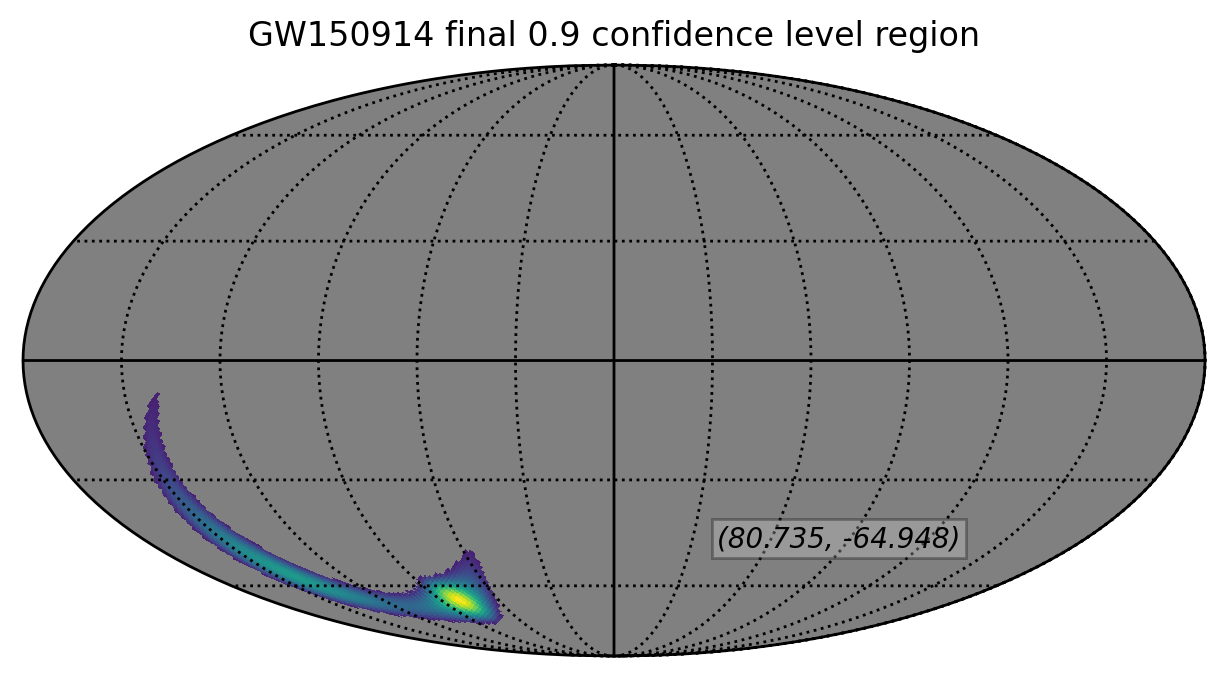}
\caption{Final sky localization for the source of the GW150914 event,
	with 0.9 confidence level region of measure $M_u$.
	The two numbers between parentheses denote the preferred central position as longitude and latitude.
}
\label{fig:loc-gw150914-v1}
\end{figure}

Applying the L2D+PMR procedure to this event, we calculate its localization 
and present the final 0.9 confidence level region for GW150914 in Fig. \ref{fig:loc-gw150914-v1}.
This result should be compared with the 90\% credible localization region obtained 
using the LIGO LALInference method.
\begin{figure}[H]
\centering
\includegraphics[clip,width=0.48\textwidth]{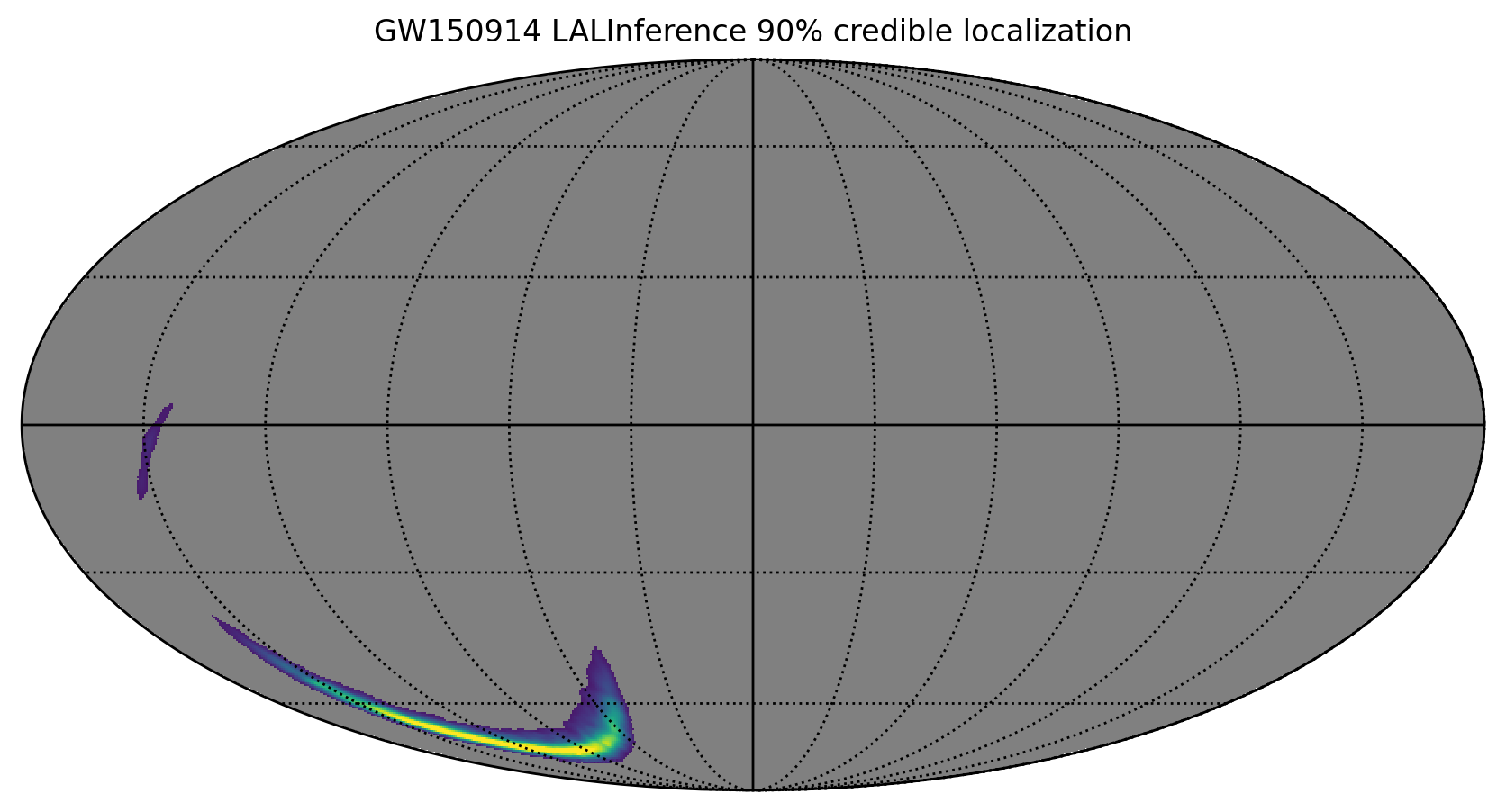}
\caption{Sky localization from LIGO team with LALInference method.
}
\label{fig:loc-ligo-LALI-GW150914}
\end{figure}
Both regions, shown in Figs. \ref{fig:loc-gw150914-v1} and \ref{fig:loc-ligo-LALI-GW150914}, 
are very similar in location and extent.

\section{Reconstruction of the spin-2 polarization modes of GW150914}\label{sec:PM-GW150914}
\begin{figure}[H]
\centering
\includegraphics[clip,width=0.48\textwidth]{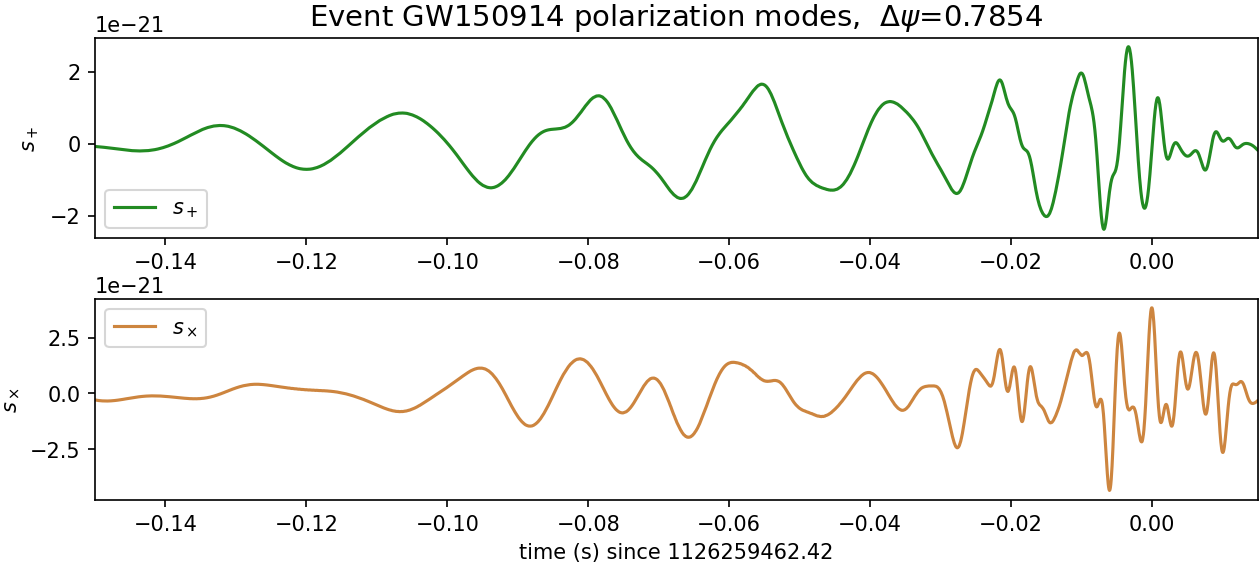}
\caption{Polarization modes + and $\times$ close to the reference time of the GW150914 event.}
\label{fig:s+_sx_gw150914}
\end{figure}

\begin{figure}[H]
\centering
\includegraphics[clip,width=0.48\textwidth]{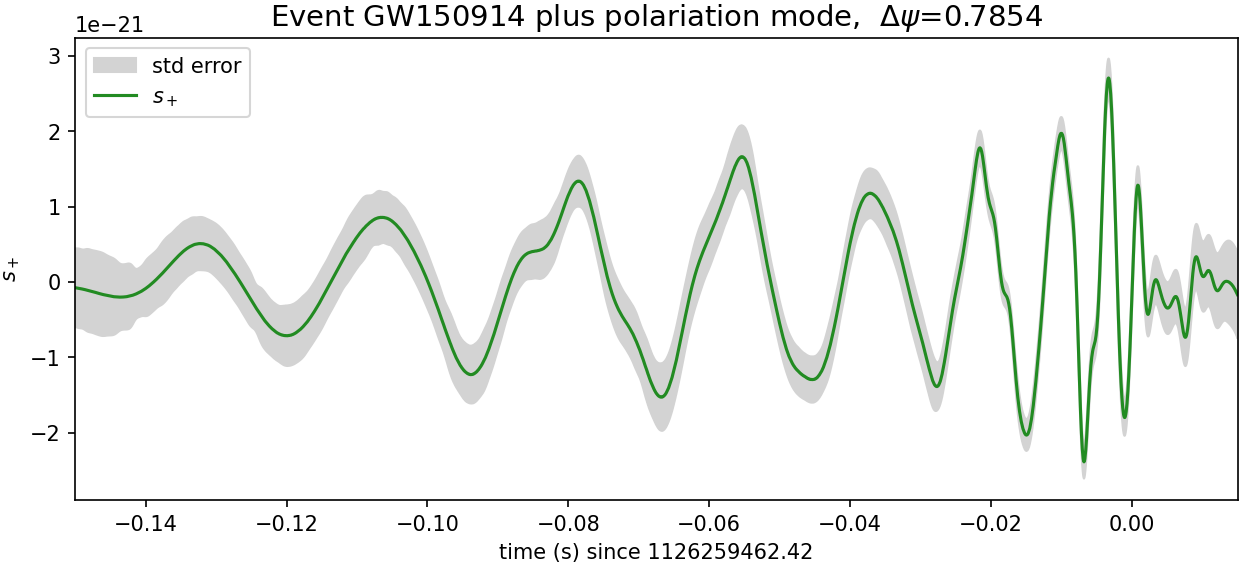}
\includegraphics[clip,width=0.48\textwidth]{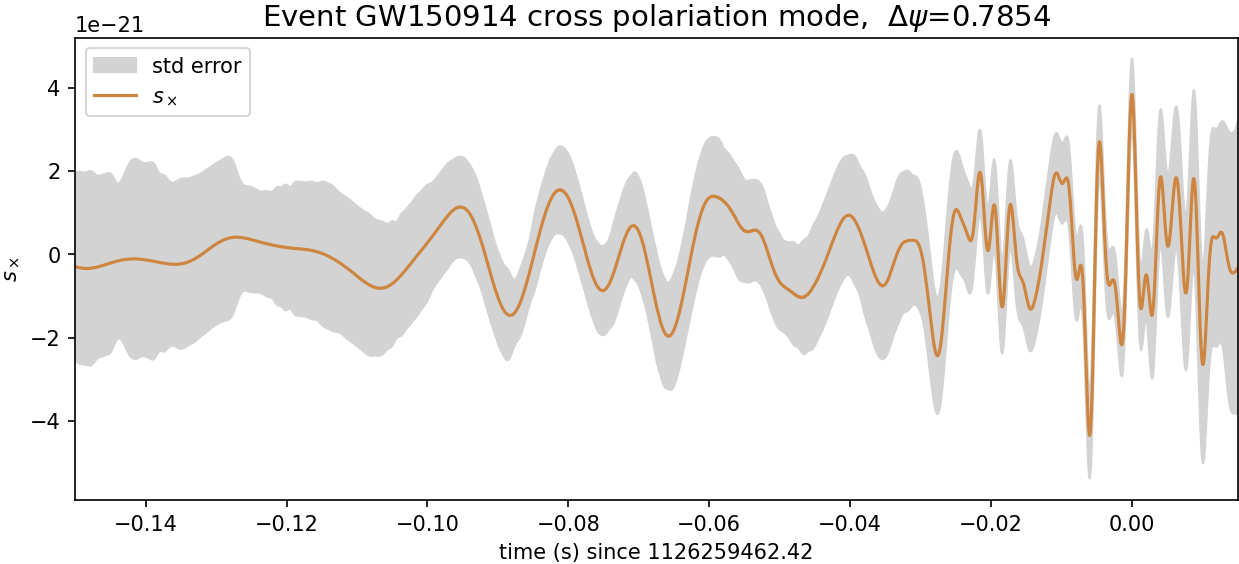}
\caption{Polarization modes of the event GW150914 with estimated error bands,
	in the region close to the nominal event time.
}
\label{fig:s+_sx_errores_gw150914}
\end{figure}

\begin{figure}[H]
\centering
\includegraphics[clip,width=0.48\textwidth]{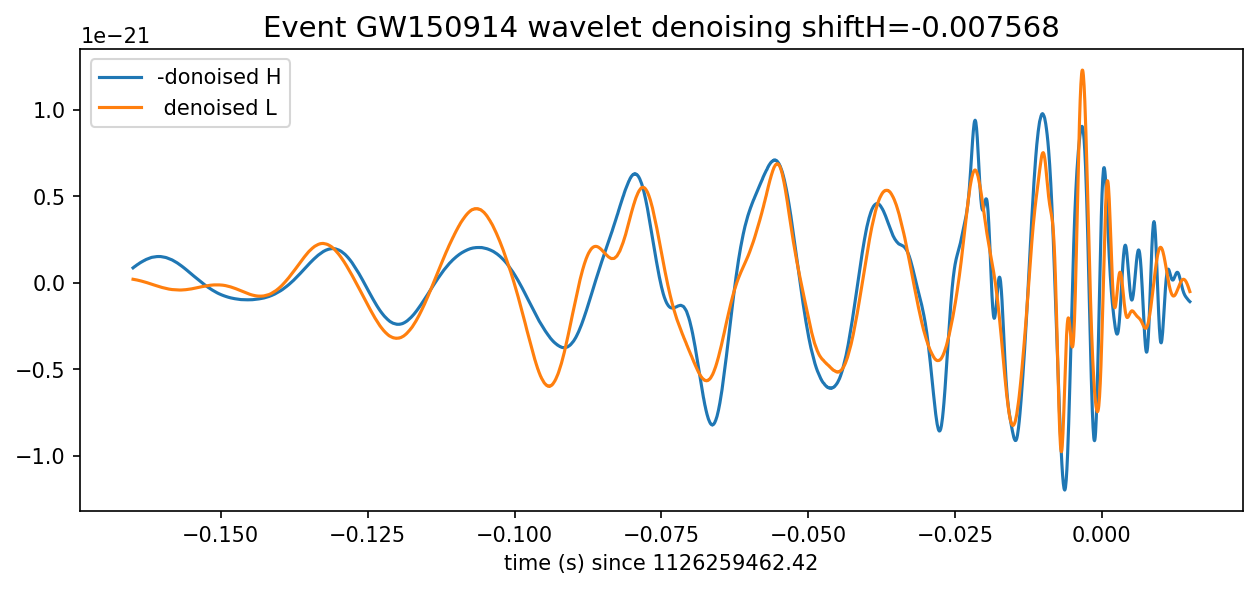}
\caption{Denoised signals of the GW150914 event.}
\label{fig:-wHywL_gw150914}
\end{figure}

The reconstructed polarization modes of GW150914 are presented in Fig. \ref{fig:s+_sx_gw150914}.
Figure \ref{fig:s+_sx_gw150914} shows that while $s_+$ exhibits a familiar 
chirp-like shape, $s_\times$ does not display a strong signal and appears to contain more noise. 
This observation is corroborated in Fig. \ref{fig:s+_sx_errores_gw150914}, which includes the error bands.
Figure \ref{fig:s+_sx_errores_gw150914} confirms that the estimated error 
for $s_\times$ is considerably larger than the error for $s_+$. 
This is consistent with the fact that both observatories recorded nearly 
identical polarization mode components of the gravitational wave. 
This can also be verified by comparing the denoised -H and L signals, 
as shown in Fig. \ref{fig:-wHywL_gw150914}.
Figure \ref{fig:-wHywL_gw150914} demonstrates the remarkable similarity between 
both denoised signals, and the similarity with the $s_+$ component shown in 
Fig. \ref{fig:s+_sx_gw150914} is also notable. 

\begin{figure}[H]
\centering
\includegraphics[clip,width=0.48\textwidth]{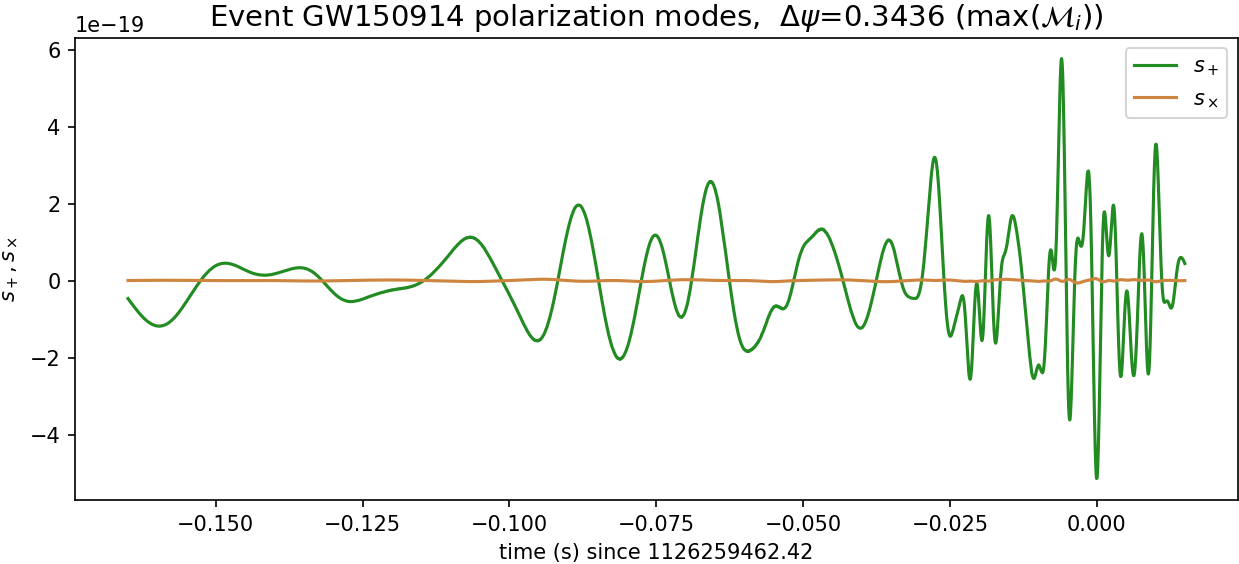}
\caption{Polarization modes of GW150914 in selected frame in which one of the components
has negligible values.
}
\label{fig:s+_sx_gw150914_maxMi}
\end{figure}

\begin{figure}[H]
\centering
\includegraphics[clip,width=0.48\textwidth]{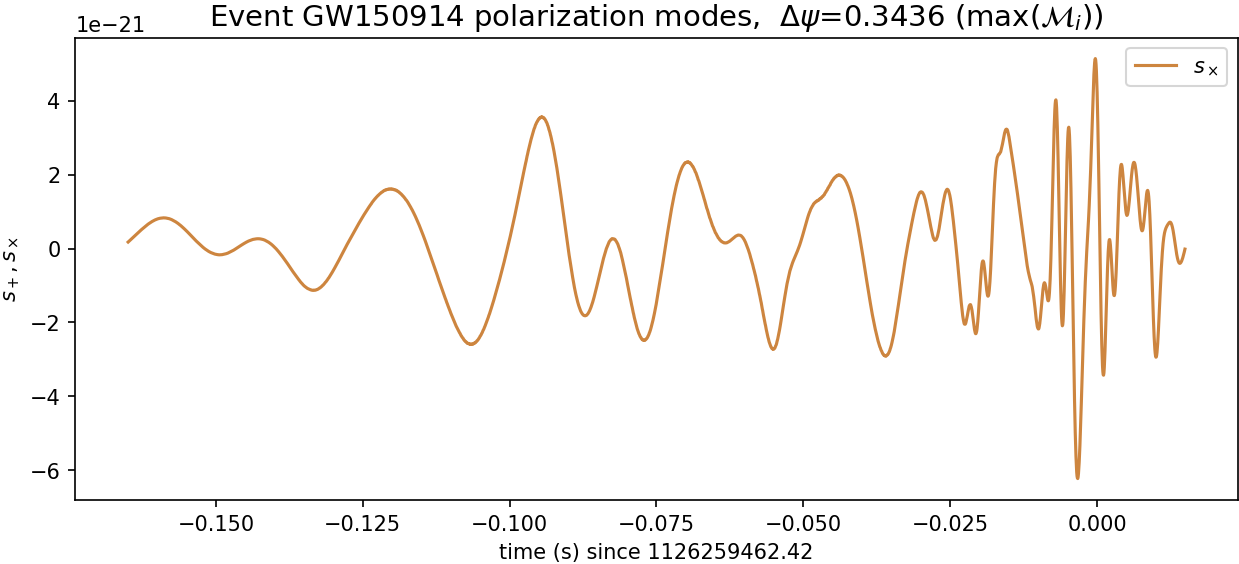}
\caption{The $\times$ polarization modes of GW150914 in selected frame.
}
\label{fig:sx_gw150914_maxMi}
\end{figure}

Due to space constraints, we do not include the spin-2 polarization modes of GW150914 
in multiple frames here, but present only the + and $\times$ components in a frame 
that accentuates the diminishing of one component, as shown in Fig. \ref{fig:s+_sx_gw150914_maxMi}.
Figure \ref{fig:s+_sx_gw150914_maxMi} shows that the $\times$ component 
is negligible compared to the + component; 
however, the $\times$ component exhibits significant noise, as demonstrated in Fig. \ref{fig:sx_gw150914_maxMi}.
It must be emphasized that to achieve the impressive cancellation shown in Fig. \ref{fig:s+_sx_gw150914_maxMi}, 
the polarization mode reconstruction must be performed at the location of the maximum of $\mathcal{M}_i$, 
near the nominal delay ring, as shown in Fig. \ref{fig:loc_1-gw150914}. 
All this indicates that in this case as well, both LIGO observatories detected nearly identical 
spin-2 components of the gravitational wave. 
Since these reconstructions were performed using the algebraic and geometrical properties 
inherent to the spin-2 nature of gravitational waves, 
the preceding graphs provide direct evidence of the spin-2 nature of the detected signals.

The remarkable cancellation of $s_\times$ shown in Fig. \ref{fig:s+_sx_gw150914_maxMi} 
warrants further discussion. 
Using strains recorded at two different gravitational wave observatories, 
each with intrinsic independent noise, we applied filtering techniques and subjected 
the data to algebraic manipulation consistent with our source localization procedure. 
We then reconstructed the polarization modes and found that one component is essentially 
zero compared to the other in a particular polarization frame. 
Since this flat response, shown in Fig. \ref{fig:s+_sx_gw150914_maxMi}, 
cannot be generated by chance, the inevitable conclusion, 
before any discussion of the content and meaning of this extraordinary cancellation, 
is that we could not have achieved this result if our source localization procedure 
and spin-2 polarization mode reconstruction technique were incorrect. 
Therefore, we interpret this impressive cancellation as strong evidence for 
the consistency of the L2D+PMR procedure.

\section{Final comments}\label{sec:final}

In order to gain confidence that a perturbation in a GW detector
has an astrophysical origin one normally requires to record some signal
in at least two detectors. However, how can one be sure that these would correspond
to a GW rather than just a mere coincidence, even if such a coincidence  
is highly unlikely? The answer 
lies in testing whether the signals behave as GWs;
which is done by representing them in terms of its polarization modes.
In this article, we have completed this program for the GW170104 and GW150914 events.
We have successfully established the reconstruction of the polarization modes
of detected GWs, in a model independent approach, using the data
recorded from only the two LIGO observatories.
This is a fundamental result in  the observational study of gravitational waves.

At this point it is important to mention that it is widely belief that
in order to perform a measurement of the polarization modes
of a GW one needs about five detectors;
so that one is able to separate the contributions from the
different possible polarization modes.
However, we remark in appendix \ref{sec:theoretical_frame_apendix-1} that
for a GW of astrophysical origin,
modeled by an asymptotically flat spacetime, the only possible
detectable polarization modes are those with spin 2.

Independently of the previous theoretical considerations,
our findings represent
the first explicit reconstruction of the spin-2 polarization modes of GWs
in terms of time series. 

It is natural to ask what are we doing differently to obtain 
localization results from data of just two observatories. 
And the answer is manyfold that we summarize in:
that we are optimizing the use of spin-2 polarization geometrical equations,
that we are using a new denoising technique and
that we are employing a convenient measure on the celestial sphere.

The reliability of our procedure can be inferred from several results.
One of them is the fact that the reconstructed polarization modes account for
all the content of the detected signal, as is shown explicitly in the
bottom graph of Fig.
\ref{fig:strain-wtilde}; 
which shows the consistency
of our method with the spin-2 nature of the detected signal.
But to consider more validation arguments for our procedure, we
also have included 10 simulations where we have
calculated the localization from the injection of simple synthetic signals at different locations
and with a nontrivial initial eccentricity.
We have also reconstructed satisfactory the polarization
modes of the original injected signal.
Since the procedure
depends only\citep{Abbott:2018utx}
on the
geometry of a gravitational wave’s strain and its direction
of propagation, not on the details of any specific theory of
gravity, we have built a very simple signal out of the spin-2 polarization modes,
that is described in appendix \ref{sec:simul-signal}.
We have chosen the position of ten synthetic sources at different sky locations
corresponding to a single different delay time.
In section \ref{sec:synth} we present several graphs showing the reasonable accuracy 
and precision properties of our localization methods.
In Figs. \ref{fig:synth_+mode} and \ref{fig:synth_xmode}
we show the reconstructed spin-2 polarization modes of the synthetic signals;
where one can check the excellent similarity with the original polarization modes,
used to construct the signals.
All these demonstrate the coherence and reliability of our method.


After the L2D+PMR procedure is applied one is in position to undertake a research on
trying to model the detected GW and polarization modes in terms of specific
parameters.
In this article we have not attempted to get into the detail astrophysical interpretation
of our results; instead we have concentrated in the presentation of a new procedure
that allows for the localization of a GW source using data from just two
detectors, and that also permits the reconstruction of the PMs from this data.

Let us note that our preliminary and final localization for GW170104
are concentrated in regions of the celestial sphere that are close to those
predicted by Bayestar and LALIinference of LIGO, that
we reproduce in Figs. \ref{fig:loc-ligo-bayestar} and \ref{fig:loc-ligo-LALI}.

The results presented above confirm the expectation that a GW of astrophysical
origin should be completely represented in terms of spin-2 polarization modes;
since we have measured no contributions from spin 1 and spin 0 polarization
modes to GW170104.
This not only authenticates the recorded signals to be actual GWs,
but also validates our understanding on the nature of them.

\add{The GW150914 event presents a case of a very strong GW signal 
accompanied by high-amplitude noise, which is characteristic of the first observing run. 
We have demonstrated in Section} \ref{sec:nomianl-time-shift-GW150914} 
\add{the subtleties 
involved in determining a nominal time shift for strain H with respect to L. 
The relationship between the preliminary and final time shifts shown in the graphs 
is consistent with the effective single-site timing accuracy for each 
detector}\citep{KAGRA:2013rdx,Bartos:2010zz}, 
\add{on the order of $10^{-4}$s, 
which we have neglected in our treatment, 
and with the phase uncertainty}\citep{LIGOScientific:2016xax} 
\add{reported by the LIGO team.}

\add{Regarding intersite timing accuracy, it was claimed}\citep{KAGRA:2013rdx} 
\add{that the absolute timing discrepancy from UTC, and therefore between detectors, 
was no larger than $10\mu$s. 
The LIGO team concluded that the total uncertainty in the signal was less than 10\% 
in magnitude and $10^o$ in phase from 20Hz to 1kHz during the GW150914 observation. 
This is related to the fact that GW150914 was detected during the engineering run 
ER8}\citep{KAGRA:2013rdx}, 
\add{which immediately preceded the formal start of the first run O1. 
Naturally, one expected that at these early stages there would be ongoing 
efforts to control systematic errors and unwanted uncertainties.}

\add{For these reasons, our strategy was to develop our procedure using data from 
later observing runs. 
The effects of different uncertainties in strain calibrations on each step of 
our procedure will be the subject of further study. 
In this article, we present our results assuming no systematic errors in strain calibration. 
The impressive anticorrelation of signals in both observatories can already 
be observed by comparing the strains shown in Fig.} \ref{fig:-HyL_gw150914}.

\add{The fact that our 0.9 confidence level region for GW150914 nearly overlaps 
with the 90\% credible localization calculated using the LALInference method 
indicates agreement between both approaches for GW source localization in this case.}

\add{The graph corresponding to GW170104 polarization modes in Fig.} \ref{fig:s+_sx} 
\add{suggests that both LIGO observatories recorded nearly identical PM components 
of the GW. 
This observation is corroborated by the PM graphs presented in other polarization frames 
(Figs.} \ref{fig:s+_sx_0_aum}-\ref{fig:s+_sx_4_aum}), 
\add{where the components become very similar in some frames while one component becomes negligible in others.
We observed the same phenomenon with GW150914, as shown in Fig.} \ref{fig:s+_sx_gw150914}. 
\add{While the + component exhibits a clean, expected waveform, the $\times$ component appears 
to be dominated by noise, a conclusion supported by the corresponding error graphs 
in Fig.} \ref{fig:s+_sx_errores_gw150914}.
\add{The clearest indication of this behavior comes from the noticeable cancellation 
of one polarization mode in a selected polarization frame (Fig.} \ref{fig:s+_sx_gw150914_maxMi}) 
\add{and through direct comparison of the denoised signals shown in Fig.} \ref{fig:-wHywL_gw150914}.
\add{The physical information contained in the spin-2 polarization modes presented here
will be analyzed in dedicated future studies.}

\add{It is curious that 
both GW events analyzed here exhibit the same characteristic: 
each observatory detected nearly identical spin-2 PM components 
of the respective GW (with one being approximately the negative of the other). 
The degree of this effect is particularly striking in Fig.} \ref{fig:s+_sx_gw150914_maxMi}, 
\add{where one polarization mode component appears to be few orders of magnitude smaller than the other.}

\add{As discussed in Section} \ref{sec:PM}, 
\add{the polarization mode reconstruction algebra 
from two-observatory strain data has a remarkable property: for highly (anti)correlated signals, 
the calculation involves subtracting the Hanford and Livingston measurements, 
thereby increasing noise susceptibility. 
This explains why the reconstructed polarization modes for GW170104 appear noise-dominated, 
while the + polarization mode component of GW150914 shows physically meaningful structure. 
The difference stems from the signal-to-noise ratio in each case, 
the cWB Search Pipeline reports a Network SNR of 13.0 for GW170104 compared 
to 25.2 for GW150914. 
The higher SNR of GW150914 provides sufficient signal strength to enable 
meaningful polarization mode reconstruction.
This represents the first direct measurement of a spin-2 polarization mode 
of a gravitational wave.}

\add{These observations encourage further investigation into applying these 
novel methods to real gravitational wave events. 
The significance of the results presented here suggests that such studies will yield valuable scientific insights.}

\add{Gravitational wave observatories undergo continuous improvements, 
with each scientific run achieving lower intrinsic noise levels and 
enhanced sensitivity to weaker signals. 
This increased sensitivity expands the observable volume range, enabling 
the detection of more events with favorable signal-to-noise ratios.
We are therefore eager to access strain data from the fourth observing run, 
which is expected to provide detailed gravitational wave observations 
from multiple observatories. 
This presentation suggests that our procedure
will prove valuable for analyzing these future detections.}


In Appendices we have included details of 
GW analysis that deserve an extended explanation
but that would enlarge unnecessary the main part of the text.

In future work we intend to apply these techniques to other GW events,
expand on the details of our methods,
and continue to improve and refine them.


\subsection*{Acknowledgments}

This work is possible thanks to the open data policy of the 
LIGO Scientific Collaboration, the Virgo Collaboration and Kagra Collaboration;
who are giving freely access to data through the Gravitational Wave Open Science
Center at \href{https://gwosc.org}{https://gwosc.org};
which is described in \cite{LIGOScientific:2019lzm} and \cite{KAGRA:2023pio}.

We have used python tools included in the project PyWavelets\citep{Lee:2019}.

We acknowledge support from  SeCyT-UNC and CONICET.

The inclusion of the GW150914 event in this article is due to the interaction
with an anonymous referee to whom we are grateful; since the discussion
of these two events becomes essential to grasp the nature of our procedure.

We are very grateful to Ezequiel F. Boero for comments and proof reading
and Emanuel Gallo for numerous deep discussions and 
indicating several improvements.
We also thank Maximiliano Isi for fruitful elaborated discussions.


\subsection*{Declarations}

Data Availability:

The time series of the calculated spin-2 polarization modes for event
GW170104 and GW150914 can be requested via email to the author.
We intend to publish this and other PMs on the web in the future.
The numerical calculation is described in the article.

\vspace{2mm}
\noindent
Ethics declaration: 

not applicable.



\appendix

\vspace{4mm}
\noindent
{\large\bf APPENDICES}

\section{Basic theoretical framework}\label{sec:theoretical_frame_apendix-1}

Up to now the most successful description of gravitational phenomena
comes from the equivalence principle\citep{Einstein11a};
which has been tested to high precision\citep{Touboul:2022yrw}.
An important implication of this principle, is that the gravitational effects
are encoded in the curvature of the spacetime metric.

In particular, present ``L'' shape gravitational-wave detectors intend to measure the characteristics
of a passing GW by observing the effects on mirrors at the ends of the arms.
The relative motion of each pair of mirrors is described by the
geodesic deviation equation, that for the deviation vector $\xi^a$
can be expressed as\citep{Wald84}
\begin{equation}\label{eq:aa}
	a^a = t^c \nabla_c v^a = t^c \nabla_c (t^b \nabla_b \xi^a)
	= R_{cbd}^{\;\;\;\;\; a} \xi^b t^c t^d
	;
\end{equation}
where $t^b$ is a unit timelike vector orthogonal to $\xi^a$ and $a^a$ is the acceleration.
For describing the components of the curvature tensor $R_{cbd}^{\;\;\;\;\; a}$
we use the conventions of \cite{Pirani64}, which agrees with \cite{Penrose84}.
It is also advantageous to use a null tetrad to describe the components of the curvature and
therefore,
the nature of the waves, since they travel along null directions.
One common nomenclature for a null tetrad, adapted to these type of situations
is to use a real null
vector ($\ell^a$) to point in the direction of the propagation; two complex 
(spacelike) null vectors ($m^a, \bar{m}^a$) are chosen perpendicular
to the propagation direction, and a fourth real null vector ($n^a$)
is chosen perpendicular to the last two and with
unit contraction with the first ($g_{ab} \ell^a n^b = 1$).
This is the basis for the Geroch-Held-Penrose(GHP) formalism\citep{Geroch73} which
is useful for the geometric discussion of a variety of situations
in which this type of basis appears naturally.
When contracting this basis with different objects one obtains
quantities with spin weight.

The local null tetrad can be related to a Cartesian frame by:
$\ell^a = \frac{1}{\sqrt{2}}(t^a + e_3^a)$,
$n^a = \frac{1}{\sqrt{2}}(t^a - e_3^a)$,
$m^a = \frac{1}{\sqrt{2}}(e_1^a + i e_2^a)$
and
$\bar{m}^a = \frac{1}{\sqrt{2}}(e_1^a - i e_2^a)$;
where the $e_j$ form the unit spacelike basis,
and $i^2 = -1$, is the complex basis number.

Then, one can express the polarization modes in terms of following curvature
components:
{\footnotesize 
\begin{equation}\label{eq:R+}
	R_+ = R_{0101} - R_{0202} = R_{0m0m} + R_{0 \bar{m} 0 \bar{m}} 
	= \Re e\{ \Psi_0 - 2 \Phi_{02} + \bar{\Psi}_4 \}
	,
\end{equation}
\begin{equation}\label{eq:R_x}
	R_\times = R_{0102} + R_{0201} 
	=  -i (R_{0m0m}  - R_{0 \bar{m} 0 \bar{m}})  
	= \Im m\{ \Psi_0 - 2 \Phi_{02} + \bar{\Psi}_4 \}
	,
\end{equation}
\begin{equation}\label{eq:R_R}
	R_R = R_{0101} + R_{0202}
	= 2 R_{0m0 \bar{m}} 
	=  \Phi_{00}  - 2 (\Re e\{\Psi_2\} + 2 \Lambda) + \Phi_{22}
	,
\end{equation}
\begin{equation}\label{eq:R_{vx}}
	R_{vx} = R_{0103} + R_{0301} = 2 R_{0103}  
	=\Re e \{ -3 (\Psi_1 + \Phi_{01} ) + ( \bar{\Psi}_3 +   \Phi_{12} ) \}
	,
\end{equation}
\begin{equation}\label{eq:R_{vy}}
	R_{vy} = R_{0203} + R_{0302} = 2 R_{0203}  
	=\Im m\{ -3 (\Psi_1 + \Phi_{01} ) + ( \bar{\Psi}_3 +   \Phi_{12} ) \}
	,
\end{equation}
\begin{equation}\label{eq:R_M}
	R_M = R_{0303}
	= \Psi_2 + \bar{\Psi}_2 + 2 \Phi_{11} - 2 \Lambda
	= 2 \big( \Re e\{\Psi_2\} + \Phi_{11} - \Lambda \big)
	.
\end{equation}
}%
It is worthwhile noting that $R_+$ and $R_\times$ are the real and 
imaginary parts of a spin 2 quantity\citep{Geroch73},
and $R_{vx}$ and $R_{vy}$ are the real and 
imaginary parts of a spin 1 quantity,
while $R_R$ and $R_M$ are spin 0 quantities.

These six polarization modes where studied in \cite{Eardley:1974nw} and \cite{Eardley:1973br}
for the specific case of plane waves.
In \cite{Eardley:1974nw} they stated: 
``General relativity permits only the two $\Psi_4$ modes.'';
which sometimes has been taken as a general absolute truth.
However, the mathematics of a Universe filled with a plane wave is very
different from the description of an isolated system, that is usually employed
for the description of a bound system, as for example a binary black hole system.

An expected quality of a relativistic theory of gravity is to be able to provide 
good representations of isolated system, so that the contribution to the 
curvature from these isolated systems approaches zero as one moves away from the 
central region of the system. This is expected even in a cosmological contexts
when one considers compact bounded systems faraway from other bodies.
That is, the full Riemann tensor can be understood as 
$R_{abc}^{\;\;\;\;\; d} = R_{abc}^{(B)\; d}+R_{abc}^{(GW)\; d}$
where $(B)$ refers to a slow varying background and $(GW)$ stands for the
rapidly varying contribution due to a passing GW(See section 1.5 of reference \cite{Jaranowski:2009zz}).
The fact is that the gravitational-wave observatories are designed to be
sensitive to the $(GW)$ contribution to the curvature;
for this reason in gravitational-wave studies one normally does not
distinguish between $R_{abc}^{\;\;\;\;\; d}$ and $R_{abc}^{(GW)\; d}$.

In an scenario in which there are no {\it a priori} cosmological assumptions,
the idea of the notion of an isolated system independent of a particular field equation  
led to the concept of general asymptotically flat
spacetimes\citep{Moreschi87}.
In this reference it was studied the general case of a spacetime which
is asymptotically flat at future null infinity($\mathscr{I}^+$) with the
basic behavior
\begin{equation}\label{eq:asymp}
R_{abc}^{\;\;\;\;\; d} = f(\Omega) \hat{R}_{abc}^{\;\;\;\;\; d}
+ \delta R_{abc}^{\;\;\;\;\; d} ;
\end{equation}
where $\Omega$ is the conformal factor used to define future null infinity,
$f(\Omega)$ is a monotonic function with the property $\lim_{\Omega\rightarrow 0} f = 0$,
$\hat{R}_{abc}^{\;\;\;\;\; d}$ is a regular tensor at $\mathscr{I}^+$
and 
$\delta R_{abc}^{\;\;\;\;\; d}$ is a tensor in a neighborhood of $\mathscr{I}^+$
that goes to zero faster than $f(\Omega)$ for $\Omega\rightarrow 0$;
for details see \cite{Moreschi87}.
One of the remarkable findings in this work was that independently of
the functional form of $f(\Omega)$, one can establish the asymptotic behavior of
the radiation field, namely
\begin{equation}\label{eq:psi4_asym}
\Psi_4 = - \Omega \, \ddot{\bar{\sigma}}_0 + O(-1)
,
\end{equation}
\begin{equation}\label{eq:psi3_asym}
\Psi_3 = - \Omega^2 \, \eth_0 \dot{\bar{\sigma}}_0 + O(-2)
,
\end{equation}
and
\begin{equation}\label{eq:phi22_asym}
\Phi_{22} =  O(-1)
;
\end{equation}
where $\sigma_0$ encodes the asymptotic shear of the bundle defined by the vector field $\ell$,
dots mean time derivatives, a quantity $h(\Omega)$ is said to be $O(q)$
if $\lim_{\Omega\rightarrow 0} \Omega^q h = 0$,
and $\eth_0$ is the edth operator\citep{Geroch73} of the unit sphere.
This is a very strong result, that it says that independently of the field equation,
if the GW is due to the gravitational radiation of an astrophysical
isolated system, then the possible observable is just $\Psi_4$;
since all the other component fields decay too fast with distance.
In any case, for the sake of completeness we describe next the
asymptotic leading order behavior of the $R_{0X0Y}$ components for the case of a
general asymptotically flat spacetime but with $f(\Omega)=\Omega=\frac{1}{r}$,
where $r$ is an affine distance:
\begin{equation}\label{eq:mm_asym}
R_{0m0m} 
=
\frac{1}{2}
\frac{\bar{\Psi}^0_4}{r} + O(-1)
,
\end{equation}
\begin{equation}\label{eq:mmb_asym}
R_{0m0\bar{m}} = 
\frac{1}{2}
\frac{\Phi^0_{22}}{r^2} + O(-2)
,
\end{equation}
\begin{equation}\label{eq:m3_asym}
\begin{split}
R_{0m03} =
\frac{1}{2\sqrt{2}} \frac{\bar{\Psi}^0_3}{r^2} + O(-2)
,
\end{split}
\end{equation}

\begin{equation}\label{eq:33_asym}
\begin{split}
R_{0303} =
2 \frac{\Re e\{ \Psi_2^0\}}{r^3}+ O(-3)
;
\end{split}
\end{equation}
where each quantity with a supra index 0 represents the leading
order behavior.
In terms of the polarization modes defined above one has:
\begin{equation}\label{eq:R+asymp}
R_+  
= \frac{\Re e\{\bar{\Psi}_4^0 \}}{r} + O(-1)
,
\end{equation}
\begin{equation}\label{eq:R_xasymp}
R_\times 
= \frac{\Im m\{ \bar{\Psi}_4^0 \}}{r} + O(-1)
,
\end{equation}
\begin{equation}\label{eq:R_Rasymp}
R_R = \frac{\Phi^0_{22}}{r^2} + O(-2)
,
\end{equation}
\begin{equation}\label{eq:R_{vx}asymp}
R_{vx} = 
\frac{ \Re e \{ \bar{\Psi}^0_3\} }{r^2} + O(-2)
,
\end{equation}
\begin{equation}\label{eq:R_{vy}asymp}
R_{vy} = 
\frac{ \Im m \{ \bar{\Psi}^0_3\} }{r^2} + O(-2)
,
\end{equation}
and
\begin{equation}\label{eq:R_Masymp}
R_M = 
2 \frac{\Re e\{ \Psi_2^0\}}{r^3}+ O(-3)
.
\end{equation}

The above discussion indicates that it is
natural to describe the polarization state
in terms of components which behave as quantities of spin weight 2, 1 or 0;
making reference in the case of spin weight 0 of the two channels,
that we call the mass channel, associated to $\Re e\{ \Psi_2^0\}$,
and the matter radiation channel, associated to $\Phi_{22}^0$.
The assignment of the name `mass channel' to the component $\Re e\{ \Psi_2^0\}$
is due to the fact that it contributes to the calculation
of total momentum\citep{Moreschi04}
at future null infinity.

It is deduced then that although the equations for the polarization modes
are algebraically compatible with those used in the weak plane wave models
of reference \cite{Eardley:1974nw},
our equations  \ref{eq:mm_asym}-\ref{eq:33_asym},
or \ref{eq:R+asymp}-\ref{eq:R_Masymp},
show that only the spin 2 modes are the astrophysical observable 
polarizations; in other words, if the other modes or channels
were detected, then they would not have an astrophysical distant origin.
It is probably worthwhile to emphasize that we have not assumed a particular
field equation, but only that isolated systems are well represented 
by general asymptotically flat spacetimes\citep{Moreschi87}.

In summary, within this set of gravitational theories, in a typical astrophysical scenario one
expects to record only spin 2 polarization gravitational-wave signals.
For this reason, in this article we concentrate only
in the presence of these two components of the polarization of the gravitational wave.

\section{Basis and coordinate systems}

Based on the results from previous efforts found in the literature,
we expect to have errors in the localization of the sources of
the order of one degree in angular coordinates.
This allows us to consider the local geometry, used in the astrophysics
determinations, to be given by a flat geometry; in agreement
with the discussions of \citep{Ashby:2003vja}.

The position and orientation of the observatories are defined in terms
of geocentric coordinates that rotate with the Earth.
More specifically one normally uses the International Terrestrial Reference System(ITRS).
We also need to use the
Earth Rotation Angle(ERA) which is the angle between the Terrestrial Intermediate Origin 
and the Celestial Intermediate Origin, positively in the retrograde direction.
Its specification is determined by 
the International Earth Rotation and Reference Systems Service(IERS)\citep{IERS_Conventions_2010} 
conventions.

\section{Basis for the calculation of the detector pattern functions}

Our conventions agree with those of \cite{Poisson2014}
with a slight change in the notation.

Let $\{ \hat{e}_1, \hat{e}_2, \hat{e}_3 \}$ be the orthonormal basis for a detector.
One can perform first a rotation around $\hat{e}_3$ of an angle $\phi$,
so that one obtains:

\begin{align}
\hat{e}_1'' =& \cos(\phi) \hat{e}_1 + \sin(\phi) \hat{e}_2 \\
\hat{e}_2'' =& -\sin(\phi) \hat{e}_1 + \cos(\phi) \hat{e}_2 \\
\hat{e}_3'' =& \hat{e}_3
;
\end{align}
then one performs a rotation $\theta$ around the new direction $\hat{e}_2''$;
so that one obtains:

\begin{align}
\hat{e}_1' =& \label{eq:pw-prime-basis-1}
\cos(\theta) \hat{e}_1'' - \sin(\theta) \hat{e}_3'' 
=
\cos(\theta) \cos(\phi) \hat{e}_1 \nonumber \\
&+ \cos(\theta) \sin(\phi) \hat{e}_2 - \sin(\theta) \hat{e}_3 \\
\label{eq:pw-prime-basis-2}
\hat{e}_2' =& \hat{e}_2'' = -\sin(\phi) \hat{e}_1 + \cos(\phi) \hat{e}_2 \\
\label{eq:pw-prime-basis-3}
\hat{e}_3' =& 
 \sin(\theta) \hat{e}_1'' + \cos(\theta) \hat{e}_3''
=\sin(\theta) \cos(\phi) \hat{e}_1 \nonumber \\
&+ \sin(\theta) \sin(\phi) \hat{e}_2  + \cos(\theta) \hat{e}_3
.
\end{align}
The final degree of freedom is to perform a rotation $\psi$ around the new axis $\hat{e}_3'$,
and we also carry out an inversion of the second and third basis vectors,
so that one obtains:

\begin{align}\label{eq:pw-mnomega-basis-1}
\tilde{e}_1      =&  \cos(\psi) \hat{e}_1' + \sin(\psi) \hat{e}_2' \\
\label{eq:pw-mnomega-basis-2}
\tilde{e}_2      =&  \sin(\psi) \hat{e}_1' - \cos(\psi) \hat{e}_2' \\
\label{eq:pw-mnomega-basis-3}
\tilde{e}_3 =&  -\hat{e}_3'
;
\end{align}
which agrees with ref. \cite{Poisson2014},
by noting for example that our $\tilde{e}_1$ is theirs $\mathbf{e}_X$.
In this way, $\tilde{e}_3$ points in the direction of the propagation of
the GW.

\section{Reference frames for the propagation of the gravitational wave}

Normally the basis for the GW is defined with respect to a
unit vector $\hat{k}=\tilde{e}_3$
in the direction of the propagation and two other orthonormal vectors defining
an oriented basis.

In our work we think of a GW as coming from the direction
$(\delta,\alpha)$ in the sky, which could be identified with standard
equatorial coordinates\citep{bradt2004astronomy}, but instead to use
the astronomical definitions, we will use the standard geometrical
angles, so that we use radians with the ranges:
$\delta \in [0, \pi]$(-declination using colatitude) and 
$\alpha \in [0, 2 \pi]$(right ascension in radians) also eastward;
and with origin at the vernal equinox.

Then, we can define the vector $\hat{k}$ by
\begin{equation}\label{eq:hat_k}
\hat{k} = - \big( \sin(\delta) \cos(\alpha), \sin(\delta) \sin(\alpha), \cos(\delta)  \big)
,
\end{equation}
and
\begin{equation}\label{eq:vec_delta}
\hat{\delta} = \big( \cos(\delta) \cos(\alpha), \cos(\delta) \sin(\alpha), -\sin(\delta)  \big)
,
\end{equation}
\begin{equation}\label{eq:vec_alpha}
\hat{\alpha} = \big( - \sin(\alpha), \cos(\alpha), 0  \big)
;
\end{equation}
with orientation $\{ \hat{\alpha}, \hat{\delta}, \hat{k} \}$,
noting that in a Mollweide projection, $\hat{\alpha}$ points
to the eastward direction (left) and $\hat{\delta}$
points to the south (down).
In this way, $\hat{k}$ points to us from the celestial coordinates $(\delta,\alpha)$.

The + and $\times$ polarization components are defined with respect to this
$\{ \hat{\alpha}, \hat{\delta} \}$ basis.

\section{Detector pattern functions for GW}\label{sec:patt-func}

Let us start by noting that in the literature, different authors use diverse
language to refer to the polarization properties of a GW.
So here we first review our approach to the subject and then relate to
other point of views found in publications.

The perturbations produced by a GW are thought as
local variations of the curvature tensor, which when expressed in terms
of a local basis can be described by field components as the ones
used in \cite{Eardley:1974nw}, namely the Weyl
components $\Psi_4$,
$\Psi_3$ and $\Psi_2$, and the Ricci component $\Phi_{22}$.
Since a gravitational-wave detector is supposed to have a linear response
to these perturbations, one can make the following decomposition
of the signal $s(t)$ recorded by an observatory:
\begin{equation}\label{eq:s-t}
s(t) = \sum_{p} F_p(\theta,\phi,\psi,t) s_p(t)
;
\end{equation}
where $p$ is used to distinguish among the different types of polarizations.

Very often in the literature one finds the discussion of polarization states
to be based directly in perturbations of the 
metric\citep{Nishizawa:2009bf,Poisson2014}; where one can map
the perturbations of the curvature to variations of the metric\citep{Poisson2014}.

The six pattern functions presented below, correspond to the complete algebraic
study of the curvature matrix appearing in the geodesic deviation equation;
however, we have presented arguments above that indicate
that in a normal astrophysical situation, the $+$ and $\times$ modes
would be the more relevant ones.
In reconstructing the polarizations $s_+$ and $s_\times$
for each event, one has 
\vspace{1mm}
{\small 
\begin{equation}\label{eq:X}
\begin{split}
s_X(t + \tau_X) =& 
F_{+X}(\theta_X,\phi_X,\psi_X,t) s_+(t) \\
&+
F_{\times X}(\theta_X,\phi_X,\psi_X,t) s_\times(t)
.
\end{split}
\end{equation}
}

Note that our choice for the angular terrestrial basis
agrees with the choice of reference \cite{Poisson2014};
which means that our signs in the calculations of the {\it detector pattern functions}
should agree with this reference.
More specifically, using prime for the quantities of reference \cite{Poisson2014};
we have that the complete set of pattern functions are

{\footnotesize
\begin{align}
F_+ =& F'_+ = \frac{1}{2}( 1 + \cos(\theta)^2) \cos(2 \phi) \cos(2 \psi)
- \cos(\theta) \sin(2 \phi) \sin(2 \psi) \\
F_\times =& F'_\times = \frac{1}{2}( 1 + \cos(\theta)^2) \cos(2 \phi) \sin(2 \psi)
+ \cos(\theta) \sin(2 \phi) \cos(2 \psi) \\
F_{vx} =& F'_{v1} = -\sin(\theta) \big(
\cos(\theta) \cos(2 \phi) \cos(\psi)
- \sin(2 \phi) \sin(\psi) \big) \\ 
F_{vy} =& F'_{v2} = -\sin(\theta) \big(
\cos(\theta) \cos(2 \phi) \sin(\psi)
+ \sin(2 \phi) \cos(\psi) \big)\\ 
F_M =& F'_L =  \frac{1}{2} \sin(\theta)^2 \cos(2 \phi) \\ 
F_R =& F'_S = - \frac{1}{2} \sin(\theta)^2 \cos(2 \phi) 
; 
\end{align}
}where beyond the spin-2 polarization $+$ and $\times$ 
we denote with $F_{vx}$ and $F_{vy}$ the spin-1 polarizations 
and for the spin-0 polarizations we use
$F_M$ to denote the mass polarization 
and $F_R$ to denote the matter radiation polarization.
We have just used the notation in terms of plane waves introduced in
reference \cite{Eardley:1974nw}; but 
we describe the modes in terms of available curvature components
to an observer of an astrophysical system emitting gravitational radiation;
that is, we do not assume that the detected GW is a plane wave.

The pattern functions have several properties, for example that:
$F_{+,\times}(\pi-\theta, \pi -\phi, \psi) = F_{+,\times}(\theta, \phi, \psi)$.
This means that in the their manipulations, having a physically 
selected direction, one should expect phantom repetitions of corresponding maximum
or minimum of functions built out of the pattern functions.

The geometrical data for the gravitational-wave observatories can be
obtained from the file {\sf LALDetectors.h} that is available on the web at several
places, for example at \url{git.ligo.org/lscsoft/lalsuite/-/blob/master/lal/lib/tools/LALDetectors.h}

\section{Fitting a universal chirp form}

The argument of the trigonometric functions in the fitting
mechanism, has the form
$\Phi(t) = \phi_c(t) + \phi_f$;
where $\phi_c(t)=-2 \big(\frac{ t_f - t}{5 t_{ch}}\big)^{p_c 5/8}$ has the
non-trivial time dependence, and the constant $\phi_f$ denotes a global phase.
This means that the fitting of the signal at each detector only requires two
parameters to relate to the dynamical dependence of $g(t)\cos(\phi_c(t))$
and $g(t)\sin(\phi_c(t))$.
Therefore we organize the calculation by first fitting the parameters
$[B_{+ cX},B_{\times cX}]$ to the chirp base, just presented,
and then infer the values of $[B_{+ X},B_{\times X}]$
for the phase $\phi_f$; which it can be shown to be given by
\begin{align}
B_{+ X} =& \cos(\phi_f) B_{+ cX} -  \sin(\phi_f) B_{\times cX} \\
B_{\times X} =& \sin(\phi_f) B_{+ cX} + \cos(\phi_f) B_{\times cX}
.
\end{align}

\section{Localization by fitting a universal chirp form}\label{sec:loc-chirp}

\begin{figure}[H]
\centering
\includegraphics[clip,width=0.48\textwidth]{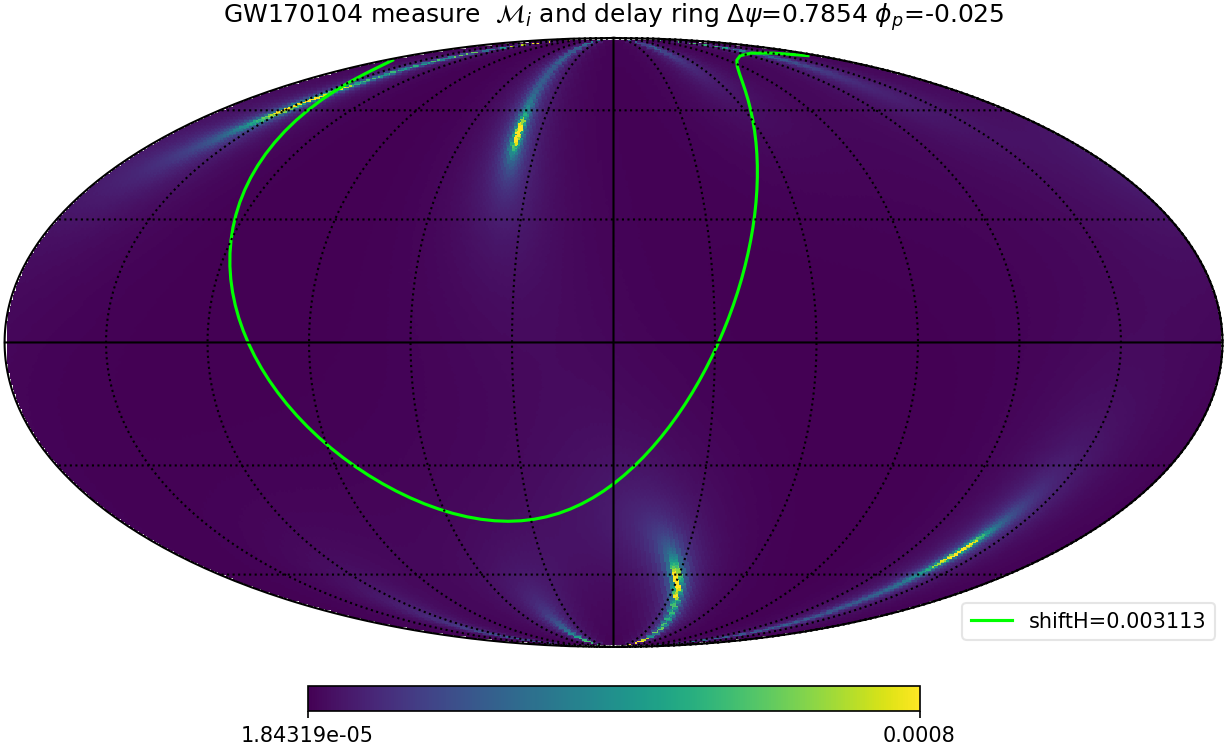}
\caption{Plot of the measure $\mathcal{M}_i$ and the delay ring.
	Sky localization for the source of the GW170104 event
shows a signal on the nominal ring on the northern east region.
}
\label{fig:loc_1-gw170104}
\end{figure}

As explained in section \ref{sec:univ_fitt_chirp} we study the maxima of $M_i$;
which best estimates are shown in Fig. \ref{fig:loc_1-gw170104}.
It can be seen that there are several local maxima, marked in yellow;
but only one of them overlap with the delay ring.
The other local maxima are considered phantom images due to effects
explained in appendix \ref{sec:patt-func}. 
The final location is obtained by the product of this signal with the Gaussian
of the delay ring; so that it only remains the maximum on the ring,
as shown in Fig. \ref{fig:loc-gw170104}.

\begin{figure}[H]
\centering
\includegraphics[clip,width=0.48\textwidth]{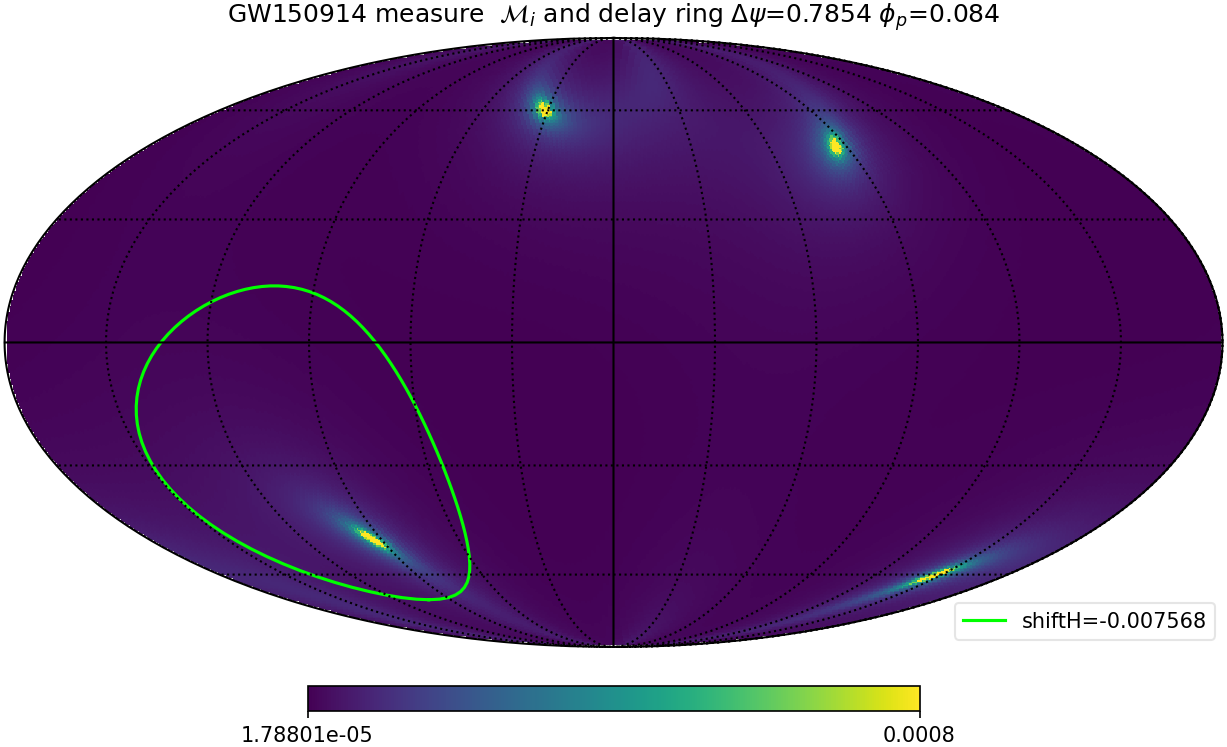}
\caption{Plot of the measure $\mathcal{M}_i$ and nominal ring.
	Sky localization for the source of the GW150914 event
	is close to the southern part of the ring at (94.16,-49.70) longitude and latitude coordinates.
}
\label{fig:loc_1-gw150914}
\end{figure}

\add{The initial localization for the event GW150914 is shown in Fig.}
\ref{fig:loc_1-gw150914}.
\noindent
\add{In the main text it was discussed the impresive anticorrelation of the
signals detected at both observatories. This has a strong effect on our procedure
which as a consequence it finds the location of the source, close to the
nominal delay ring, but separated from it; as it can be seen in the graph
of Fig.}
\ref{fig:loc_1-gw150914}.

\section{Characteristics of the strain}

The event GW170104 is presented in the 
\href{gwosc.org/eventapi/html/GWTC-1-confident/GW170104/v2/}{gwosc.org}
page as a GWTC-1-confident type, with GPS event time 1167559936.6;
corresponding to UTC time: 2017-01-04 10:11:58.6.
It was assigned a network SNR of 13.8 and a sky localization area of 1000
square degrees. We have used version v2 of the provided LIGO data.

The one-sigma calibration uncertainties for the strains is 
informed\citep{Abbott:2017vtc}
to be better than 5\% in amplitude and 
$3^\circ$ in phase over the frequency
range 20-1024Hz.

In Fig. 1 of reference \cite{Abbott:2017vtc} they show a signals
with a time span of approximately 0.11s, and it has been reported
that the Livingston data has been shifted by -0.003s,
and the sign of its amplitude has been inverted.

\section{Determination of the appropriate time delay for the GW170104 event}\label{sec:timedelay}

The initial step in the procedure is to apply the pre-processing filtering techniques
to the strains as explained in \cite{Moreschi:2019vxw};
for which we have used a bandpass of [27,1003]Hz.

In our preliminary study of GW170104 in \cite{Moreschi:2024njx}
we have used a window time span of 0.28s, and applying 
the measure OM we could assign a level of significance $\alpha=5.5 \times 10^{-8}$
to the detection of similar signals in the two LIGO observatories.

Here we carryout a more detailed analysis on the data of this  event.
\add{In this study we have used a bandpass in the range [30,350]Hz.}
The measure OM can be used to study a variety of topics.
Its result and sensitivity depends on several parameter,
in particular in the time length of the window($wl$) that is being used.
Although in the preliminary study we used a window of 0.28s;
in order to study the behavior of the strains close to
the time of maximum amplitude we use also a window
of 0.14s as a function of the time shift of the Hanford(H) strain
with respect to the Livingston(L) strain.
We observe
that while for 
$wl=0.28$s the maximum is at $t_{d0} = 0.003235$s;
for $wl=0.14$s the maximum is at $t_d= 0.003113$s.
We will use this last one as the reference time delay between the observatories;
since it gives a better representation of the coincidence time for the maximum
at high amplitudes.

\begin{figure}[H]
\centering
\includegraphics[clip,width=0.47\textwidth]{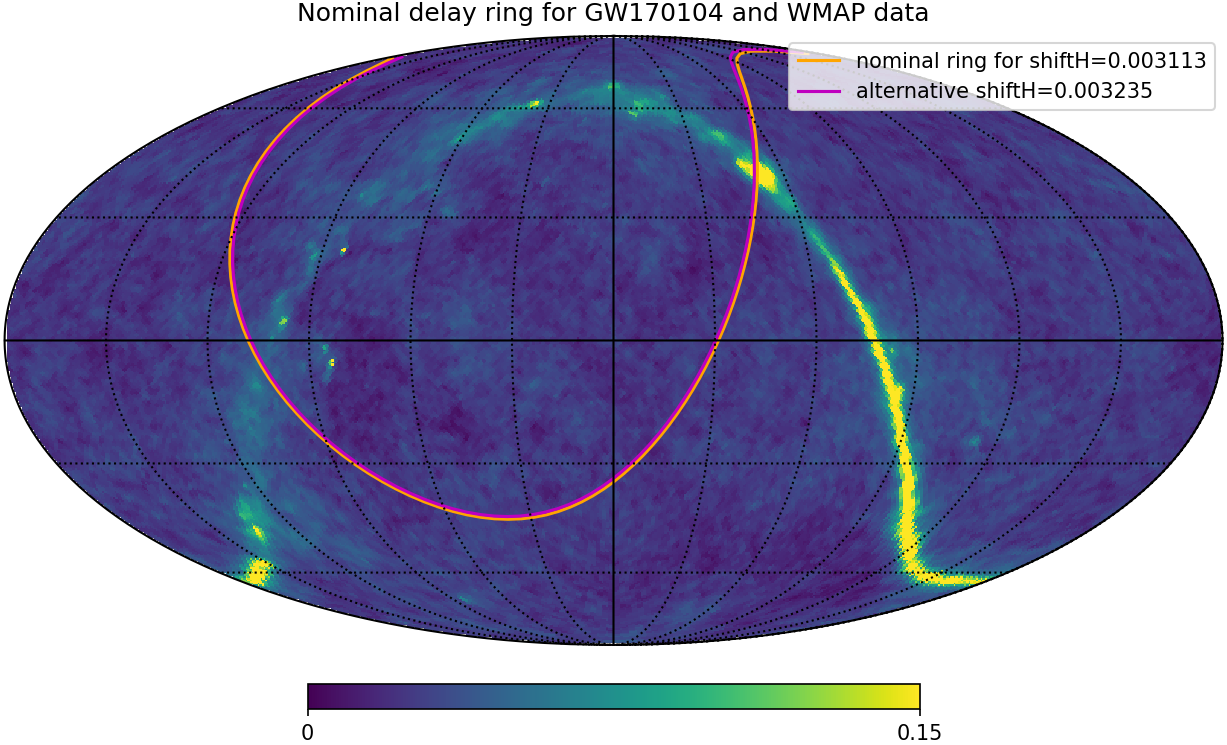}
\caption{Nominal delay ring and alternative ring in the sky along with a WMAP
	graph of the CMB, indicating the position of the galaxy,
	in order to show the nature of the Mollweide projection in
	equatorial coordinate system with origin at the center
	and East towards left. 
}
\label{fig:delay-ring+Wmap}
\end{figure}

The corresponding delay ring in the sky is shown in Fig. \ref{fig:delay-ring+Wmap}
where we have included also a CMB map from the \href{science.nasa.gov/mission/wmap/}{WMAP} team, in order to show
the position of the galaxy and so clarify the origin of the equatorial coordinate system.
The determination of the delay ring has an inherent error
which we will estimate next.
To begin with, the two previous time delays yield two different angles
to determine the delay ring. Let us call them $\varTheta_{d0}$ and $\varTheta_d$;
associated with the delay times $t_{d0}$ and $t_d$ respectively.
These two delay times already provide us with an operational
time delay error  $\delta t_{op} = |t_{d0} - t_d| = 0.000122$s.
An estimate of the time delay error can also be calculated
from the approximate high frequency contribution
at maximum signal.
For instance, for this event we have estimated from the scalograms
that the maximum relevant contributions close to the chirp time are around $\nu_{max} = 300$Hz.
From this, in the case in which there is no information on the orientation of the observatories 
and neither there is no information on the location of the source nor on its polarization angle,
one could estimate a general time delay error of the order of $\delta t = \frac{T_4}{2}=0.0004167$s;
where $T_4= \frac{T}{4}= \frac{1}{4  \nu_{max}}$ is a quarter of the period determined by $\nu_{max}$.
This in turn induces an error in the determination of the delay ring angle
of $\delta \varTheta = | \varTheta_d - \arccos( \frac{t_d + \delta t}{t_{LH}})|$,
where $\varTheta_d$ is the angle that determines the delay ring and 
$t_{LH}$ is the time of flight of a signal between the two observatories.
We construct a Gaussian distribution around the delay ring considering both
contributions to the estimate of the error in the angle that determines the ring and so
we take the variance of the ring as 
$\sigma_r = |\varTheta_{d0}- \varTheta_d| +  \delta \varTheta$;
which intends to provide a coarse estimate on our error in the determination of
the delay ring.

\section{Other studies of GW170104}\label{subsec:other-studies}

Considering other efforts on the localization for GW170104 with electromagnetic counterparts
we mention here some of those works.
In reference \cite{Fermi-GBM:2017soa}
they presented the Fermi Gamma-ray Burst Monitor and Large Area Telescope observations of 
the event GW170104 and claimed that no candidate electromagnetic counterpart was
detected by either of the two instruments; although they report upper bounds for the fluxes.

The results from the analysis of hard X-ray and
gamma-ray data of the AGILE mission on the localization of GW170104 
were reported in \cite{AGILE:2017vou};
and claimed that no transient gamma-ray source was detected
over timescales of 2, 20 and 200s starting at the time of the event.
However they reported an event E2 occurring at $0.46\pm0.05$s
before the event time which they claim is significant.
They  could no determine the position of E2, but they obtained
the sky region where is the B arc, 
which is the southern arc reported by the LIGO LALInference 
method, to which our localization belongs.

In \cite{Savchenko:2017dyg} the authors reported on the
data from the International Gamma-Ray Astrophysics Laboratory; which allowed them 
to set upper limits on the $\gamma$-ray and hard X-ray emission associated with the GW170104 event
challenging the possible association of this event with electromagnetic counterparts.

An all-sky high-energy neutrino follow-up
search using data from the Antares neutrino
telescope has been reported in \cite{ANTARES:2017fqy}.
They found no neutrino candidates
within $\pm500$s around the event time nor any time 
clustering of events over an extended time window of $\pm3$ months.

In \cite{Stalder:2017qic} they reported the peculiar optical transient, ATLAS17aeu;
which was discovered 23.1 hr after GW170104 and rapidly faded over the next three
nights.
They claimed that the observations indicate that ATLAS17aeu is plausibly a normal 
GRB afterglow at significantly higher redshift than the distance constraint 
for GW170104 and therefore a chance coincidence.

\section{Spin-2 polarization modes of GW170104 for different polarization angles}\label{sec:spin2-polar}

\begin{figure}[H]
\centering
\includegraphics[clip,width=0.48\textwidth]{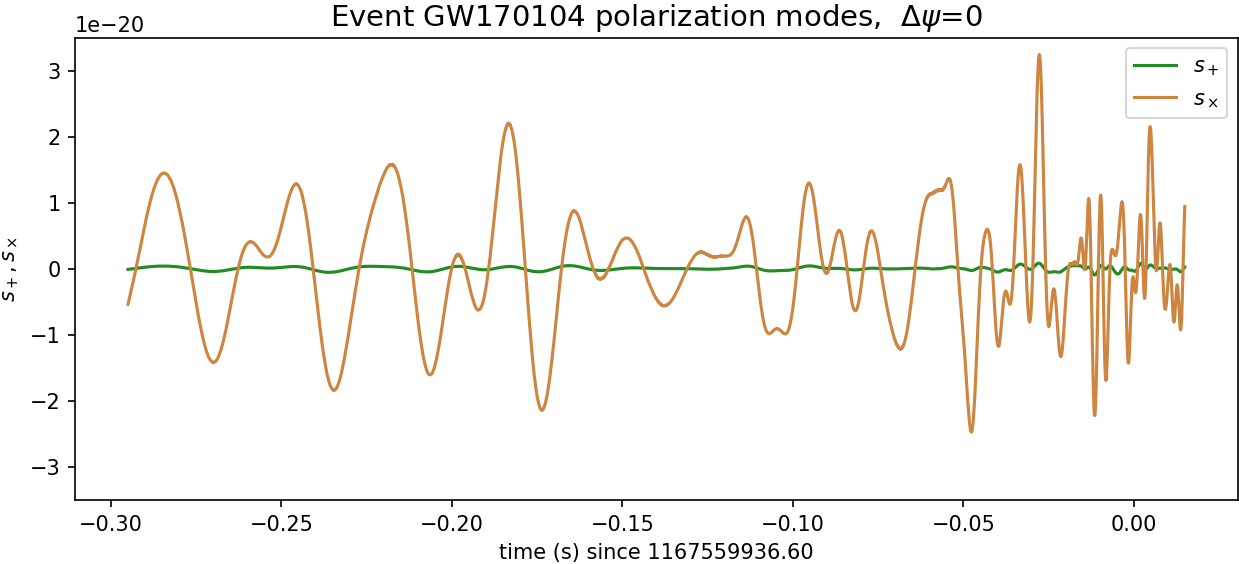}
\caption{Polarization modes of GW170104.
}
\label{fig:s+_sx_0_aum}
\end{figure}

\begin{figure}[H]
\centering
\includegraphics[clip,width=0.48\textwidth]{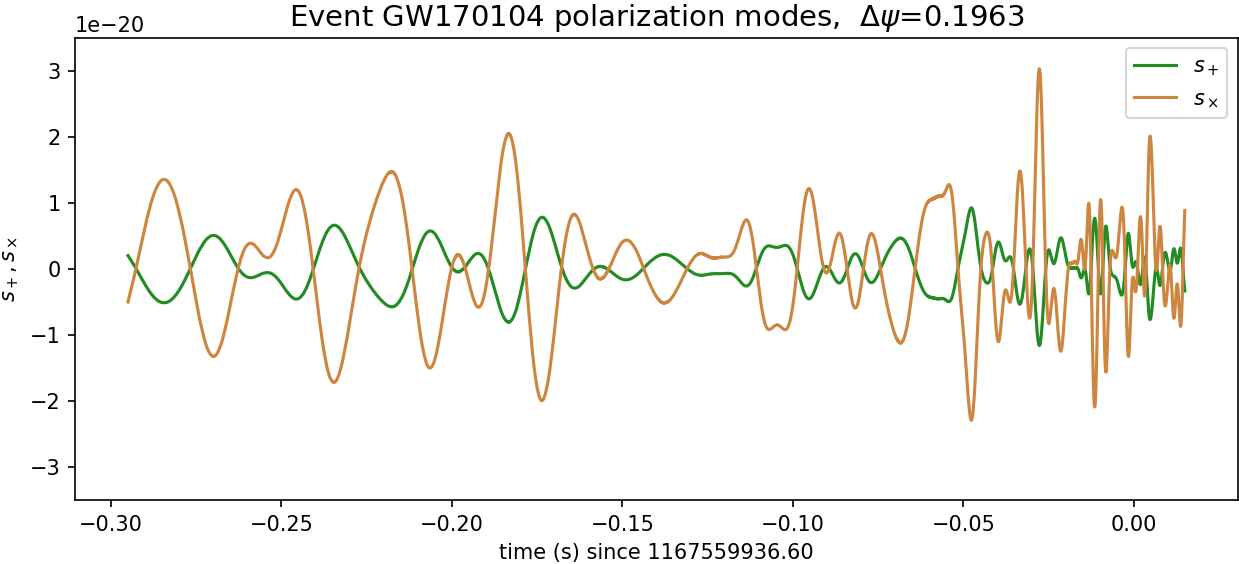}
\caption{Polarization modes of GW170104.
}
\label{fig:s+_sx_1_aum}
\end{figure}

\begin{figure}[H]
\centering
\includegraphics[clip,width=0.48\textwidth]{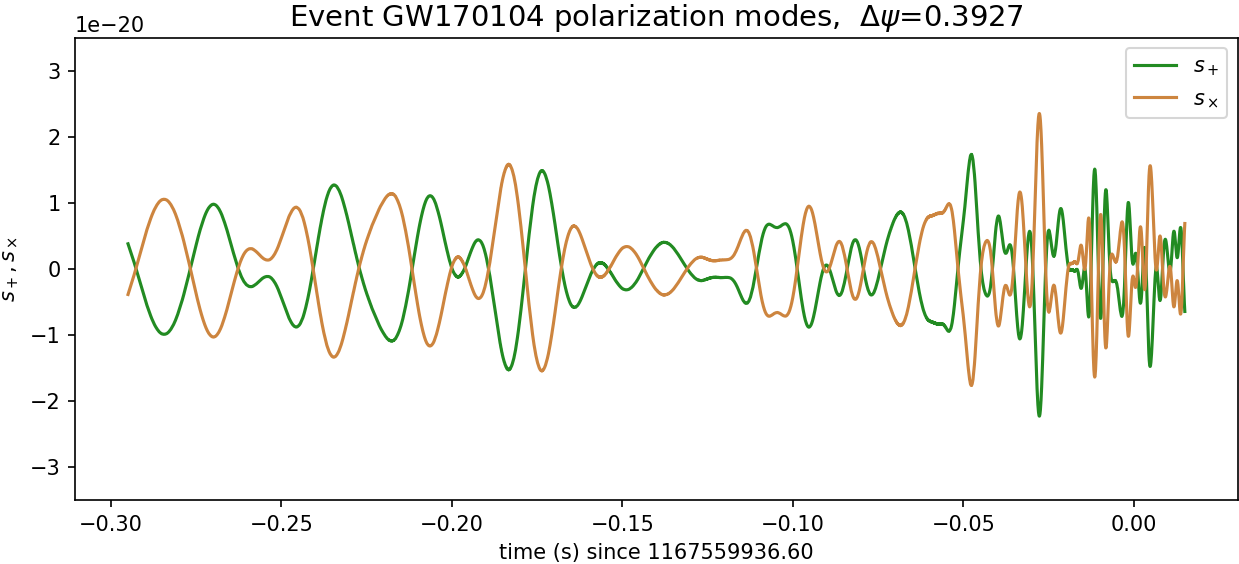}
\caption{Polarization modes of GW170104.
}
\label{fig:s+_sx_2_aum}
\end{figure}

\begin{figure}[H]
\centering
\includegraphics[clip,width=0.48\textwidth]{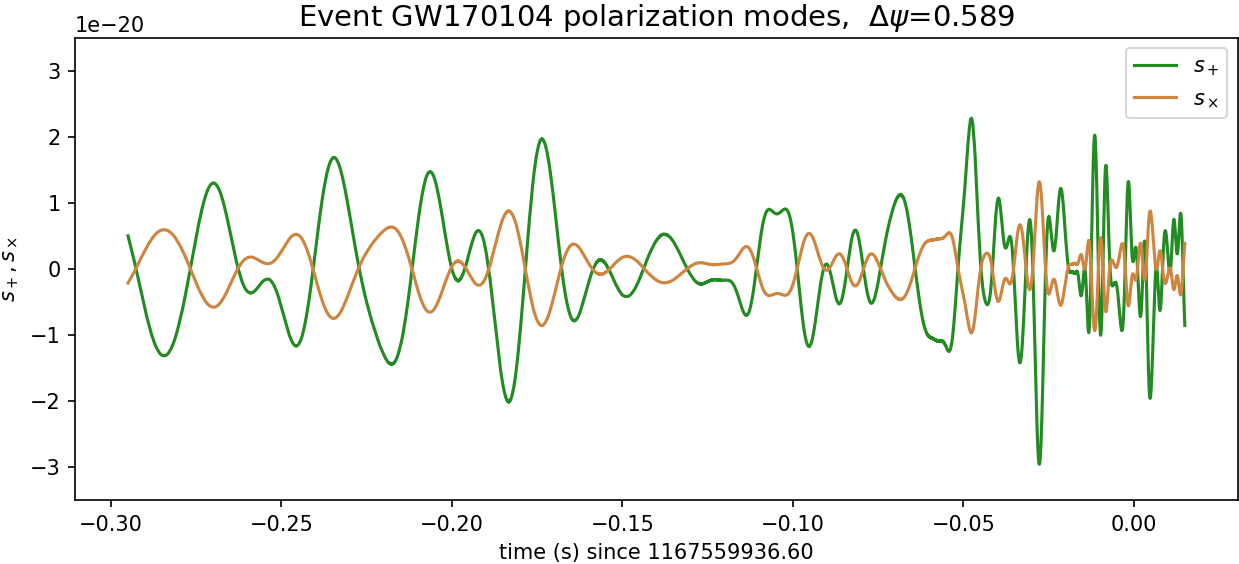}
\caption{Polarization modes of GW170104.
}
\label{fig:s+_sx_3_aum}
\end{figure}

\begin{figure}[H]
\centering
\includegraphics[clip,width=0.48\textwidth]{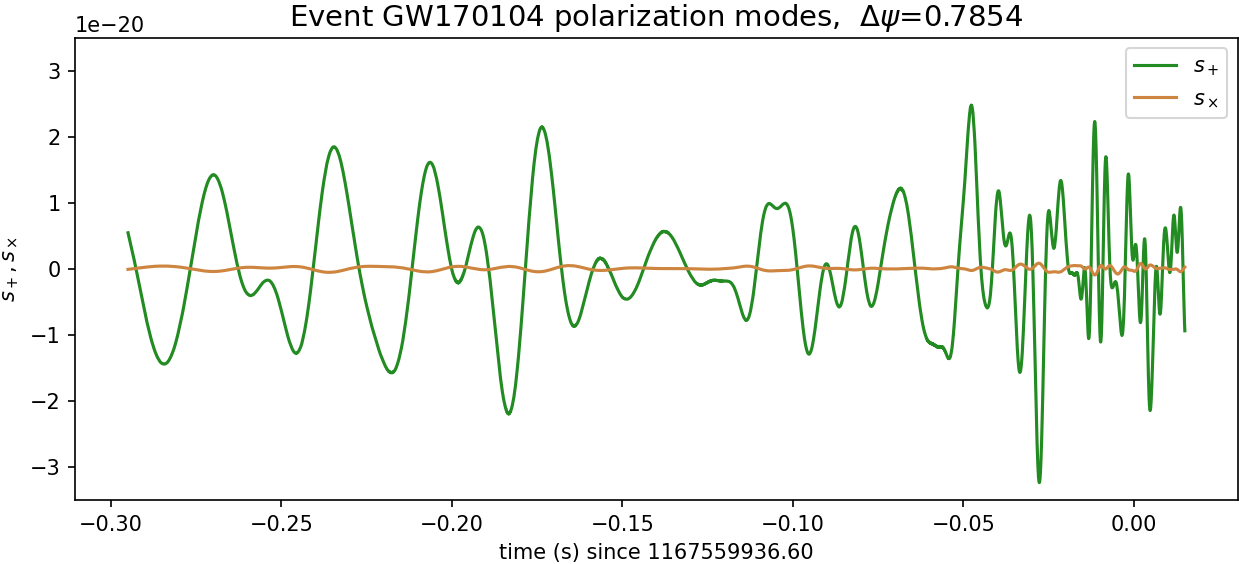}
\caption{Polarization modes of GW170104.
}
\label{fig:s+_sx_4_aum}
\end{figure}

For completeness we also present the polarization modes in the frames with $\Delta\psi=0$,
$\Delta\psi=\frac{\pi}{16}$, $\Delta\psi=\frac{2\pi}{16}$, $\Delta\psi=\frac{3\pi}{16}$ and $\Delta\psi=\frac{\pi}{4}$ 
close to the nominal event time
in Figs. \ref{fig:s+_sx_0_aum}-\ref{fig:s+_sx_4_aum}. 
It can be seen that for $\Delta\psi=\frac{\pi}{4}$ the polarization modes return to
the original values according to the transformation properties of the modes.
This indicates that in fact the time series we have calculated with the names
$s_+$ and $s_\times$ behave as spin-2 quantities; within the numerical precision
of our calculation.

\section{Simulation using an injected simple signal}\label{sec:simul-signal}

In order to further validate our procedure, we here consider the injection of a simple
synthetic signal for a binary black hole coalescence.
We use the chirp wave form model\citep{Peters:1964zz} described in the textbook on analysis of 
gravitational-wave data
\cite{Jaranowski:2009zz} 
for binary
polarization waveforms with radiation reaction effects, 
with a notation closer to that of 
\cite{Cutler:1994ys}
that we express as:
\begin{align}
s_+ =& \frac{s_0}{a (1-e^2)}  \frac{1 + \cos(\iota)^2}{2} \label{eq:inj-s+}
\Bigg(
-\cos(2 \phi(t)) \\
&+ e \bigg(- \frac{1}{4} \Big( 5 \cos( \phi(t) + \phi_0) +  \cos(3\phi(t) - \phi_0 )\Big) \\
& \quad\quad         + \frac{\sin(\iota)^2}{2 (1 + \cos(\iota)^2)} \cos(\phi(t) - \phi_0)
    \bigg) \\
&+ e^2 \bigg(-\frac{1}{2} \cos(2 \phi_0) + \frac{\sin(\iota)^2}{2 (1 + \cos(\iota)^2)}
    \bigg)
    \Bigg) \\
s_\times =& \frac{s_0}{a (1-e^2)} \cos(\iota)  \label{eq:inj-sx}
\Bigg(
- \sin(2 \phi(t)) \\
&- \frac{e}{4} \bigg( 5 \sin( \phi(t) +\phi_0) + \sin(3 \phi(t) -\phi_0)  \bigg) \\
&- \frac{e^2}{2} \sin(2\phi_0)
\Bigg)
,
\end{align}
where $a$ is the instantaneous semi-mayor axis, $e$ the instantaneous eccentricity,
$\iota$ is the angle between
the orbital angular momentum vector of the binary and the line of sight
and $\phi_0$ the angle of the orbital periapsis which we assume to occur at $t=0$.
Also $s_0 = \frac{4 G^2 M \mu}{c^4 R}$ where for individual masses $m_1$ and $m_2$,
$M=m_1+m_2$ is the total mass and $\mu=\frac{m_1 m_2}{M}$ is the reduced mass.
In the case of a circular orbit these expressions reduce to
\begin{align}
s_+ &= - s_0 \frac{1 + \cos(\iota)^2}{2 a} \label{eq:inj-s+0}
\cos\bigg(- 2\Big(\frac{|t_c - t |}{5 t_{ch}}\Big)^{5/8} + \phi_c\bigg) \\
s_\times &= - s_0 \frac{\cos(\iota)}{a}   \label{eq:inj-sx0}
\sin\bigg(- 2\Big(\frac{|t_c - t |}{5 t_{ch}}\Big)^{5/8} +\phi_c\bigg)
,
\end{align}
where $t_c$ is the `coalescence time', $t_{ch}$ is the `chirp time',
$\phi_c$ is a reference constant phase,
and in this case we use
\begin{equation}\label{dq:uno_a-de-t}
\frac{s_0}{a(t)} = \frac{ A_{ch}}{ ( |t_c - t |^{1/4} + \Delta t_0^{1/4} ) }
,
\end{equation}
where, we integrate in $A_{ch}$ all the amplitude dependence,
 and $\Delta t_0$ are parameters that are chosen
to adjust the amplitude and to limit the divergent pure chirp 
behavior.
We choose the coalescence time to be $t_e - 0.002$s
where $t_e$ is the published reference time for this event.

Let us note that in the literature normally appears the reference to the
chirp mass $\mathscr{M} \equiv \mu^{3/5} M^{2/5}$; which is related to the
chirp time by $t_{ch} = G \mathscr{M}/c^3$.

For the chirp time we use $t_{ch}=1.5\text{e-}4$s. 
Also we use $\iota=\frac{\pi}{6}$, $\phi_c=-\pi$ and $\Delta t_0=1.\text{e-}5$s.
Since our method makes use of the data mainly in the inspiral phase
and we are testing our procedure with a simplistic synthetic signal;
we do not attempt to model the merger and ringdown stages,
which occur in a very short time, when compared with the 
working window in the inspiral phase.

The signal is injected by choosing the angles in the celestial sphere
to agree with a different delay ring,
and for simplicity we have chosen arbitrarily the orientation of the GW
to be given by the frame determined by $\Delta \psi = 0$.

For the general case $e\neq0$ we follow \cite{Peters:1964zz} and express
\begin{equation}\label{eq:adee}
a(e) = a_0 (\frac{e}{e_0})^{12/19} 
\big(\frac{1 - e_0^2}{1 - e^2}\big)
\bigg( \frac{1 + \frac{121}{304} e^2}{1 + \frac{121}{304} e_0^2} \bigg)^{870/2299}
,
\end{equation}
and
\begin{equation}\label{eq:edot}
\frac{de}{dt} = -\frac{19}{12} \frac{\beta}{c_0^4} 
\frac{e^{-29/19} (1-e^2)^{3/2} }{\big( 1 + \frac{121}{304} e^2\big)^{1181/2299}}
,
\end{equation}
with
\begin{equation}\label{eq:c0}
c_0 = a_0 \Bigg(\frac{1 - e_0^2}{e_0^{12/19} \big( 1 + \frac{121}{304} e_0^2 \big)^{870/2299}} \Bigg) 
,
\end{equation}
and
\begin{equation}\label{eq:beta}
\beta = \frac{64 G^3 m_1 m_2 (m_1+m_2)}{5 c^5}
;
\end{equation}
where $a_0$ and $e_0$ are the corresponding initial values.

One also needs to integrate
\begin{equation}\label{eq:fidot}
\frac{d\phi}{dt} =  
\sqrt{G M} \frac{\big(1+ e \cos(\phi - \phi_0) \big)^2}{\sqrt{ a^3(1-e^2)^3}} 
,
\end{equation}
and for completeness we also mention that the radial coordinate is
\begin{equation}\label{eq:r}
r = \frac{a (1 - e^2)}{1 + e \cos(\phi - \phi_0)}
,
\end{equation}
the orbital period is
\begin{equation}\label{eq:orb-period}
P = 2 \pi \sqrt{\frac{a^3}{G M}}
;
\end{equation}
and the instantaneous orbital frequency can be written as
\begin{equation}\label{eq:orb-frec}
\nu \equiv \dot{\phi} = \frac{2 \pi}{P}\frac{\big(1+ e \cos(\phi - \phi_0) \big)^2}{\sqrt{(1-e^2)^3}} 
.
\end{equation}

\begin{figure}[H]
\centering
\includegraphics[clip,width=0.48\textwidth]{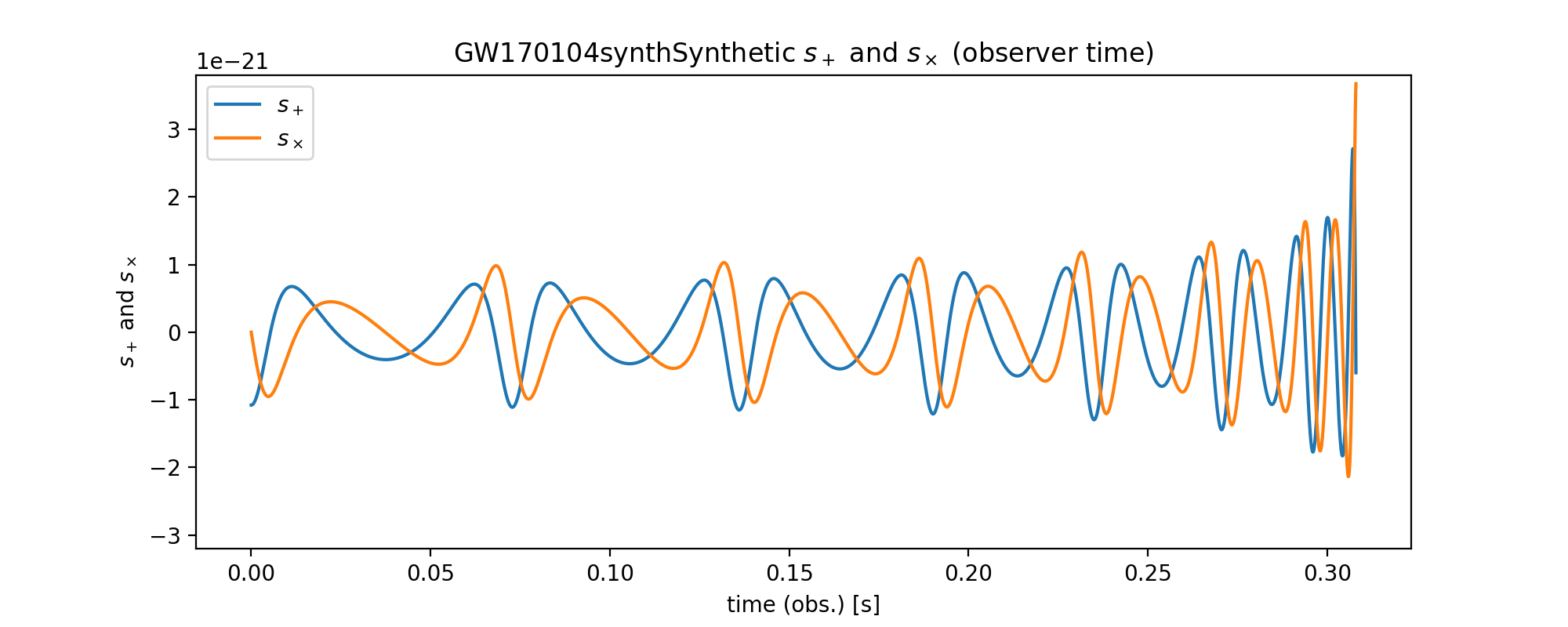}
\caption{Injected + and $\times$ polarization modes. }
\label{fig:synth_+mode_xmode}
\end{figure}

As a validation of our procedure, for the location of a GW source with spin-2 polarization,
we have applied it to this peculiar case of synthetic signal with $a_0=12(m_1+m_2)$ and 
$e_0=0.35$, where the resulting synthetic polarization modes are shown in Fig. \ref{fig:synth_+mode_xmode}.

Also, as a validation of our procedure, for the fidelity in the measurement of the PM,
the results in reconstructing the gravitational-wave polarization modes of this synthetic signal are
shown in the main text in Figs. \ref{fig:synth_+mode} and \ref{fig:synth_xmode};
where one can check an excellent agreement.




\begin{thebibliography}{65}
\providecommand{\natexlab}[1]{#1}
\providecommand{\url}[1]{{#1}}
\providecommand{\urlprefix}{URL }
\providecommand{\doi}[1]{\url{https://doi.org/#1}}
\providecommand{\eprint}[2][]{\url{#2}}
\bibcommenthead

\bibitem[{Abbott et~al(2016{\natexlab{a}})}]{TheLIGOScientific:2016qqj}
Abbott BP, et~al (2016{\natexlab{a}}) {GW150914: First results from the search
	for binary black hole coalescence with Advanced LIGO}. Phys Rev
D93(12):122003. \doi{10.1103/PhysRevD.93.122003},
{\href{https://arxiv.org/abs/1602.03839}{{arXiv:1602.03839}}} {[gr-qc]}

\bibitem[{Abbott et~al(2016{\natexlab{b}})}]{Abbott:2016blz}
Abbott BP, et~al (2016{\natexlab{b}}) {Observation of Gravitational Waves from
	a Binary Black Hole Merger}. Phys Rev Lett 116(6):061102.
\doi{10.1103/PhysRevLett.116.061102},
{\href{https://arxiv.org/abs/1602.03837}{{arXiv:1602.03837}}} {[gr-qc]}

\bibitem[{Abbott et~al(2016{\natexlab{c}})}]{LIGOScientific:2016lio}
Abbott BP, et~al (2016{\natexlab{c}}) {Tests of general relativity with
	GW150914}. Phys Rev Lett 116(22):221101.
\doi{10.1103/PhysRevLett.116.221101}, [Erratum: Phys.Rev.Lett. 121, 129902
(2018)], {\href{https://arxiv.org/abs/1602.03841}{{arXiv:1602.03841}}}
{[gr-qc]}

\bibitem[{Abbott et~al(2017{\natexlab{a}})}]{LIGOScientific:2016xax}
Abbott BP, et~al (2017{\natexlab{a}}) {Calibration of the Advanced LIGO
	detectors for the discovery of the binary black-hole merger GW150914}. Phys
Rev D 95(6):062003. \doi{10.1103/PhysRevD.95.062003},
{\href{https://arxiv.org/abs/1602.03845}{{arXiv:1602.03845}}} {[gr-qc]}

\bibitem[{Abbott et~al(2017{\natexlab{b}})}]{Abbott:2017vtc}
Abbott BP, et~al (2017{\natexlab{b}}) {GW170104: Observation of a 50-Solar-Mass
	Binary Black Hole Coalescence at Redshift 0.2}. Phys Rev Lett 118(22):221101.
\doi{10.1103/PhysRevLett.118.221101},
{\href{https://arxiv.org/abs/1706.01812}{{arXiv:1706.01812}}} {[gr-qc]}

\bibitem[{Abbott et~al(2017{\natexlab{c}})}]{Abbott:2017oio}
Abbott BP, et~al (2017{\natexlab{c}}) {GW170814: A Three-Detector Observation
	of Gravitational Waves from a Binary Black Hole Coalescence}. Phys Rev Lett
119(14):141101. \doi{10.1103/PhysRevLett.119.141101},
{\href{https://arxiv.org/abs/1709.09660}{{arXiv:1709.09660}}} {[gr-qc]}

\bibitem[{Abbott et~al(2018)}]{Abbott:2018utx}
Abbott BP, et~al (2018) {Search for Tensor, Vector, and Scalar Polarizations in
	the Stochastic Gravitational-Wave Background}. Phys Rev Lett 120(20):201102.
\doi{10.1103/PhysRevLett.120.201102},
{\href{https://arxiv.org/abs/1802.10194}{{arXiv:1802.10194}}} {[gr-qc]}

\bibitem[{Abbott et~al(2020)}]{KAGRA:2013rdx}
Abbott BP, et~al (2020) {Prospects for observing and localizing
	gravitational-wave transients with Advanced LIGO, Advanced Virgo and KAGRA}.
Living Rev Rel 23(3):69. \doi{10.1007/s41114-020-00026-9}

\bibitem[{Abbott et~al(2021)}]{LIGOScientific:2019lzm}
Abbott R, et~al (2021) {Open data from the first and second observing runs of
	Advanced LIGO and Advanced Virgo}. SoftwareX 13:100658.
\doi{10.1016/j.softx.2021.100658},
{\href{https://arxiv.org/abs/1912.11716}{{arXiv:1912.11716}}} {[gr-qc]}

\bibitem[{Abbott et~al(2023)}]{KAGRA:2023pio}
Abbott R, et~al (2023) {Open Data from the Third Observing Run of LIGO, Virgo,
	KAGRA, and GEO}. Astrophys J Suppl 267(2):29. \doi{10.3847/1538-4365/acdc9f},
{\href{https://arxiv.org/abs/2302.03676}{{arXiv:2302.03676}}} {[gr-qc]}

\bibitem[{Albert et~al(2017)}]{ANTARES:2017fqy}
Albert A, et~al (2017) {All-sky search for high-energy neutrinos from
	gravitational wave event GW170104 with the Antares neutrino telescope}. Eur
Phys J C 77(12):911. \doi{10.1140/epjc/s10052-017-5451-z},
{\href{https://arxiv.org/abs/1710.03020}{{arXiv:1710.03020}}} {[astro-ph.HE]}

\bibitem[{Allen et~al(2012)Allen, Anderson, Brady, Brown, and
	Creighton}]{Allen:2005fk}
Allen B, Anderson WG, Brady PR, et~al (2012) {FINDCHIRP: An Algorithm for
	detection of gravitational waves from inspiraling compact binaries}. Phys Rev
D85:122006. \doi{10.1103/PhysRevD.85.122006},
{\href{https://arxiv.org/abs/gr-qc/0509116}{{arXiv:gr-qc/0509116}}} {[gr-qc]}

\bibitem[{Ashby(2003)}]{Ashby:2003vja}
Ashby N (2003) {Relativity in the Global Positioning System}. Living Rev Rel
6:1. \doi{10.12942/lrr-2003-1}

\bibitem[{Bacon et~al(2023)Bacon, Trovato, and Bejger}]{Bacon:2022lsm}
Bacon P, Trovato A, Bejger M (2023) {Denoising gravitational-wave signals from
	binary black holes with a dilated convolutional autoencoder}. Mach Learn Sci
Tech 4(3):035024. \doi{10.1088/2632-2153/acd90f},
{\href{https://arxiv.org/abs/2205.13513}{{arXiv:2205.13513}}} {[gr-qc]}

\bibitem[{Bartos et~al(2010)Bartos, Bork, Factourovich, Heefner, Marka, Marka,
	Raics, Schwinberg, and Sigg}]{Bartos:2010zz}
Bartos I, Bork R, Factourovich M, et~al (2010) {The Advanced LIGO timing
	system}. Class Quant Grav 27:084025. \doi{10.1088/0264-9381/27/8/084025}

\bibitem[{Bradt(2004)}]{bradt2004astronomy}
Bradt H (2004) Astronomy Methods: A Physical Approach to Astronomical
Observations. Cambridge Planetary Science, Cambridge University Press

\bibitem[{Casella and Berger(2002)}]{Casella2002}
Casella G, Berger RL (2002) {Statistical Inference}, 2nd edn. {DUXBURY}

\bibitem[{Chatterjee and Jani(2024)}]{Chatterjee:2024alf}
Chatterjee C, Jani K (2024) {Reconstruction of Binary Black Hole Harmonics in
	LIGO Using Deep Learning}. Astrophys J 969(1):25.
\doi{10.3847/1538-4357/ad4602},
{\href{https://arxiv.org/abs/2403.01559}{{arXiv:2403.01559}}} {[gr-qc]}

\bibitem[{Chatterjee et~al(2023)Chatterjee, Kovalam, Wen, Beveridge,
	Diakogiannis, and Vinsen}]{Chatterjee:2022ggk}
Chatterjee C, Kovalam M, Wen L, et~al (2023) {Rapid Localization of
	Gravitational Wave Sources from Compact Binary Coalescences Using Deep
	Learning}. Astrophys J 959(1):42. \doi{10.3847/1538-4357/ad08b7},
{\href{https://arxiv.org/abs/2207.14522}{{arXiv:2207.14522}}} {[gr-qc]}

\bibitem[{Chatterji et~al(2006)Chatterji, Lazzarini, Stein, Sutton, Searle, and
	Tinto}]{Chatterji:2006nh}
Chatterji S, Lazzarini A, Stein L, et~al (2006) {Coherent network analysis
	technique for discriminating gravitational-wave bursts from instrumental
	noise}. Phys Rev D 74:082005. \doi{10.1103/PhysRevD.74.082005},
{\href{https://arxiv.org/abs/gr-qc/0605002}{{arXiv:gr-qc/0605002}}}

\bibitem[{Cohen et~al(1992)Cohen, Daubechies, and
	Feauveau}]{cohen1992biorthogonal}
Cohen A, Daubechies I, Feauveau JC (1992) Biorthogonal bases of compactly
supported wavelets. Communications on pure and applied mathematics
45(5):485--560

\bibitem[{Cornish and Littenberg(2015)}]{Cornish:2014kda}
Cornish NJ, Littenberg TB (2015) {BayesWave: Bayesian Inference for
	Gravitational Wave Bursts and Instrument Glitches}. Class Quant Grav
32(13):135012. \doi{10.1088/0264-9381/32/13/135012},
{\href{https://arxiv.org/abs/1410.3835}{{arXiv:1410.3835}}} {[gr-qc]}

\bibitem[{Cutler and Flanagan(1994)}]{Cutler:1994ys}
Cutler C, Flanagan EE (1994) {Gravitational waves from merging compact
	binaries: How accurately can one extract the binary's parameters from the
	inspiral wave form?} Phys Rev D49:2658--2697. \doi{10.1103/PhysRevD.49.2658},
{\href{https://arxiv.org/abs/gr-qc/9402014}{{arXiv:gr-qc/9402014}}} {[gr-qc]}

\bibitem[{Donoho and Johnstone(1994)}]{Donoho:1994}
Donoho DL, Johnstone IM (1994) {Ideal spatial adpatation by wavelet shrinkage}.
Biometrika 81(3)(3):425--455. \doi{10.1093/biomet/81.3.425}

\bibitem[{Eardley et~al(1973{\natexlab{a}})Eardley, Lee, and
	Lightman}]{Eardley:1974nw}
Eardley DM, Lee DL, Lightman AP (1973{\natexlab{a}}) {Gravitational-wave
	observations as a tool for testing relativistic gravity}. Phys Rev
D8:3308--3321. \doi{10.1103/PhysRevD.8.3308}

\bibitem[{Eardley et~al(1973{\natexlab{b}})Eardley, Lee, Lightman, Wagoner, and
	Will}]{Eardley:1973br}
Eardley DM, Lee DL, Lightman AP, et~al (1973{\natexlab{b}}) {Gravitational-wave
	observations as a tool for testing relativistic gravity}. Phys Rev Lett
30:884--886. \doi{10.1103/PhysRevLett.30.884}

\bibitem[{Einstein(1911)}]{Einstein11a}
Einstein A (1911) {{\"U}ber den Einflu{\ss} der Schwerkraft auf die Ausbreitung
	des Lichtes}. Annalen der Physik 340(10):898--908

\bibitem[{Fairhurst(2009)}]{Fairhurst:2009}
Fairhurst S (2009) {Triangulation of gravitational wave sources with a network
	of detectors}. New J of Physics 11:123006.
\doi{10.1088/1367-2630/11/12/123006},
{\href{https://arxiv.org/abs/0908.2356}{{arXiv:0908.2356}}} {[gr-qc]}

\bibitem[{Fairhurst(2011)}]{Fairhurst:2010is}
Fairhurst S (2011) {Source localization with an advanced gravitational wave
	detector network}. Class Quant Grav 28:105021.
\doi{10.1088/0264-9381/28/10/105021},
{\href{https://arxiv.org/abs/1010.6192}{{arXiv:1010.6192}}} {[gr-qc]}

\bibitem[{Ferguson(1967)}]{Ferguson67}
Ferguson TS (1967) Mathematical Statistics: A Decision Theoretic Approach.
Academic Press

\bibitem[{Geroch et~al(1973)Geroch, Held, and Penrose}]{Geroch73}
Geroch R, Held A, Penrose R (1973) A space-time calculus based on pairs of null
directions. J Math Phys 14:874--881

\bibitem[{Goldstein et~al(2017)}]{Fermi-GBM:2017soa}
Goldstein A, et~al (2017) {Fermi Observations of the LIGO Event GW170104}.
Astrophys J Lett 846(1):L5. \doi{10.3847/2041-8213/aa8319},
{\href{https://arxiv.org/abs/1706.00199}{{arXiv:1706.00199}}} {[astro-ph.HE]}

\bibitem[{G\"ursel and Tinto(1989)}]{Gursel:1989}
G\"ursel Y, Tinto M (1989) {Near optimal solution to the inverse problem for
	gravitational-wave bursts}. PhysRevD 40(12):3884--3938

\bibitem[{Helstrom(1975)}]{Helstrom75}
Helstrom CW (1975) Statistical theory of signal detection, 2nd edn. Pergamon
Press

\bibitem[{Hu et~al(2021)Hu, Zhou, Peng, Wen, Chu, and Kovalam}]{Hu:2021nvy}
Hu Q, Zhou C, Peng JH, et~al (2021) {Semianalytical approach for sky
	localization of gravitational waves}. Phys Rev D 104(10):104008.
\doi{10.1103/PhysRevD.104.104008},
{\href{https://arxiv.org/abs/2110.01874}{{arXiv:2110.01874}}} {[gr-qc]}

\bibitem[{Jaranowski and Krolak(2009)}]{Jaranowski:2009zz}
Jaranowski P, Krolak A (2009) {Analysis of gravitational-wave data}. Cambridge
Univ. Press, Cambridge, \doi{10.1017/CBO9780511605482}

\bibitem[{Klimenko et~al(2011)Klimenko, Vedovato, Drago, Mazzolo, Mitselmakher,
	Pankow, Prodi, Re, Salemi, and Yakushin}]{Klimenko:2011hz}
Klimenko S, Vedovato G, Drago M, et~al (2011) {Localization of gravitational
	wave sources with networks of advanced detectors}. Phys Rev D 83:102001.
\doi{10.1103/PhysRevD.83.102001},
{\href{https://arxiv.org/abs/1101.5408}{{arXiv:1101.5408}}} {[astro-ph.IM]}

\bibitem[{Klimenko et~al(2016)}]{Klimenko:2015ypf}
Klimenko S, et~al (2016) {Method for detection and reconstruction of
	gravitational wave transients with networks of advanced detectors}. Phys Rev
D 93(4):042004. \doi{10.1103/PhysRevD.93.042004},
{\href{https://arxiv.org/abs/1511.05999}{{arXiv:1511.05999}}} {[gr-qc]}

\bibitem[{Kolmus et~al(2022)Kolmus, Baltus, Janquart, van Laarhoven, Caudill,
	and Heskes}]{Kolmus:2021buf}
Kolmus A, Baltus G, Janquart J, et~al (2022) {Fast sky localization of
	gravitational waves using deep learning seeded importance sampling}. Phys Rev
D 106(2):023032. \doi{10.1103/PhysRevD.106.023032},
{\href{https://arxiv.org/abs/2111.00833}{{arXiv:2111.00833}}} {[gr-qc]}

\bibitem[{Lee et~al(2019)Lee, Gommers, Waselewski, Wohlfahrt, and
	O’Leary}]{Lee:2019}
Lee GR, Gommers R, Waselewski F, et~al (2019) {PyWavelets: A Python package for
	wavelet analysis}. Journal of Open Source Software 4(36)(1237):1--2.
\doi{10.21105/joss.01237}

\bibitem[{LIGO(2017)}]{LIGO_loc_GW170104}
LIGO (2017) {LIGO Document T1700179-v1, GW170104 sky localization}.
\url{https://dcc.ligo.org/LIGO-T1700179/public}, [Online; accessed
28-May-2024]

\bibitem[{Mallat(2009)}]{Mallat2009}
Mallat S (2009) {A Wavelet Tour of Signal Processing: The Sparse Way}. Elsevier
Inc.

\bibitem[{McDonough and Whalen(1995)}]{McDonough1995}
McDonough RN, Whalen AD (1995) {Detection of Signals in Noise}, 2nd edn.
Academic Press

\bibitem[{Moreschi(1987)}]{Moreschi87}
Moreschi OM (1987) General future asymptotically flat spacetimes. Class Quantum
Grav 4:1063--1084

\bibitem[{Moreschi(2004)}]{Moreschi04}
Moreschi OM (2004) Intrinsic angular momentum and center of mass in general
relativity. ClassQuantum Grav 21:5409--5425.
\doi{10.1088/0264-9381/21/23/008}

\bibitem[{Moreschi(2019)}]{Moreschi:2019vxw}
Moreschi OM (2019) {Convenient filtering techniques for LIGO strain of the
	GW150914 event}. JCAP 1904:032. \doi{10.1088/1475-7516/2019/04/032},
{\href{https://arxiv.org/abs/1903.00546}{{arXiv:1903.00546}}} {[gr-qc]}

\bibitem[{Moreschi(2024)}]{Moreschi:2024njx}
Moreschi OM (2024) {Comparison of unknown gravitational-wave signals in two
	detectors}. Astrophys Space Sci 369(1):12. \doi{10.1007/s10509-024-04276-9}

\bibitem[{Moreschi(2025)}]{Moreschi:2025zde}
Moreschi OM (2025) {Localization of GW190521 and reconstruction of the spin-2
	gravitational-wave polarization modes}
{\href{https://arxiv.org/abs/2504.00207}{{arXiv:2504.00207}}} {[gr-qc]}

\bibitem[{Nishizawa et~al(2009)Nishizawa, Taruya, Hayama, Kawamura, and
	Sakagami}]{Nishizawa:2009bf}
Nishizawa A, Taruya A, Hayama K, et~al (2009) {Probing non-tensorial
	polarizations of stochastic gravitational-wave backgrounds with ground-based
	laser interferometers}. Phys Rev D 79:082002.
\doi{10.1103/PhysRevD.79.082002},
{\href{https://arxiv.org/abs/0903.0528}{{arXiv:0903.0528}}} {[astro-ph.CO]}

\bibitem[{Penrose and Rindler(1984)}]{Penrose84}
Penrose R, Rindler W (1984) Spinors and Space-Time, vol~1. Cambridge University
Press, Cambridge

\bibitem[{Peters(1964)}]{Peters:1964zz}
Peters PC (1964) {Gravitational Radiation and the Motion of Two Point Masses}.
Phys Rev 136:B1224--B1232. \doi{10.1103/PhysRev.136.B1224}

\bibitem[{Petit et~al(2010)Petit, Luzum, and (Eds.)}]{IERS_Conventions_2010}
Petit G, Luzum B, (Eds.) (2010) {International Earth Rotation and Reference
	Systems Service (IERS)}.
\urlprefix\url{https://iers-conventions.obspm.fr/content/tn36.pdf}, technical
Note No. 36

\bibitem[{Pirani(1965)}]{Pirani64}
Pirani FAE (1965) Introduction to gravitational radiation theory. In: Trautman
A, Pirani FAE, Bondi H (eds) Brandeis Summer Institute in Theoretical Physics
1964, Volume One: Lectures on general relativity. Prentice-Hall Inc., p
249--373

\bibitem[{Poisson and Will(2014)}]{Poisson2014}
Poisson E, Will CM (2014) Gravity: Newtonian, Post-Newtonian, Relativistic.
Cambridge Univeristy Press

\bibitem[{Savchenko et~al(2017)}]{Savchenko:2017dyg}
Savchenko V, et~al (2017) {INTEGRAL observations of GW170104}. Astrophys J Lett
846(2):L23. \doi{10.3847/2041-8213/aa87ae},
{\href{https://arxiv.org/abs/1707.03719}{{arXiv:1707.03719}}} {[astro-ph.HE]}

\bibitem[{Singer and Price(2016)}]{Singer:2015ema}
Singer LP, Price LR (2016) {Rapid Bayesian position reconstruction for
	gravitational-wave transients}. Phys Rev D 93(2):024013.
\doi{10.1103/PhysRevD.93.024013},
{\href{https://arxiv.org/abs/1508.03634}{{arXiv:1508.03634}}} {[gr-qc]}

\bibitem[{Stalder et~al(2017)}]{Stalder:2017qic}
Stalder B, et~al (2017) {Observations of the GRB afterglow ATLAS17aeu and its
	possible association with GW170104}. Astrophys J 850(2):149.
\doi{10.3847/1538-4357/aa95c1},
{\href{https://arxiv.org/abs/1706.00175}{{arXiv:1706.00175}}} {[astro-ph.HE]}

\bibitem[{Takeda et~al(2018)Takeda, Nishizawa, Michimura, Nagano, Komori, Ando,
	and Hayama}]{Takeda:2018uai}
Takeda H, Nishizawa A, Michimura Y, et~al (2018) {Polarization test of
	gravitational waves from compact binary coalescences}. Phys Rev
D98(2):022008. \doi{10.1103/PhysRevD.98.022008},
{\href{https://arxiv.org/abs/1806.02182}{{arXiv:1806.02182}}} {[gr-qc]}

\bibitem[{Touboul et~al(2022)}]{Touboul:2022yrw}
Touboul P, et~al (2022) {Result of the MICROSCOPE weak equivalence principle
	test}. Class Quant Grav 39(20):204009. \doi{10.1088/1361-6382/ac84be},
{\href{https://arxiv.org/abs/2209.15488}{{arXiv:2209.15488}}} {[gr-qc]}

\bibitem[{Tsutsui et~al(2021)Tsutsui, Cannon, and Tsukada}]{Tsutsui:2020sml}
Tsutsui T, Cannon K, Tsukada L (2021) {High speed source localization in
	searches for gravitational waves from compact object collisions}. Phys Rev D
103(4):043011. \doi{10.1103/PhysRevD.103.043011},
{\href{https://arxiv.org/abs/2005.08163}{{arXiv:2005.08163}}} {[astro-ph.HE]}

\bibitem[{Veitch et~al(2015)}]{Veitch:2014wba}
Veitch J, et~al (2015) {Parameter estimation for compact binaries with
	ground-based gravitational-wave observations using the LALInference software
	library}. Phys Rev D 91(4):042003. \doi{10.1103/PhysRevD.91.042003},
{\href{https://arxiv.org/abs/1409.7215}{{arXiv:1409.7215}}} {[gr-qc]}

\bibitem[{Verrecchia et~al(2017)}]{AGILE:2017vou}
Verrecchia F, et~al (2017) {AGILE Observations of the Gravitational Wave Source
	GW170104}. Astrophys J Lett 847(2):L20. \doi{10.3847/2041-8213/aa8224},
{\href{https://arxiv.org/abs/1706.00029}{{arXiv:1706.00029}}} {[astro-ph.HE]}

\bibitem[{Wald(1984)}]{Wald84}
Wald R (1984) General Relativity. The Chicago University Press

\bibitem[{Wen and Chen(2010)}]{Wen:2010cr}
Wen L, Chen Y (2010) {Geometrical Expression for the Angular Resolution of a
	Network of Gravitational-Wave Detectors}. Phys Rev D 81:082001.
\doi{10.1103/PhysRevD.81.082001},
{\href{https://arxiv.org/abs/1003.2504}{{arXiv:1003.2504}}} {[astro-ph.CO]}

\bibitem[{You et~al(2021)You, Ashton, Zhu, Thrane, and Zhu}]{You:2021eeq}
You ZQ, Ashton G, Zhu XJ, et~al (2021) {Optimized localization for
	gravitational waves from merging binaries}. Mon Not Roy Astron Soc
509(3):3957--3965. \doi{10.1093/mnras/stab2977},
{\href{https://arxiv.org/abs/2105.04263}{{arXiv:2105.04263}}} {[gr-qc]}
	
\end{thebibliography}
%
%

\end{document}